\documentclass[twocolumn,final,traditabstract]{aa}
\usepackage[utf8]{inputenc} 
\usepackage{multirow}
\usepackage{verbatimbox}

\usepackage{amsmath,amsfonts,amssymb}
\usepackage{natbib}
\usepackage{graphicx}
\usepackage{subfig}
\usepackage{empheq}
\usepackage{color}
\usepackage{enumitem}
\usepackage[linktocpage=true,colorlinks=true,linkcolor=blue,citecolor=blue,urlcolor=blue]{hyperref}

\definecolor{mygreen}{RGB}{34,139,34}
\definecolor{myviolet}{RGB}{90,0,83}
\graphicspath{{figs/}}

\begin{document}

\title{Dependence of Type Ia supernova luminosities on their
  local environment} \titlerunning{Dependence of Type Ia
  supernova luminosities on their local environment}
\authorrunning{M.~Roman {\it et al.}}

%\subtitle{Version 1.4-3}
\author{
M.~Roman\inst{1},
D.~Hardin\inst{1},
M.~Betoule\inst{1},
P.~Astier\inst{1},
C.~Balland\inst{1},
R.~S.~Ellis\inst{2,3},
S.~Fabbro\inst{4},
J.~Guy\inst{1},
I.~Hook\inst{5},
D.~A.~Howell\inst{6,7},
C.~Lidman\inst{8},
A.~Mitra\inst{1},
A.~M\"{o}ller\inst{9,10},
A.~M.~Mour\~{a}o\inst{11},
J.~Neveu\inst{12},
N.~Palanque-Delabrouille\inst{13},
C.~J.~Pritchet\inst{14},
N.~Regnault\inst{1},
V.~Ruhlmann-Kleider\inst{13},
C.~Saunders\inst{1,15},
M.~Sullivan\inst{16}
}

\institute{
{\tiny\textit{Sorbonne Universit\'e, Universit\'e Paris Diderot, CNRS/IN2P3, Laboratoire de Physique Nucl\'eaire et de Hautes Energies, LPNHE, 4 Place Jussieu, F-75252 Paris, France\goodbreak
\and
European Southern Observatory (ESO), Karl-Schwarzschild Strasse 2, 85748 Garching, Germany\goodbreak
\and
Department of Physics and Astronomy, University College London, Gower Street, London, WC1E 6BT, UK\goodbreak
\and
NRC Herzberg Institute for Astrophysics, 5071 West Saanich Road, Victoria V9E 2E7, British Columbia, Canada
\and
Department of Physics, Lancaster University, Lancaster, Lancs, LA1 4YB, U.K.\goodbreak
\and
Las Cumbres Observatory Global Telescope Network, 6740 Cortona
Dr., Suite 102, Goleta, CA 93117\goodbreak
\and
Physics Department, University of California, Santa Barbara, CA 93106, USA\goodbreak
\and
Australian Astronomical Observatory, PO Box 915, North Ryde,
NSW 1670, Australia
\and
Research School of Astronomy and Astrophysics, Australian National University, Canberra, Australia\goodbreak
\and
ARC Centre of Excellence for All-sky Astrophysics (CAASTRO), Canberra, Australia\goodbreak
\and
CENTRA e Departamento de F\'isica, Instituto Superior T\'ecnico, ULisboa, Avenida Rovisco Pais, 1049-001 Lisboa, Portugal\goodbreak
\and
LAL, Univ. Paris-Sud, CNRS/IN2P3, Universit\'e Paris-Saclay, Orsay, France\goodbreak
\and
IRFU, CEA, Universit\'e Paris-Saclay, F-91191 Gif-sur-Yvette, France\goodbreak
\and
Department of Physics and Astronomy, University of Victoria
\and
Sorbonne Universit\'e, Institut Lagrange de Paris (ILP), 98 bis Boulevard Arago, 75014 Paris, France\goodbreak
\and
Department of Physics and Astronomy, University of Southampton, Southampton, SO17 1BJ, UK
}}
}

%\offprints{toto\@@titi.com}

\date{Received Mont 06, 2017; accepted Mont DD, YYYY}

\abstract {
%We present fully consistent catalogs of Type Ia supernov\ae\ (SNIa) local and global host galaxy properties across the whole Hubble diagram and corresponding to the final SNLS 5 year sample.
We present a fully consistent catalog of local and global properties of host galaxies of 882 Type Ia supernov\ae\ (SNIa)
that were selected based on their light-curve properties,
spanning the redshift range $0.01 < z < 1.\text{}$   
This catalog
corresponds to a preliminary version of the compilation sample and includes Supernova Legacy Survey (SNLS) 5-year data, Sloan Digital Sky Survey (SDSS), and low-redshift surveys.
We measured low- and moderate-redshift host galaxy photometry in SDSS stacked and single-epoch images and used spectral energy distribution (SED) fitting techniques to derive host properties such as stellar mass and $U-V$ rest-frame colors; the latter are an indicator of the luminosity-weighted age of the stellar population in a galaxy. We combined these results with high-redshift host photometry from the SNLS survey and thus obtained a consistent catalog of host stellar masses and colors across a wide redshift range. We also estimated the local observed fluxes at the supernova location within a  proper distance radius of 3 kpc,
corresponding to the SNLS imaging resolution,
and transposed them into local $U-V$ rest-frame colors. This is the first time that local environments surrounding SNIa have been measured at redshifts spanning the entire Hubble diagram.
Selecting SNIa based on host photometry quality, we then performed cosmological fits using local color as a third standardization variable, for which we split the sample at the median value. We find a local color step significance of $-0.091\pm0.013$ mag ($7\sigma$), which effect is as significant as the maximum mass step effect.
This indicates that the remaining luminosity variations in SNIa samples can be reduced with a third standardization variable
that takes the environment into account.
Correcting for the maximum mass step correction of $-0.094\pm0.013$ mag, we find a local color effect of $-0.057\pm0.012$ mag ($5\sigma$), which shows that additional information is provided by the close environment of SNIa. Departures from the initial choices were investigated and showed that the local color effect is still present, although less pronounced.
%We find that local $U-V$ color 
%brightness dependence is at least as significant as for
%properties of the host galaxy as a whole (host stellar mass or global $U-V$ rest-frame color). Once selection requirements are chosen, we perform cosmological fits using local color as a third standardization variable and find its significance at the level of $7\sigma$, indicating that the remaining luminosity variations in SNIa samples can be reduced using a third standardization variable
%taking the local environment into account.
%Correcting for mass step, we find a remaining $4.6\sigma$ correlation with the corresponding Hubble diagram residuals. It is worth noting that a correction for a local $U-V$ color step also yields a small correlation of residuals with host stellar mass ($2.3\sigma$).
%Moreover, Hubble diagram residuals
%after regular standardization have a stronger correlation
%with local color than any other variable we measured.
We discuss the possible implications for cosmology and find that using the local color in place of the stellar mass results in a change in the measured value of the dark energy equation-of-state parameter of 0.6\%. Standardization using local $U-V$ color in addition to stretch and color reduces the total dispersion in the Hubble diagram from 0.15 to 0.14 mag.
This will be of tremendous importance for the forthcoming SNIa surveys, and in particular for the Large Synoptic Survey Telescope (LSST), for which
uncertainties on the dark energy equation of state will be comparable to the effects reported here.% in this work.
%$w(z)$ will be accessible in bins of redshift with a constraining power several orders of magnitude greater than that of current surveys.
}
% {intro} {methods} {results}  {}
%\keywords{image processing: photometry lightcurves -- image processing: astrometry}
%}
\maketitle
%\linenumbers
%\switchlinenumbers
%\setpagewiselinenumbers
%\tableofcontents
%
%________________________________________________________________
\section{Introduction}
\label{sec:introduction}

The first Hubble diagrams that were constructed using both high- and low-redshift Type Ia supernov\ae\ (SNIa) revealed that the expansion of the Universe recently entered a 
phase of acceleration \citep{Perlmutter99,Riess98,Schmidt98}, which is inconsistent with a matter-dominated Universe, and which in the simplest cosmology model requires the existence of a now dominant, supplementary component that is dubbed ``dark energy''.
%In an effort to further constrain the underlying measurement of this acceleration, dubbed ``dark energy'', other cosmological probes have also measured this effect.

The past decade saw the confirmation of this effect by the combination of different cosmological probes based on large-scale structures, at the recombination epoch, or later on through Cosmic Microwave Background (CMB) emission, Baryonic Acoustic Oscillations (BAO), and lensing measurements. The deeper, better-quality supernova samples of increasing size still provide the most sensitive probe for the nature of dark energy, which is phenomenologically described as a perfect fluid with the equation of state $w=p/(\rho c^2)$.

%A number of supernov\ae\ experiments have also reproduced these results with increasing precision,
The Supernova Legacy Survey (SNLS) has
regularly published constraints on the dark energy fluid characterization as it
updated both its data sample and analysis methods
\citep{Astier06,Guy10,Sullivan11,Conley11}. The latest such release
was a joint analysis between the SNLS and ``Sloan Digital Sky Survey''
(SDSS) teams that is called the ``Joint Light-curve Analysis'' \citep[JLA,][]{Betoule14}, which provided a 6\% measurement
of the dark energy equation-of-state parameter.

%Type Ia supernov\ae\ have then long proved to be a successful probe of dark energy thanks to their property of \textit{standardizable} candle.
The success of the Type Ia supernov\ae\  as a cosmological probe is based on them being {\em{standardizable}} candles.
The standardization process relies on correcting the supernova rest-frame B-peak magnitude using the empirical light-curve width (brighter-slower effect, or stretch), with relations shown by \cite{pskovskii77,pskovskii84}, \cite{rust74} and \cite{phillips93}. Later on, color-dependence laws (brighter-bluer relation) were established by \cite{hamuy96_calantololo} and \cite{tripp98}. This correction process, as well as similar approaches, reduces the {\em {standard}} SNIa peak magnitude dispersion to $\sim0.15$ mag. However, the
remaining dispersion in the SNIa population reflects the fact that the
standardization procedure does not entirely capture the physical
processes at play during the explosion and leaves room for
at least one more
variable that correlates with supernova brightness. Our theoretical
understanding suggests that a thermonuclear explosion of a
carbon-oxygen white dwarf that accretes mass from a companion star or from another white dwarf and reaches
the Chandrasekhar limit gives rise to SNIa \citep{Iben84,Whelan73}. The exact
mechanism for how the progenitor system accretes mass is still poorly
known and could be different between galaxy types or local stellar
populations.

Lately, numerous studies on a few individual Type Ia supernov\ae\
gave birth to different conclusions regarding the type of stellar
progenitor. {\em{Hubble Space Telescope}} (HST) imaging of an SNIa
remnant in the Large Magellanic Cloud \citep{Schaefer12} strongly
favored a progenitor system consisting of two white dwarf stars. This
is known as the {\em{double-degenerate}} scenario \citep{Iben84}. It has received further support by recent simulations \citep{pakmor12,kromer12} and by various analyses. The recent nearby
SNIa 2011fe observed in M101 \citep{Nugent11} fits this scenario. The analysis of archival HST imaging ruled out any kind of
main-sequence companion star. However, it has been established that other individual SNIa have
interacted with circumstellar material \citep[for example,][]{gonzalez12}, thus favoring the
{\em{single-degenerate}} scenario of a low-mass main-sequence star
accompanying a white dwarf star \citep{Whelan73}. In reality, it is probable that a large variety of explosion mechanisms exists
that defines a continuum between the single- and double-degenerate models.

This debate on the
progenitor type and explosion mechanism for a few individual supernov\ae\ 
is one of the concerns calling for further investigations of the luminosity dependence of SNIa. Other concerns include metallicity effects through the nickel mass that is synthesized in the explosion \citep{Hoeflich98,Dominguez01,Timmes03,morenoraya16,piersanti17} or the varying properties of the dust surrounding the explosion region. One way to answer the question of the progenitor type is a
statistical study of host galaxy properties, such as stellar age or
metallicity, in order to address the question of the progenitor
influence on the SNIa luminosity \citep[e.g.,][]{foley13}. Although the evidence of cosmic
acceleration is not questioned, this might affect the cosmological
parameters measurement and bias the estimation of the dark energy
equation-of-state $w$ and its redshift evolution.

All this strongly motivates the search for correlations between SNIa characteristics and the
properties of their parent galaxies.
The SNIa progenitor system and explosion scenarii can be constrained by studying the supernova environment:  \cite{mannucci05,sullivan06} showed evidence that Type Ia supernov\ae\ explosions occur more often in young environments.
Early studies also found  significant evidence that the SNIa intrinsic  luminosity could vary with galaxy morphology. \cite{hamuy00}
confirmed this trend for a sample of 44 galaxies that hosted the best observed low-redshift ($z<0.1$) SNIa. With a larger sample spanning a wider range of redshifts, \cite{sullivan06} supported this result. 
Significant correlations were soon brought to evidence between SNIa light-curve parameters and luminosities and the properties of their hosts \citep{Hamuy1996_abs_luminosities,hamuy00,gallagher05, gallagher08,Howell09,Neill09}: bluer, more slowly declining, and thus intrinsically brighter supernov\ae\ occurred in galaxies
with a lower mass, and also in bluer galaxies, in galaxies with a stronger star formation or in galaxies with a lower metallicity. These results have  been confirmed as the SNIa sample increased in size and quality.

SNIa  spectroscopic properties were also studied regarding the host photometric and/or spectroscopic properties: in the framework of the SNLS survey, \cite{Bronder08} and \cite{Balland09} pointed out the correlation between  the equivalent width of the SiII feature at $\sim 4000$~\AA\ and the host type, which is similar at low and high redshift.
Correlations of low-redshift SNIa properties with host spectroscopic properties were studied in \cite{pan15}, showing evidence that high-velocity SNIa preferably explode in more massive galaxies and in the inner region of their host galaxies. SNIa with a stronger high-velocity component of the Ca II near-IR absorption are located in low-mass star-forming hosts, which suggests a young progenitor population.

The observed differences in SNIa intrinsic luminosities with respect to the host caracteristics, however, was thought to be captured by the slower-brighter relations, so that the {\em{standardized}} supernova-{\em{corrected}} luminosity -- and the use of these supernovae as standard candle -- remained unaffected \citep{Perlmutter99,riess99_cfa1,sullivan03}.
By refining the analysis
techniques and increasing the size
and quality of their data sample, various studies were finally able to address the question of the variation of the {\em standardized} SNIa luminosity.
%giving birth to different conclusions depending on the light-curve correction.
\cite{hicken09b}
exploited the CfAIII sample to highlight  that the {\em{standardized}} SNIa luminosity when corrected for stretch and color dependence is brighter in elliptical galaxies than those in spiral galaxies at the 2$\sigma$ level. Note that SNIa are on the whole fainter in elliptical galaxies because their stretch is smaller.
Using a sample of about
166 nearby SNIa, \cite{Neill09}, hereafter N09, found evidence that
events exploding in star-forming galaxies decline more slowly and are thus intrinsically brighter, and that fainter, faster declining SNIa preferentially occur in more massive hosts with presumably higher metallicity, which confirmed the results of \cite{hamuy00} and \cite{sullivan06}. N09 also demonstrated that when the supernov\ae\ peak brightnesses were corrected for stretch and color, the {\em standardized} SNIa in older hosts were brighter. This observed residual trend corroborates the conclusions of \cite{hicken09b}. 
Similar behaviors have also been obtained with the whole SDSS-II supernova survey \citep{Lampeitl10}. Concerning the low-redshift restricted SDSS-II subset, \cite{konishi11} and \cite{DAndrea11} obtained that after correction for stretch and color, the {\em {standardized}} SNIa are $\sim0.1$ magnitude brighter in hosts with higher metallicity. In these analyses, SNIa are also found to be fainter in passive hosts before any brightness correction, but a residual trend suggests that they are fainter in active hosts after corrections
for stretch and color at a significance of between 2 and $3\sigma$.
%One possible explanation for this is that younger hosts contain more dust than older ones.

Several analyses also tried to correlate SNIa luminosities with other properties of their host galaxies, such as the galaxy stellar mass, which was directly estimated as a function of the galaxy luminosity. 
Considering nearby SNIa, \cite{Kelly10}
computed a positive correlation at $2.5\sigma$ significance between the observed supernova magnitude residuals on the Hubble diagram (where the laws for color and stretch correction were fitted along with the cosmological model) and the stellar masses of
the hosts: the corrected luminosity was 10\% brighter for SNIa that were
located in massive parent galaxies.
Furthermore, for the first time, a sample that included distant SNIa
observed by the SNLS after a three-year survey and intermediate-redshift events has been gathered and
analyzed by \cite{Sullivan10}. Completing the sample with nearby SNIa
from N09, this work found at a $4\sigma$ confidence level that SNIa exploding in
massive and passive galaxies are on average brighter after stretch and color correction are applied. A similar trend
has been observed using the SDSS-II supernova survey for SNIa at
intermediate redshift \citep{Gupta11}.  The most recent release of
dark energy equation-of-state parameter constraints using SNIa in
\cite{Betoule14}, simultaneously obtained with distant (SNLS),
intermediate (SDSS), and nearby events, showed a non-zero correlation at $5\sigma$ confidence level between observed Hubble diagram residuals and the stellar
mass of host galaxies. Based on a complete SDSS-II Supernova Survey of
345 SNIa, 
\cite{wolf16}  recovered the now well-known relation between Hubble residuals and host-galaxy mass and found a $3.6\sigma$ significance of a non-zero linear slope. They also confirmed the correlations between Hubble residuals and host-galaxy gas-phase metallicity and specific star formation rate, estimated through spectroscopy of the host galaxy.

The galaxy stellar mass is strongly correlated with many other
properties of the stellar population (age, stellar formation rate and history,
metallicity, and color) and is likely to  act as a
proxy for other local galactic characteristics. This idea has been
exploited with SNIa located in the local Universe by
\cite{Childress13}. Examining the different physical effects that
could drive the observed trend of Hubble diagram residuals with host
properties, they showed that metallicity and progenitor age are the most
probable factors. A direct hint was provided by results from \cite{Rigault13}, hereafter R13, which were obtained using observations from the
Nearby Supernova Factory: they  showed evidence that SNIa standardized magnitudes depend on the star
formation of the supernova environment within a radius of 1 kpc,
as traced by H$\alpha$ surface brightness. This study established that SNIa
exploding in locally passive environments are on average brighter, after correction, than
in locally active star-forming regions at $3\sigma$ significance.
\cite{rigault15}, hereafter R15, exploited the nearby SNIa from the Constitution data set compiled by \cite{hicken09b} and the Far Ultra-Violet (FUV) flux  measured by the Galaxy Evolution Explorer satellite (GALEX) as a star formation indicator within a few kiloparsec of SNIa positions: they  upheld their conclusion that  SNIa from locally star-forming environments are fainter after standardization than those from locally passive environments.
%% and consequently revised \cite{Riess 2011} $H_0$ value, obtaining  a compatible at the 1-$\sigma$ level with the indirect measurement of the Hubble constant for a flat $\Lambda$CDM cosmology from \cite{Planck2014}. 
Conversely, however, with a significant increase of the SNIa sample based on a mixture of
hosts whose distances come from both JLA and HST and using different
selection criteria, \cite{jones15} found little
evidence for a star-formation bias.

Studying the dependence of SNIa luminosities on their local
environment involves a difficult measurement that depends on the resolution
and quality of the available galaxy data and is subject to systematic errors whose
characterization is challenging. As cosmological consequences are important (measurement
of the present expansion rate of the Universe, nature of dark energy),
it is fundamental to shed new light on the environmental dependence
of supernova luminosity.

We here report on measuring the local environment of a large sample of SNIa over the whole Hubble diagram and compare the results with the properties of the host galaxies as a whole.
We demonstrate, as claimed by R13, that local information is crucial and constitutes an important systematic effect that needs
to be taken into account in on-going and future surveys.
Our study is based on a composite sample of about 900 classified Type Ia
supernov\ae\ that all have high-quality multiband light curves and spectroscopic redshift spanning a wide range ($0.01 < z < 1.1$). We were
able to gather accurate photometry of both host galaxies and local SNIa environment of the whole sample. For the high-redshift end, we took advantage
of the good image quality of the Canada-France-Hawa\"{i} Telescope (CFHT) imaging that yielded the SNLS
sample.
We first describe our supernova sample and the global and local
photometry of associated host galaxies and environments in Sections~\ref{sec:data} and \ref{sec:photo}. The conversion of the raw
photometric measurements into rest-frame quantities and global
host galaxy properties is detailed in Section~\ref{sec:mass}, together with the presentation of our host property catalog and a direct comparison with published results. We then
proceed to extend previous analyses of correlations between Hubble
residuals and environment properties using our diverse and larger sample in Section~\ref{sec:analysis}. In Section~\ref{sec:robust} we assess the robustness of the results. Finally, we summarize our results and discuss cosmological implications in Section~\ref{sec:discussion}.

% In this paper, we present a first consistent catalog of host galaxy
% {\em{global}} (the galaxy as a whole) and {\em{local}} (region
% close to the explosion location) properties corresponding to the new
% SNLS 5 years release. We extend the previous analyses to a much wider
% redshift range and show correlations between host variables and Hubble
% diagram residuals for a large fraction of our host sample.
% \todo{[D\'evelopper + plan papier]}

\section{Type Ia supernov\ae\ and host galaxy data}
\label{sec:data}
\subsection{Supernova sample}
\label{subsec:sn_sample}

The discovery and follow-up of nearby and distant events
requires different instruments and survey strategies. We base this
work on a compilation of data from two low-redshift surveys,
the Carnegie Supernova Project (CSP) and the Center for
Astrophysics Supernova survey (CfA), the intermediate-redshift SDSS supernova survey, and the high-redshift SNLS
survey.
%This compilation prefigures the data sample that is assembled to provide the final cosmological results of the SNLS5.
This compilation has been assembled as part of an effort to derive cosmological constraints
 from the full SNLS spectroscopic sample. In all cases, the Spectral Adaptive Light curve Template (SALT2) model \citep{Guy07} was adjusted to
 the available multiband photometry, and consistent selection
 requirements were applied to the whole sample in order to obtain the results that are listed in the first column of Table~\ref{tab:tot}.
%  (see Section~\ref{subsec:cuts}).

A few modifications were applied compared to the most recent SNIa compilation from JLA \citep{Betoule14}. First, all
supernov\ae\ whose photometry was not measured in the natural system were discarded. This corresponds to
{\em{historical}} nearby supernov\ae\, and SNIa observed by the CfAI and CfAII programs. In our SNLS5 sample, CfAIV SNIa are included, as is the full SNLS spectroscopic sample. Compared to JLA, a slightly different selection procedure was applied to obtain the SNLS5 sample. 
In JLA, some SNIa from SDSS and low-redshift surveys were discarded by a requirement on the quality of the SALT2 fit.
This selection cut affects SNIa differently as a function of their reported signal-to-noise ratio (S/N). It is thus sensitive to the accuracy of the reported uncertainties and to the remaining dispersion around the SALT2 model.
%%This selection cut thus depends on the accuracy of measurement errors as well as on the quality of measurements themselves.
%For example, CSP exhibits better sampling and measurement quality than other surveys, but the fit quality is relatively poor, which is not a good enough reason to discard CSP events.
%\todo{Indeed, 14/28 CSP SNIa are discarded which correspond to very high SNR events.}
%We thus found that this selection requirement 
%%does a poor job discriminating between good and bad events,
%\todo{discards some of the best measured events without guarding the analysis froma any visible bias,}
%so we chose not to apply it and rely solely on spectral typing. As a consequence, the SNLS5 sample contains more of SDSS and low redshift SNIa than the JLA sample.
Most cosmological analyses additionally require a minimum
light-curve fit quality \citep{Betoule14,scolnic17}. Applying such a
requirement to our sample, for example, imposing that the
probability of the SALT2 fit is better than 0.001, would discard 132
of our 994 remaining supernov\ae, which is more than the statistically
expected $\text{one}$ event, approximately. This discrepancy could have three different origins, either underestimated
measurement uncertainties, underestimated intrinsic
variability of supernov\ae\ around the light-curve model as modeled in
the SALT2 framework, or very many peculiar SNIa in the
spectroscopic sample. It is noticeable that most of the very high
S/N events are discarded by this cut (e.g., 14/28 CSP
SNIa).
To distinguish between these assumptions, we
gathered the residuals to the blinded Hubble fit and compared
statistics for the subsample of 132 SNIa that were rejected by the fit probability cut
and the entire sample. We found that the dispersion of Hubble
residuals in the rejected sample is smaller than in the entire sample
(0.161/0.185), while their reduced $\chi^2$ is similar
(1.054/1.033). In addition, the rejected sample displays no evidence
for bias: its mean residual is $-0.021\pm0.014$. We thus concluded
that the main origin of the large number of SNIa that were rejected by the
fit probability cut was a slight underestimation of the intrinsic variability
of the SNIa population in the current SALT2. While forthcoming
cosmological analyses will have to consider this modeling
issue properly, preferentially by improving the modeling of the SNIa
population, we did not apply this cut to our sample because it discarded some of the best-measured events without preventing any visible bias in the analysis.

The sample presented in this work is very similar to the sample that will be used to derive cosmological constraints. However, a new calibration awaits the current measurement of SNLS filters. It is therefore probable that the cosmological sample differs from the sample used in this paper. A similar approach regarding environmental effects will also be presented together with the cosmological publication.

The data were blinded with respect to the cosmology by warping the
fluxes with a smooth function of the redshift $f(z)$ while photometric
calibration and supernova evolution issues were
investigated.
%While the cosmological results is extremely sensitive to the relative calibration of flux measurements in the different bands, the results presented here are essentially insensitive to calibration issues or other systematics as illustrated in Section~\ref{sec:robust}.
Cosmological
parameters measurements are extremely sensitive to the relative calibration of
flux measurements in the different bands. In contrast, the results
presented here are essentially insensitive to calibration issues or
other systematics, as discussed in Section~\ref{sec:robust}, so that we do not
expect them to be altered by any of the calibration adjustments.
%still being sorted.

% When discussing the complete available Type Ia supernova sample
% \todo{[Pour Marc]}, it is important to distinguish 3 redshift ranges
% corresponding to events found in different surveys. Numbers are
% described in more details in Table~\ref{tab:tot}.

\subsubsection{Low-redshift SNIa}
\label{subsubsec:low-z}
The low-redshift part of the compilation is dominated by SNIa from the
third and fourth releases of photometric data acquired at the
F. L. Whipple Observatory of the Harvard-Smithsonian Center for
Astrophysics (\citealt[][hereafter CfA3 and
CfA4]{hicken09_cfa3,hicken12_cfa4}). The data were acquired between
2001 and 2011 using three different CCD cameras (Keplercam, Minicam and
4Shooter2). We also included high-quality photometric data from the
first and second release of the Carnegie Supernova Project
(\citealt[hereafter CSP1 and
CSP2]{contreras10_csp,stritzinger11}). These data were acquired in
each of the six photometric bands $ugriBV$ of the Swope instrument at
the Las Campanas Observatory, with a typical two-day cadence.

In both cases, we used the photometry available in the natural
system of the photometric instrument. We did not consider
photometric data that were color-transformed, because this procedure, ill-suited to supernov\ae, introduces significant
scatter and bias in the measurement (e.g., \citealt[Appendix B]{Betoule14}).

\subsubsection{SDSS survey}
\label{subsubsec:inter-z}

The intermediate-redshift part of our sample is provided by the
SDSS-II supernova survey \citep[][hereafter S14]{Sako14}, which scanned a 300~deg$^2$
region during three months (September, October and November) in each of three years (2005 to 2007). It
delivered light curves for 10,258 variable and transient sources, as
well as host galaxy identification for thousands of transients,
photometric classifications for the candidates with good multicolor
light curves, dedicated spectroscopic observations for a subset of 889
transients, and host galaxy redshifts obtained using spectra from the
original SDSS spectrograph, the SDSS-III BOSS spectrograph, and the
telescopes used to obtain SDSS SN spectra. The survey used the SDSS
camera \citep{1998AJ....116.3040G} on the SDSS 2.5 m telescope
\citep{2000AJ....120.1579Y,2006AJ....131.2332G} at the Apache Point
Observatory (APO) to provide simultaneous light curves in the five
broad passbands, $ugriz$ \citep{Fukugita96,2010AJ....139.1628D}, with a
typical cadence of observations of once every four nights. These
observations resulted in the largest sample of supernova candidates
ever compiled, with 4607 likely supernov\ae, about 500 of which have
been confirmed as SNIa by the spectroscopic follow-up. We restrict
our analysis to SNIa in this spectroscopically confirmed subsample.

The SNIa photometry for SDSS is based on the scene modeling photometry (SMP)
described in \citet{Holtzman08}.  The basic approach of the SMP
is to simultaneously model the ensemble of survey images covering an
SNIa candidate location as a time-varying point source (the SN) and sky
background plus a time-independent galaxy background and nearby
calibration stars, all convolved with a time-varying point-spread
function (PSF).  The fitted parameters are the SN position, the
SN flux for
each epoch and passband, and the host galaxy intensity distribution in
each passband.  The galaxy model for each passband is a $20 \times 20$
grid of CCD pixels (approximately $8\arcsec \times 8\arcsec$), and
each of the $15\times15 \text{(pixels)}\times 5 \text{(passbands)} = 1125$ galaxy
intensities is an independent fit parameter.  As there is no pixel
resampling or convolution of the observed images, the procedure
yields reliable statistical error estimates.

The
complete data release of SDSS supernova light curves, spectra,
host galaxy identification, and estimate of their properties are
described in S14.

\subsubsection{SNLS survey}
\label{subsubsec:high-z}

Our sample includes spectroscopically confirmed high-redshift
SNIa from the full five-year SNLS survey, built from the deep
component of the CFHT Legacy Survey, which covers four patches of
sky, called D1, D2, D3, and D4, corresponding to a single pointing of the
1~deg$^2$ MegaCam camera on the 3.6~m Canada-France-Hawaii Telescope
(CFHT). The patches are located high over the galactic plane so as to
minimize local extinction of the incoming light (see Table
\ref{tab:cfht_deep}). They are also chosen such that at least two of them are
visible to the CFHT at any given time of the year. Images were taken
in four passbands similar to those used by the SDSS: $g_M, r_M, i_M,
z_M$, where the subscript $M$ denotes the MegaCam system.  Each field
and passband was repeatedly imaged four or five times per lunation,
with exposure times of $\sim 1$~hr (the total observing time for each
band is provided in Table~\ref{tab:deep_bands}, see
\citealt{2006AJ....131..960S} for details).  About 1000 supernov\ae\
were discovered in the redshift range $0.2 < z < 1$, and 427 of them
were confirmed as SNIa with their redshift measured by massive spectroscopic follow-up
programs
\citep{Howell05,Bronder08,Ellis08,Balland09,Walker11,balland17}. In July 2007, the CFHT $i_M$ filter broke and was replaced by a new filter a few months later. Since this corresponds to the end of the survey, we did not include images from the new filter in the coadded images.

In the same spirit as for the method developed for the SDSS survey, the SNLS supernova photometry consists of fitting a time-variable point source in addition to a time-independent galaxy image to the SNLS image series, without resampling the images before the fit. This method delivers  unbiased SN magnitudes at the mmag
accuracy level, with systematic uncertainty less than 1.5 mmag \citep{Astier13}.

\begin{table}%[!htbp]
\centering
\begin{tabular}{ c| c| c| c }
Field & RA & DEC & $E(B-V)$ \\ \hline\hline
D1 & 02:26:00.00 & -04:30:00.0 & 0.033 \\
D2 & 10:00:28.60 & +02:12:21.0 & 0.023 \\
D3 & 14:19:28.01 & +52:40:41.0 & 0.013 \\
D4 & 22:15:31.67 & -17:44:05.0 & 0.027 \\
\end{tabular}
\caption{Central coordinates of the CFHT Deep Survey fields. An estimated average value of the Milky Way $E(B-V)$ is given using the \cite{planck_dust_13} maps.}
\label{tab:cfht_deep}
\end{table}

\begin{table}%[!htbp]
\centering
\begin{tabular}{ r | c |c |c |c| c }
band & $u$ & $g$ & $r$ & $i$ & $z$ \\ \hline\hline
time (in hours) & 33 & 33 & 66 & 132 & 66 \\
\end{tabular}
\caption{Total integration time of the deep survey in different bands, {\em{for each field,}} that is, the total time allocated to the survey is four times as long as what is shown.}
\label{tab:deep_bands}
\end{table}

\subsection{Available host photometric data}
\label{subsec:available_host_photo}
% We present the host galaxies photometry that we found in an online
% database and investigated the differences between photometries
% coming from various catalogs. This allows a coherent description of
% the sample in terms of photometry.

\begin{table*}[!htbp]
\centering
\begin{tabular}{c | c |c |c |c}
Survey & SNIa & Host photometry & Reference & Filters/Instrument\\
\hline\hline
CSP
 & 19
 & 7
& \texttt{SIMBAD} 
& $ugriz$/SDSS \& $JHK$/2MASS\\
CfAIII
 & 84
 & 55
& \texttt{SIMBAD} 
& $ugriz$/SDSS \& $JHK$/2MASS\\
CfAIV
 & 53
 & 34
& \texttt{SIMBAD} 
& $ugriz$/SDSS \& $JHK$/2MASS\\
SDSS
 & 441
 & 389
& 1
& $ugriz$/SDSS\\
SNLS
 & 397
 & 397
& 2
& $ugriz$/MegaCam\\
\hline\hline
Total
 & 994
 & 882
 & \ldots & \ldots\\
\end{tabular}
\tablebib{(1) \cite{Sako14}; (2) \cite{Hardin16}.}
\caption{Number of available supernov\ae\ for the whole SNLS five-year release %($2^{\mathrm{nd}}$ column)
and the corresponding amount of host photometry. %($3^{\mathrm{rd}}$ column)
References and information on instruments and filters are also presented in the last columns.}
\label{tab:tot}
\end{table*}

%& \# SN & \# Host photometry & Reference & Filters/Instrument\\
%\hline
%\hline
%SNLS & 392 & 346 & \cite{Hardin16} & $ugriz$/MegaCam\\
%SDSS & 330 & 291 & \cite{Sako14} & $ugriz$/SDSS\\
%low-$z$ & 137 & 117 & SIMBAD & $ugriz$/SDSS \& $JHK$/2MASS\\
%\hline
%\hline
%\textbf{Total} & \textbf{859} & \textbf{754} & $-$ & $-$\\

%\begin{table*}[!htbp]
%\centering
%\begin{tabular}{c c c c c c c}
%& \# SN & \# $z$ match & \# in SDSS &\# in 2MASS & \# only 2MASS & no photo\\
%\hline
%\hline
%CSP & 18 & 16 & 8 & 16 & 8 & 0\\
%CfAIII & 79 & 68 & 46 & 59 & 19 & 3\\
%CfAIV & 40 & 38 & 24 & 33 & 12 & 2\\
%%Others & 48 & $-$ & $-$ & $-$ & $-$ & $-$ & $-$\\
%\hline
%\hline
%\textbf{Total} & \textbf{137} & \textbf{122} & \textbf{78} & \textbf{108} & \textbf{39} & \textbf{5}\\
%\end{tabular}
%\caption{Photometry available for host galaxies of nearby supernov\ae\ in the SIMBAD database. The numbers of hosts with good redshift match are shown, along with their repartition in the different available photometric catalogs.}
%\label{tab:lowz}
%\end{table*}

For the purpose of the analysis presented in this paper, we need to distinguish between the number of available SNIa in our sample and the supernov\ae\ for which we can perform our own host photometry. We relied on SNLS data for the corresponding SNIa sample and on available SDSS imaging for the SDSS and nearby SNIa samples. This requirement defined the number of supernov\ae\ for this analysis. In
Table~\ref{tab:tot} we list the corresponding numbers for each
survey. For the high-redshift SNIa from SNLS, we gathered the host
galaxy photometry for the whole available sample.
%Among the 392 supernov\ae\ being part of the SNLS5 release, there is no available host photometry for 46 of them, corresponding to hostless supernov\ae\ or SNIa whose parent galaxies are too faint to be extracted from MegaCam images.
The details describing host identification and photometric techniques used for the SNLS sample are presented in more detail in Section~\ref{subsubsec:global_snls} and also in our companion paper \citep{Hardin16}.

Redshift, position, and magnitudes (among others) of intermediate-redshift supernov\ae\ observed with the
$ugriz$ filters of the SDSS instrument and for their parent galaxies
are found in the catalogs produced by S14. In this redshift range and for this specific study, we find host galaxy photometry in S14 for approximately 80\% of the total SDSS sample. This enables
us to compute and compare local and global properties of host galaxies for the  SNLS and SDSS surveys.

The situation is more complex for low-redshift surveys.
%The low redshift SNIa total sample contains events coming from CSP, CfAI, CfAII, CfAIII, CfAIV and Cal\'an-Tololo surveys (Section~\ref{subsubsec:low-z}) which amounts up to 186 supernov\ae. Starting from this subsample, we choose to focus on events detected by the CSP, CfAIII and CfAIV surveys.
%, losing about 26\% of nearby SNIa.
The low-redshift SNIa total sample includes events from CSP and
the CfAIII and CfAIV SNIa surveys (see Section~\ref{subsubsec:low-z}),
which amounts to 156 supernov\ae. 
To determine which SNIa is included in the SDSS footprint, we checked  in the SIMBAD database\footnote{\url{http://simbad.u-strasbg.fr/simbad/}} for events with SDSS $ugriz$ data associated with their host.
We also checked for JHK photometry from the Two Micron All-Sky Survey catalogs \citep[2MASS,][]{2MASS} because we used the K magnitude as a stellar mass proxy (see Section~\ref{subsec:intercalib_2mass}). Only a fraction (35\%) of  the nearby SNIa from the CSP survey are SDSS supernov\ae,\, and about 60\% of those from the CfAIII and CfAIV surveys are located in the SDSS footprint.
When they were not recovered in the SDSS footprint, we nonetheless retained the $K$-magnitude photometry information from 2MASS for to study the host galaxy properties.

As a result,  we used the number of supernov\ae\ described in
Table~\ref{tab:tot} to study the local environment around the supernova explosion and to strictly compare local and global properties. For the global properties, we used additional information from the 2MASS catalog.

\section{Global and local photometry of the SNIa hosts}
\label{sec:photo}
The photometry of the whole host galaxy and of the region close to the SNIa explosion traces information on the age and evolution of their stellar population. 
We present in this section a reliably calibrated photometric method from which the host properties, both local and global, are derived. This provides a consistent measurement across the
wide redshift range of the supernova sample for the global photometry  of the host galaxy and for the photometry at the supernova location (local photometry).

%Photometry of the whole host galaxy and of the region close to the
%SNIa explosion reveals a lot of information about the age and
%evolution of their stellar population. As a result, we need a robust
%calibrated photometric method. In addition to local host photometry and in order to remain fully consistent, we will also measure host galaxy global photometry for all low redshift and intermediate-redshift SNIa in the SDSS images, though part of it already exists in the literature. We will then obtain a consistent photometric catalog of objects spanning a wide redshift range, and get consistent host properties. 

\subsection{Global photometry}
\label{subsec:global}

\subsubsection{Host galaxies in SNLS images}
\label{subsubsec:global_snls}

The SNLS host galaxy photometry was performed on deep stacked images in
the $ugriz$ filters.
The CFHT-LS $ugriz$ observations of the Deep fields were pre-processed
through the Elixir pipeline to correct for the instrumental effects
that affect the pixel flux. More information on the Elixir pipeline can be found in \cite{magnier04} or on the CFHT website\footnote{\url{http://www.cfht.hawaii.edu/Science/CFHTLS-DATA/dataprocessing.html}}. It includes flat-fielding and
fringe correction in $i$ and $z$ data.

Individual CCD images were then processed through the SNLS pipeline
\citep{Astier2013}. A weight map was generated that incorporates bad
pixels information (set to a zero-value weight) and was normalized with
the image sky-variance. Each CCD image catalog was built using
\texttt{SExtractor} \citep{Bertin1996}, and a smooth sky-background map
was subtracted. Each image was calibrated by performing a large-diameter
($D=15 \times \sigma_{\rm seeing}$) aperture photometry on the
tertiary stars catalog from \cite{Regnault2009} and computing a
Vega photometric zero-point. The astrometric calibration was derived
using the SNLS internal astrometric catalog.

The $ugriz$ 36-CCDs frames were selected using cuts on the
photometric zero-point, the mean seeing (FWHM$<$1.1'' for $griz$
bands, 1.3" in $u$), the sky variance, the mean star shapes, the
number of saturated stars, and the amount of fringing. The cut was
performed solely on the central CCD of each frame. About 60\% of the
best-quality images were kept in $griz$, which corresponds to 300 to
400 36-CCDs frames.  For the $u$ band, only 8\% of the data were
rejected.

The selected 36-CCDs frames then enter the stacking
procedure. Each field was observed during a season of six consecutive
months. Depending on the field, between five to six observing seasons were available.  The selected 36-CCDs frames were combined on a per-season
basis to construct 1-square-degree $griz$ ``per-season'' deep stacked
images.

Individual CCD images, rescaled to a common photometric zero-point
$zp=30$, were co-added using the SWarp V2.17.1
package\footnote{\url{http://terapix.iap.fr/soft/swarp}} with the
median-filter option. This permitted us to reject remaining satellite trails and cosmic
rays. A total weight map was also produced.

To avoid supernova light contamination, the supernova host photometry
was performed on the ``excluded-season'' stacked images. They were
obtained by coadding all seasons except for the supernova season stacks
through a weighted average, using the per-season stack weight map
scaled with the sky variance and the image quality: $w \propto
1/(\sigma_{\rm sky} \sigma_{\rm seeing})^2$.

For the $u$-band, all season stacks were produced and
used because the supernova light contamination is reduced in the SNLS
redshift range $z>0.2$. For the D2 field, the Terapix T0006 D2-u stack
was used because it incorporates COSMOS CFHT-MegaCam $u$ data that were
overlapping the D2 field.

The detection of the sources and the related photometry were performed
using \texttt{SExtractor} in the dual-image mode, with a detection
threshold level of $2.5\sigma$ for 3 contiguous pixels. The
detection was realized on the $i$ frame and the photometry on the
$ugriz$ frames.

%DH ajouter niveau deblending

Magnitudes were estimated within \texttt{SExtractor} Kron \citep{Kron1980}
elliptical-aperture magnitudes (AUTO-magnitudes), which provided a
measurement of the total galaxy flux. Then, flux uncertainties were
modeled by a two-component (added in quadrature) model: $\delta f =
a f \oplus e_{\rm empty}$.  The first term $a$ corresponds to the
magnitude uncertainty for bright sources, which is about 0.01-0.015
mag depending on the band. The second term $e_{\rm empty}$ denotes the RMS
of the ``empty'' flux, that is, the aperture flux measured at random
positions on each frame, in regions without sources. The local
residual background variation, correlated at the aperture spatial
scale, provided the major contribution to the ``empty-flux'' dispersion.
% en fait c'est plus complique que cela, bien sur
% err_flux = a R + b R^2, avec R rayonouverture
%Labbe 2003 et Gawiser 2006

The reproducibility of the magnitude measurement and the flux error
model parametrization were tested on computed on-purpose stacked
frames, combining the first two seasons S0 and S1 images, and the two
following seasons S2 and S3.  At the limiting magnitude $i\simeq
24.8$, this corresponds to a magnitude uncertainty of $0.12$ mag.
%DH ajouter : histo des magnitude des SNIa, visuel d'une sans hote avec une HST image, eg

\subsubsection{Host galaxies in SDSS images}
\label{subsubsec:global_sdss}
%Regarding the host galaxies we find in SDSS images, we perform our own global photometry for 369 low and intermediate-redshift parent galaxies of Type Ia supernov\ae, considering the 291 available hosts identified in S14 and in the SDSS DR12 databese, and the 78 nearby hosts observed by independent surveys but which are located in the SDSS footprint.

%Therefore,
We used the Sloan Digital Sky Survey DR12 legacy, which provides uniform, well-calibrated images in $ugriz$
bands of more than 31600 square degrees in the sky, comprising the
North Galactic Cap and three stripes in the South Galactic Cap \citep{alam15}. The DR12
release observed about 940000 individual fields separated into images of
size $1361\times2048$ with a pixel scale of 0.396 square arcseconds
on the sky and with a median PSF FWHM of 1.3 arcsecond. These images
are dubbed {\em{single-epoch images}} in the following. Astrometry is specified at a level of 70 mas at the limit of $r=16$, with systematic errors smaller than 30 mas.

\begin{figure}[!htbp]
  \begin{center}
   \begin{tabular}{c}
    \includegraphics[width=0.9\columnwidth]{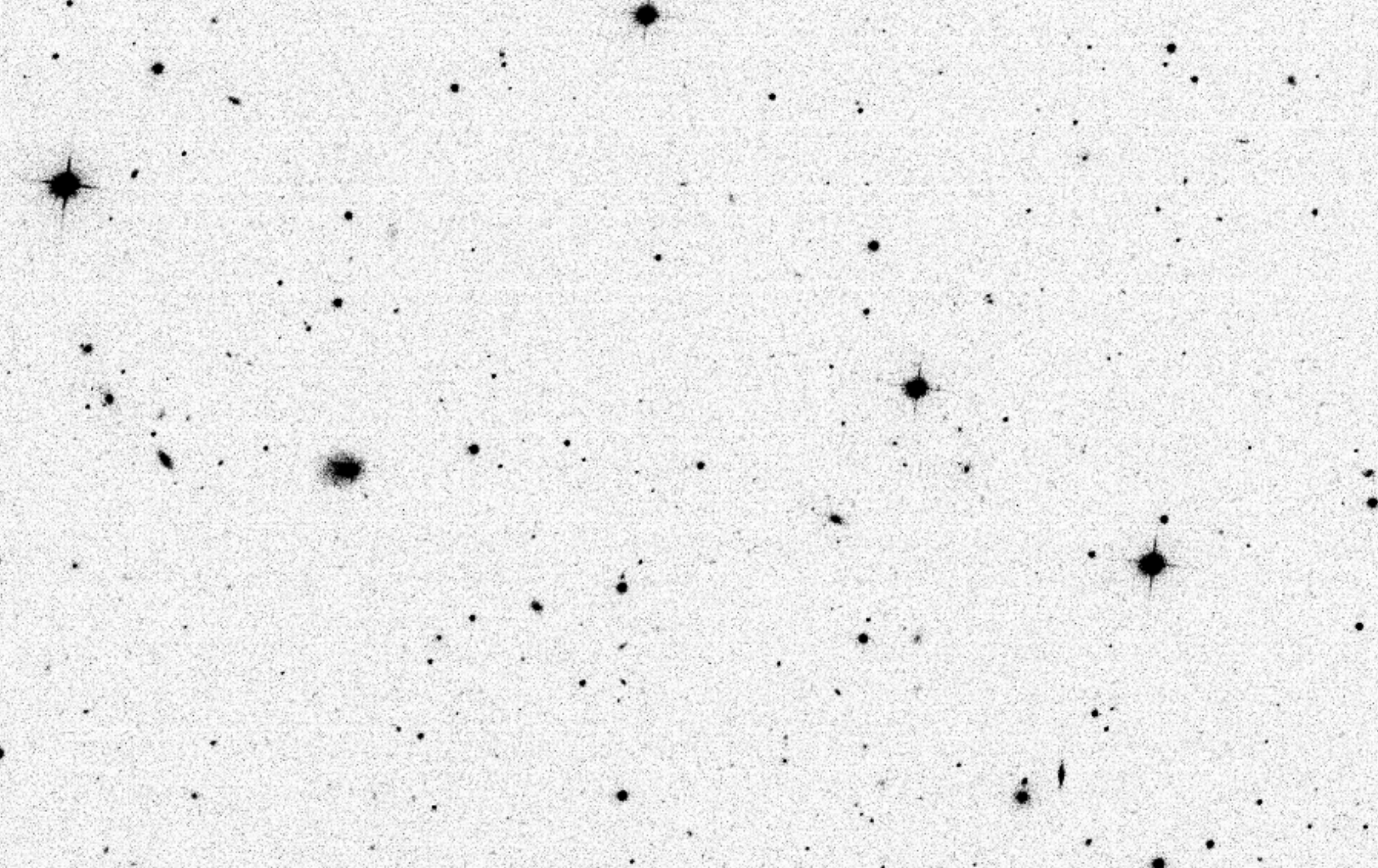} \\
    \includegraphics[width=0.9\columnwidth]{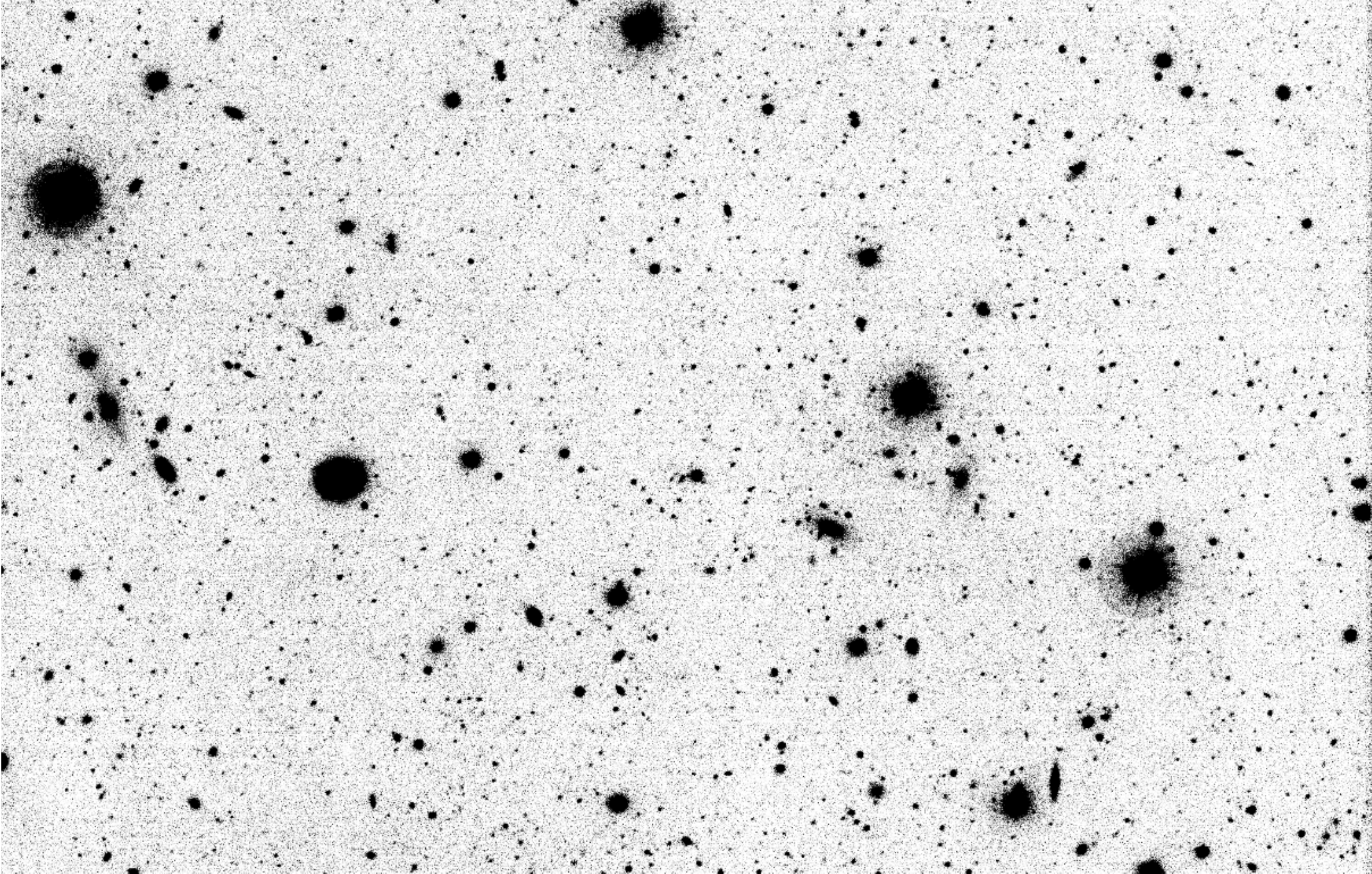}
   \end{tabular}
  \end{center}
  \caption{Examples of the two types of SDSS images on the same
    portion of the sky in the $g$ band. The DR12 single-epoch image is shown in the
    top panel, and the Stripe~82 coadded image is illustrated
in the
    bottom panel.}
  \label{fig:s82_vs_se}
\end{figure}

The Sloan Digital Sky Survey DR7 release
\citep{abazajian09} also includes all images from the Stripe~82 (S82) along the Celestial
Equator in the Southern Galactic Cap. It contains a coaddition of 47 South strip and 55
North strip scans into a pair of SDSS runs in which objects are detected
at about 2 magnitudes deeper than in any individual Stripe~82 scan. The
270 square degrees area of Stripe~82, between
$-50^\circ<\mathrm{RA}<59^\circ$ and
$-1.25^\circ<\mathrm{DEC}<1.25^\circ$, has been covered approximately
80 times \citep{jiang14}. The corresponding images are referred to as
{\em{stacks}} in the following. In Fig.~\ref{fig:s82_vs_se} we
present examples of the two types of images that show the
observation of the same portion of the sky. It is clear from that
figure and from comparing the two panels that stacked images from Stripe~82 are deeper and allow the detection of more
objects. 

As stack images are much deeper than single-epoch images,
we used them whenever possible, that is, for all supernov\ae\
exploding in the S82 footprint outside the time range of the stack
images (in practice, after December 2005; see Table~\ref{tab:s82_vs_se} for the exact survey repartition).
%We will use them whenever possible.
%Nevertheless, as coaddition of Stripe~82 images corresponds to observations
%taken before fall 2005, we decide to choose single-epoch images to
%measure global and local photometry of parent galaxies of SNIa
%exploding before fall 2005 in order to avoid supernova contamination.
%Stacks are utilized for photometry of parent galaxies of SNIa exploding after that
%date, if they are found in the coadded images. The repartition between
%stacks and single-epoch images is represented in Table
%\ref{tab:s82_vs_se}.
As expected, we mostly found SDSS host galaxies in
the coadded images, except for those that were contaminated by the
supernov\ae. Conversely, a large majority of host galaxies from
low-redshift surveys were found in single-epoch images since the SNIa
included in them exploded at random places in the sky and were not necessarily found in the Stripe~82 footprint.

\begin{table}[!htbp]
\centering
\begin{tabular}{ c | c c c c }
Image type & SDSS & CSP & CfAIII & CfAIV\\
\hline\hline
Stacks
 & 285
 & 0
 & 0
 & 1\\
Single-epoch
 & 104
 & 7
 & 55
 & 33\\
\hline\hline
Total
 & 389
 & 7
 & 55
 & 34
\\
\end{tabular}
\caption{Repartition of SNIa host galaxies in the different SDSS images. The stacks, which have been observed multiple times and which allow deeper detections, mostly correspond to SDSS SNIa. Nearby SNIa detected by CSP, CfAIII, or CfAIV are not necessarily on the Stripe~82 footprint and are mostly found in single-epoch images.}
\label{tab:s82_vs_se}
\end{table}

The sources were detected with \texttt{SExtractor,} with a detection threshold level of $2.5\sigma$ for single-epoch images ($4\sigma$ for stacked images) with at least three contiguous pixels. Each time an object was extracted, the connected pixels passed through a filter, which split them into components that might overlap. At low redshift ($z<0.05$) and at SDSS imaging resolution, larger galaxies appear as extended and sometimes clumpy objects, so that their outer and central regions might be erroneously identified as separate sources. On the other hand, at higher redshift ($z \simeq 0.2$) and for a locally denser area, neighboring objects should be identified as distinct sources. This can be handled with the \texttt{SExtractor} deblending parameter \texttt{DEBLEND\_MINCOUNT,} which governs the way in which local flux maxima are identified as  individual overlapping sources, or conversely, merged  into an extended single object.
This parameter reaches from 0, where even the faintest peaks in the detection profile are considered as single independent objects (high-deblending), to 1 (no deblending).

\begin{figure*}
  \begin{center}
   \begin{tabular}{c c c c}
    \includegraphics[width=0.4\columnwidth]{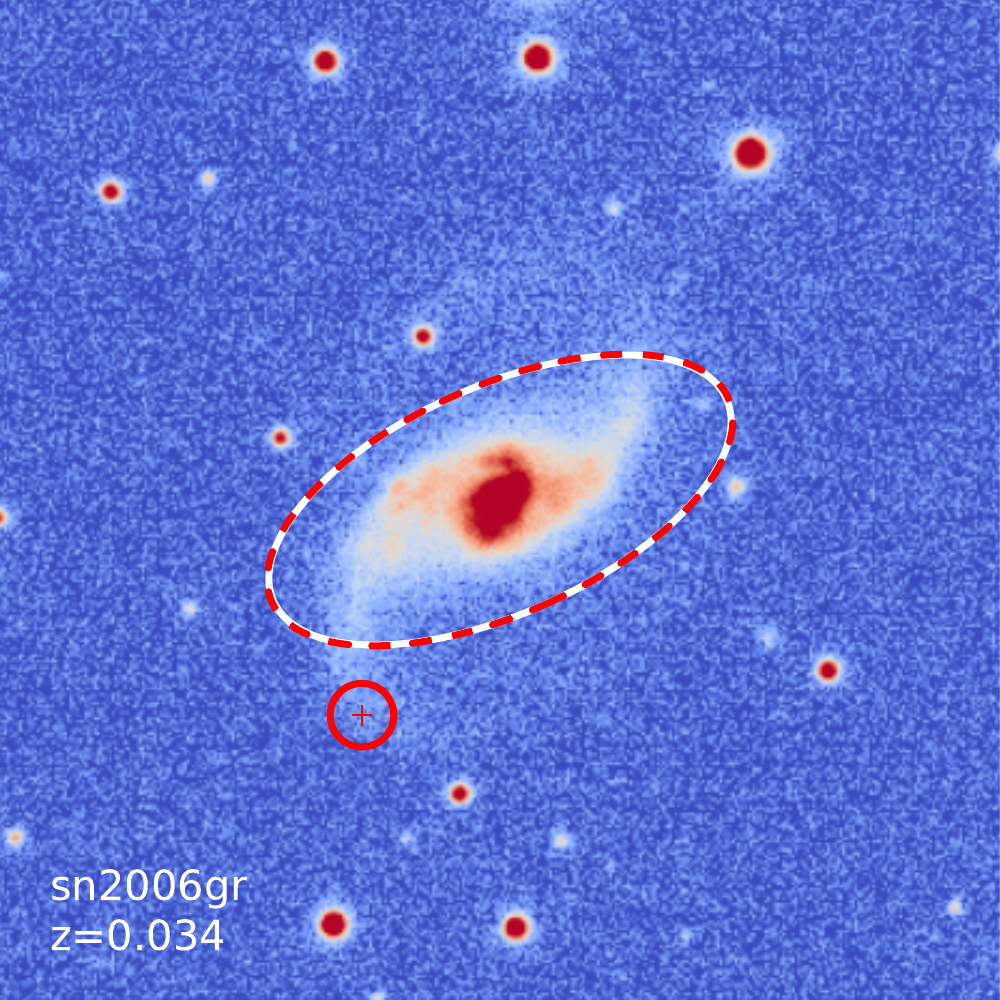}&
    \includegraphics[width=0.4\columnwidth]{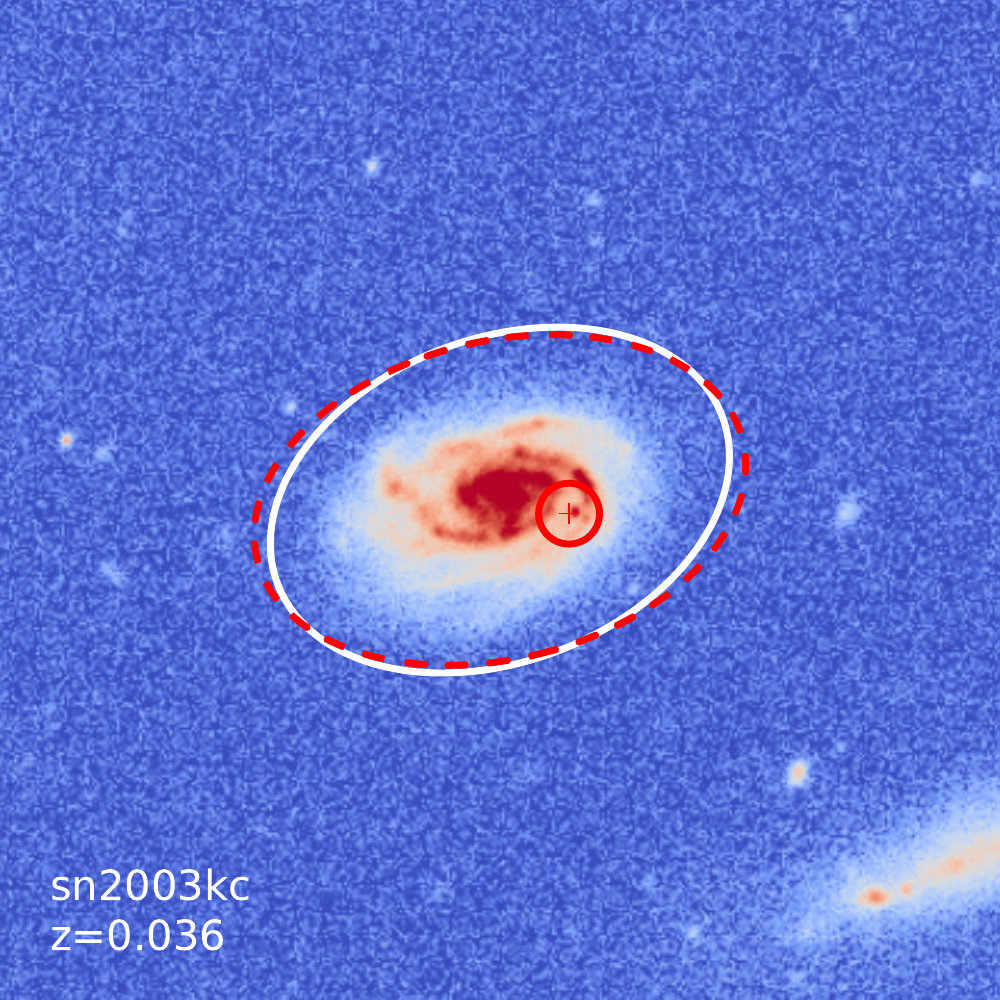}&
    \includegraphics[width=0.4\columnwidth]{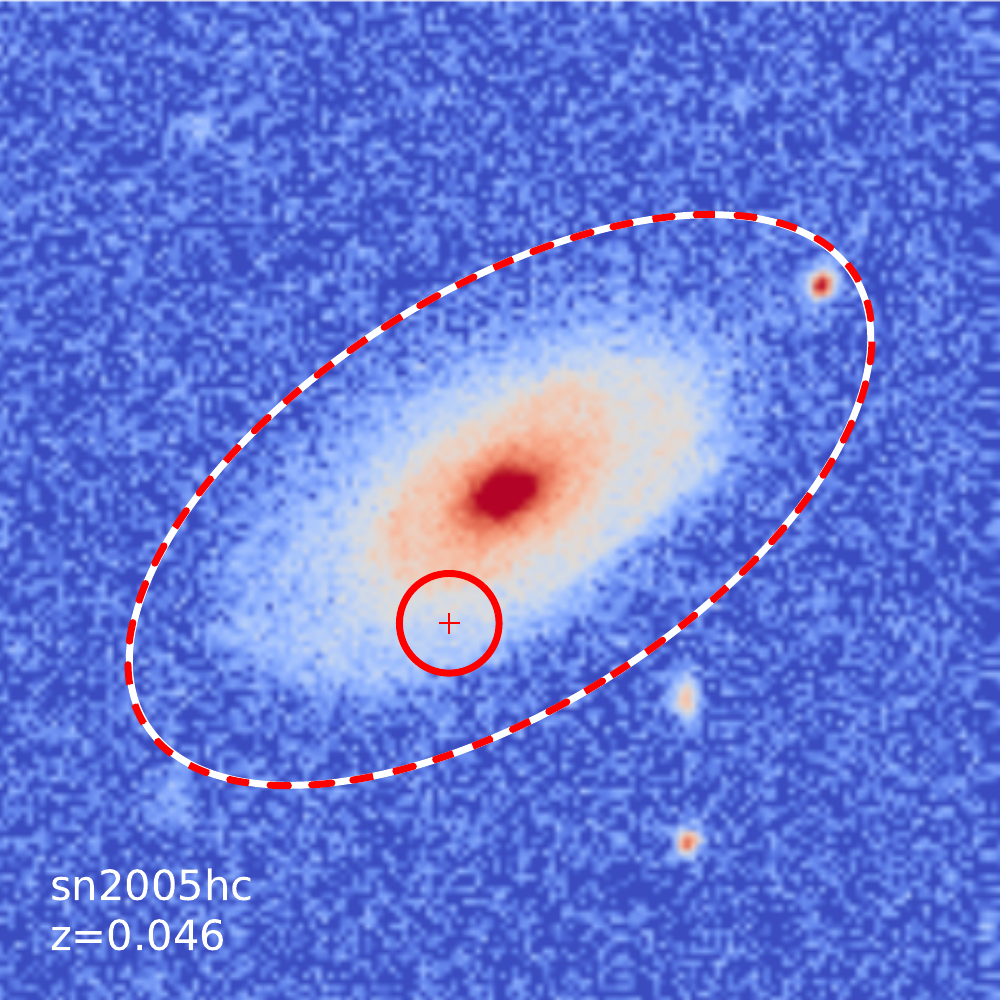} &
    \includegraphics[width=0.4\columnwidth]{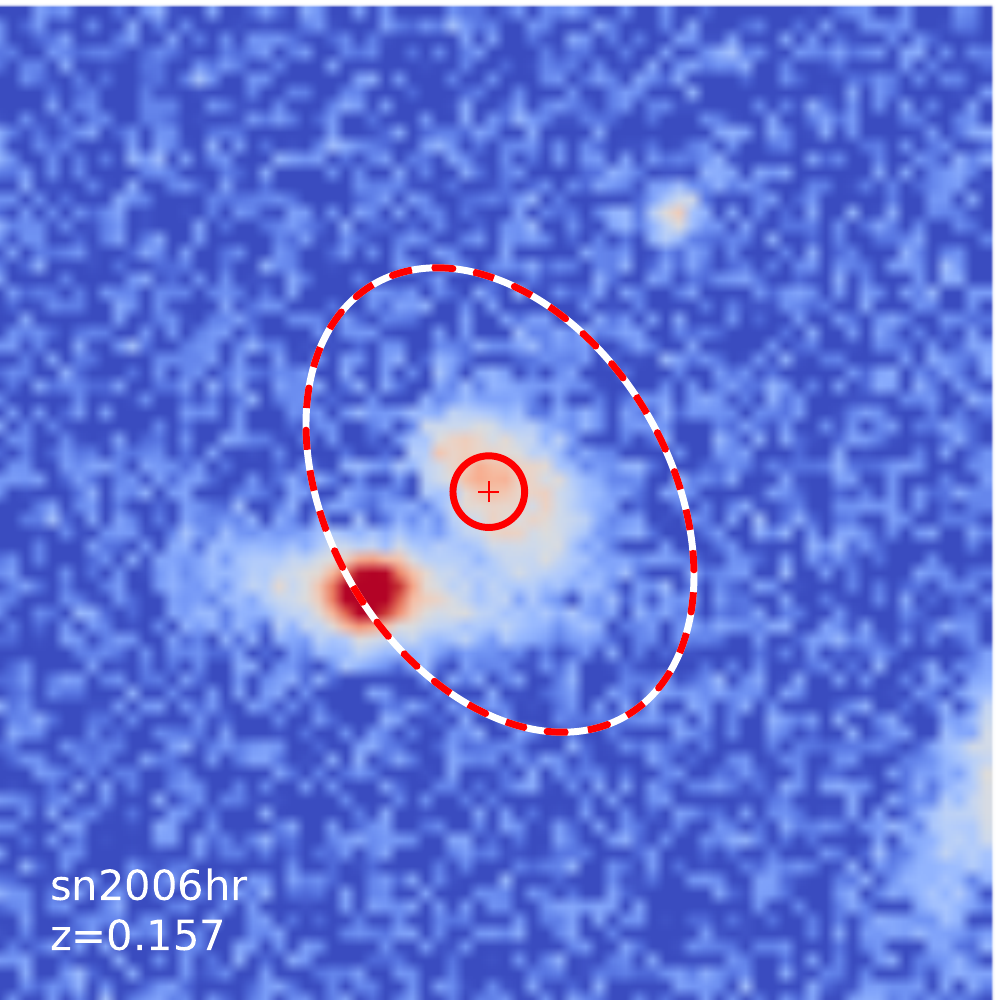}\\[0.3cm]
    \includegraphics[width=0.4\columnwidth]{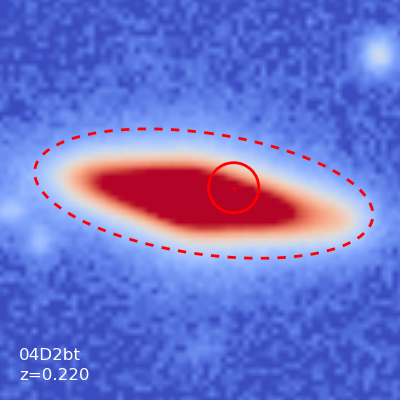} &
    \includegraphics[width=0.4\columnwidth]{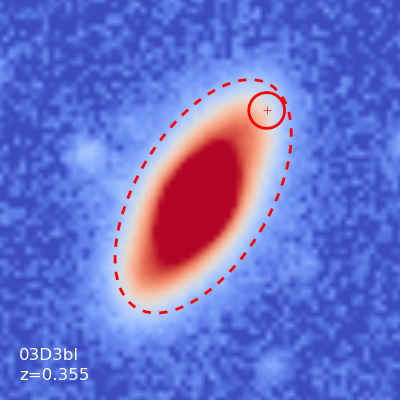} &
    \includegraphics[width=0.4\columnwidth]{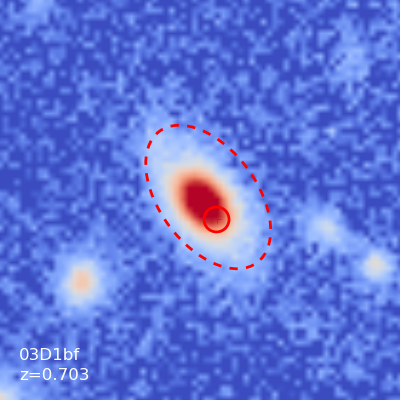} &
    \includegraphics[width=0.4\columnwidth]{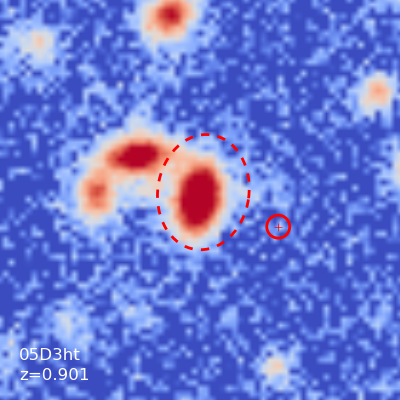}\\
   \end{tabular}
  \end{center}
  \caption{Examples of eight $g$-band images of SNIa host galaxies ordered
    in redshift. The top row corresponds to SDSS images with
    {\em{sn2006gr}}, {\em{sn2003kc}}, {\em{sn2005hc,}} and
    {\em{sn2006hr}}. For this survey, the dashed red and solid white ellipses
    represent the region where the global photometry is measured for a low and high level of deblending. The solid red
    circle symbolizes the 3 kpc region where the local photometry is
    computed. 
The second row displays SNLS images corresponding to {\em{04D2bt}}, {\em{03D3bl}}, {\em{03D1bf,}} and {\em{05D3ht}}, using one single \texttt{SExtractor} configuration.
    This illustrates the great variety of cases, from SNIa
    exploding in the inner galaxy core to SNIa observed far away from
    the center, in the galactic arms.}
  \label{fig:examples_vignettes}
\end{figure*}

\begin{figure}%[!htbp]
  \begin{center}
    \includegraphics[width=\columnwidth]{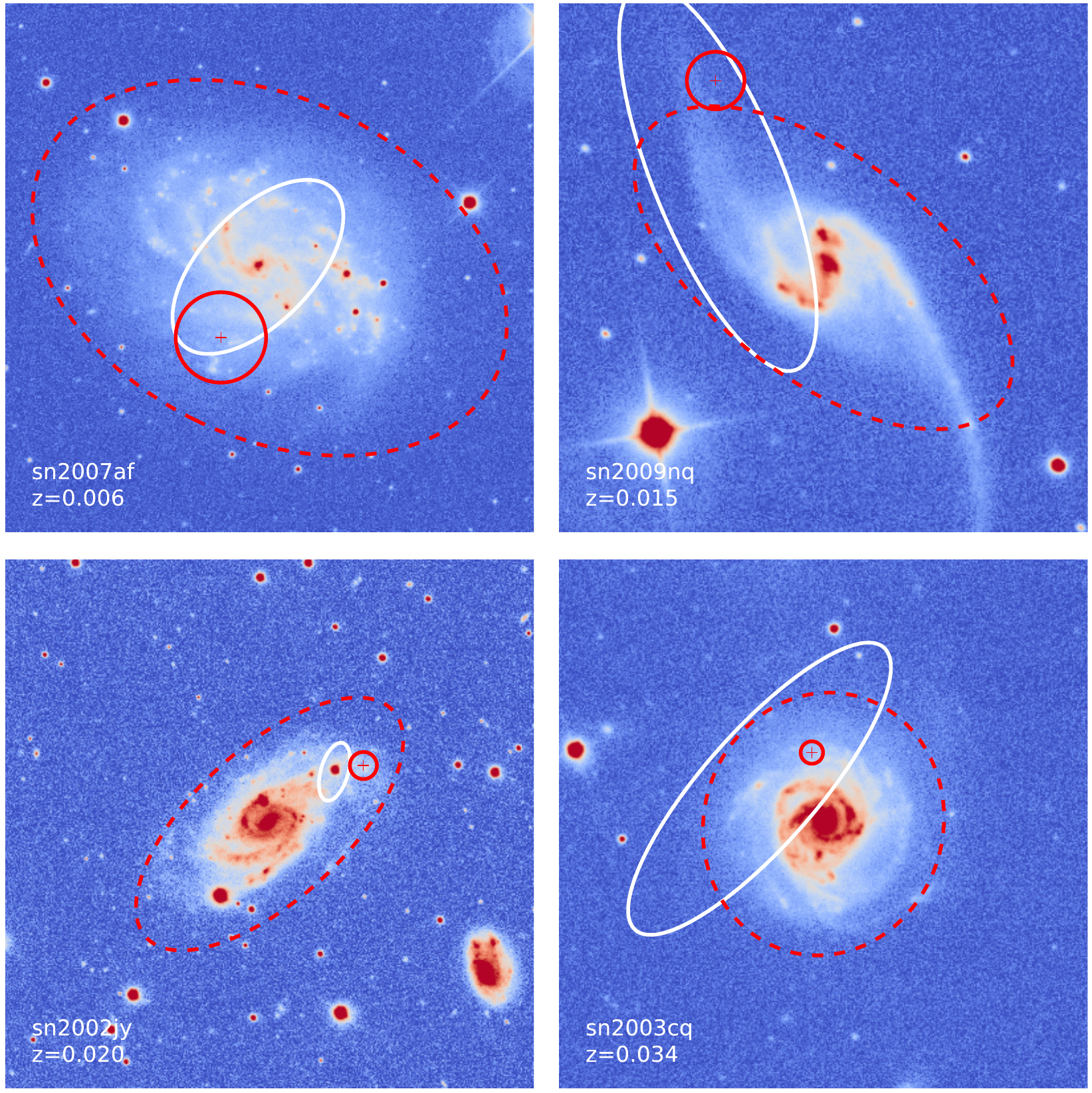}
  \end{center}
  \caption{Examples of nearby bright galaxies for
    which the low-deblending mode (red dashed ellipse) gives
    much better results than the high-deblending mode (white solid ellipse). The solid red circle represents the local photometric radius described in Section~\ref{subsec:local}.}
  \label{fig:sex_config}
\end{figure}

We thus produced two different source catalogs: a highly deblended catalog (corresponding to a deblending parameter of $2.10^{-6}$, the {\em high-deblend} configuration), so that galaxies at higher redshift are well identified; and a catalog with a deblending parameter of 0.05 (the {\em low-deblend} configuration), so that large, nearby galaxies below $z = 0.05$ are not broken up into sub-regions.
In the top row of Fig.~\ref{fig:examples_vignettes}, the red dashed ellipse corresponds to the low-deblending configuration, while the white solid ellipse represents the high-deblending case. For the vast majority of host galaxies, both extraction configurations yield similar host characteristics. Examples of low-redshift bright galaxies for which the two configurations give different results are displayed in Fig.~\ref{fig:sex_config}. Using field galaxies from SDSS images, we find that the magnitude difference using the two extraction configurations is significant for objects below $z=0.04$. Therefore, the limit we set for the transition at $z=0.05$ is reasonable.

%Objects whose signal is above a given threshold are detected on all
%images with \texttt{SExtractor}.  Each time an object extraction is
%performed, the connected pixels pass through a filter that splits them
%into eventual overlapping components. As galaxies are extended
%objects, it can give rise to multiple objects detected inside the same
%galactic area, depending on the level of deblending we choose.  We can
%influence the level of deblending through a contrast parameter in
%\texttt{SExtractor} configuration files. This parameter goes from 0,
%where even the faintest peaks in the detection profile are considered
%as single independent objects to 1 (no deblending).
%As the effect described is visible on a several low redshift cases, we redo the photometry to solve this issue making use of \texttt{SExtractor} in two different
%configuration modes: a low contrast parameter for deblending ($2\times
%10^{-6}$, thus denominated \em{high-deblend} in the following) and a
%contrast parameter of 0.05 so that large, nearby galaxies below
%$z=0.05$ are not broken up into sub-regions (\em{low-deblend}).
%An illustration of the effect of a change in deblending is shown in
%the context of the comparison of our observed photometry with the one
%published by the SDSS (Section~\ref{subsubsec:dr12}, bottom panel of
%Fig.~\ref{fig:comparison_dr12_hosts}).

For each image, a zero-point was estimated by computing the difference
between magnitudes of field stars inside a radius of 4 arcseconds and their reference magnitudes in the SDSS
catalog. We estimated the measurement errors in two different ways,
depending on the available type of image.
\begin{itemize}
\item For SNIa in {\em{single-epoch}} images $I$: the sky image
  $I_\mathrm{sky}$, the dark current $N_\mathrm{D}$ , and the gain $G$
  are available. We therefore constructed a variance map defined as\begin{equation}
\sigma_\mathrm{map}=\sqrt{\dfrac{I}{G}+\dfrac{I_\mathrm{sky}}{G}+N_\mathrm{D}}\ ,
\label{eq:sigma_map}
\end{equation}
and we measured the error by integration inside the corresponding
photometric ellipse:
\begin{equation}
\sigma^2_\phi=\sum_p\sigma_{\mathrm{map},p}^2\ ,
\label{eq:sigma_phi}
\end{equation}
going from flux error to magnitude error by computing
\begin{equation}
\sigma_\mathrm{mag}=\left(\dfrac{2.5}{\mathrm{ln}10}\right)\dfrac{\sigma_\phi}{\phi};
\label{eq:sigma_mag}
\end{equation}
where $\phi$ is the measured flux.
\item For SNIa in {\em{stacks}}: since the available weight image is a relative weight map, we resorted to building
  a $\sigma_\phi=f(\phi)$ relation, or equivalently, a $\sigma_\mathrm{mag}=f(\mathrm{mag})$ relation, 
using the magnitude errors published in the SDSS stack catalogs for the field galaxies to compute the errors for the global photometry
of the host galaxy.
\end{itemize}

In the following, the
{\em{global}} photometry corresponds to the flux integrated in the
\texttt{SExtractor} ellipse that is associated with the host galaxy using the Python tool
\texttt{aperture\_photometry} from the \texttt{photutils} module of
\texttt{Astropy}. This allows us to remain consistent throughout the analysis since the same tool is used for local photometry.

\subsection{Local photometry}
\label{subsec:local}
\begin{figure}%[!htbp]
  \begin{center}
    \includegraphics[width=0.9\columnwidth]{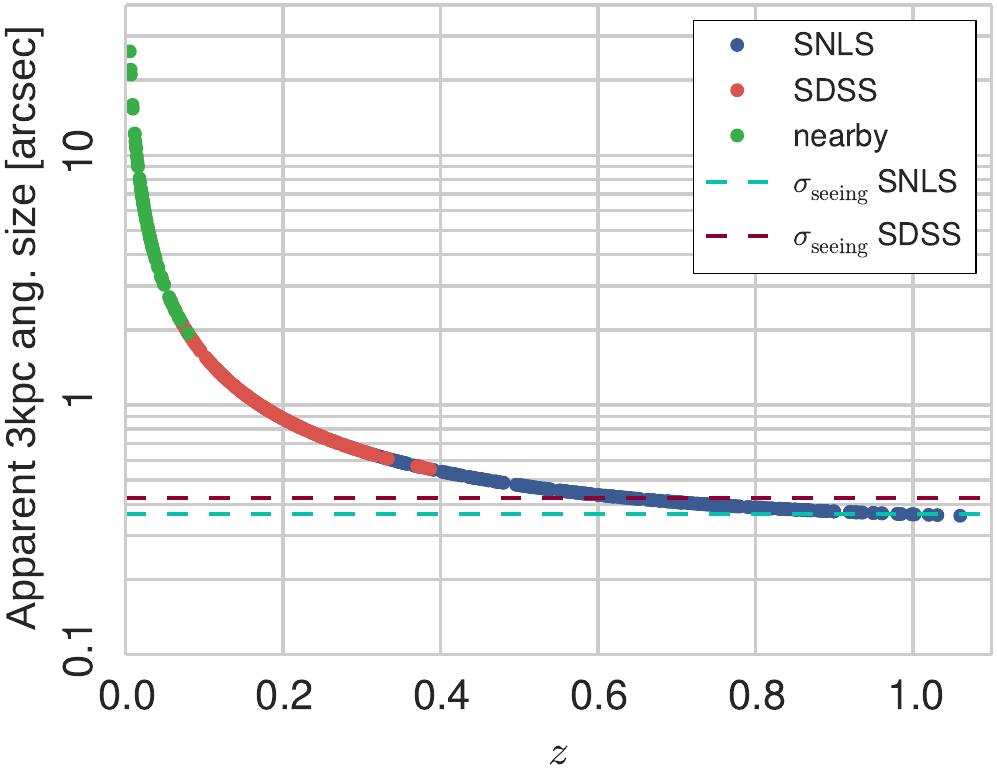}
  \end{center}
  \caption{Evolution of the local photometric radius with a size
of 3 kpc with redshift in arcseconds for
    host galaxies of SNIa observed by the SNLS (blue), SDSS (red), and low-redshift
    surveys (green points). The $1\sigma$ seeing of SNLS and SDSS images is illustrated with dashed lines using the same color code. A $1\sigma$ seeing corresponds to a diameter of about 85\% of the FWHM.}
  \label{fig:radius_vs_z}
\end{figure}

The local photometry was performed at the SN location, inside a radius with a proper size of 3 kpc. This corresponds to the smallest achievable physical size for high-redshift SNIa.
Because the true FWHM seeing of SNLS images is about 0.85'', we defined the smallest physical size as $1\sigma$ seeing in radius (0.364'', equivalent to 2 SNLS pixels) at the highest redshifts. At $z\sim0.7$, where most of SNLS SNIa are located, the radius is greater than $1.12\sigma$. The integrated flux within a $1\sigma$ radius is $\sim40$\%, so that local information can still be
extracted.
Fig.~\ref{fig:radius_vs_z} illustrates the evolution of the local photometric radius with redshift. It varies from $\sim1.9$ pixel for SNLS supernov\ae\ at the highest redshifts to $\sim26$ pixels for very low redshift SNIa. The corresponding $1\sigma$ seeing for the different surveys is displayed as dashed lines. We discuss the results we obtained when probing smaller areas in a limited redshift range in Section~\ref{subsec:more_local}. For intermediate and low redshifts, the local radius is always greater than $1.3\sigma$.

The SNLS local photometry was performed within these apertures using the SNLS data C++ analysis tool, which handles fractional pixels. The flux error was computed using the error model described in Section~\ref{subsubsec:global_snls}.
We also performed our own local photometry on SDSS images  for host galaxies with low and intermediate redshift. The local photometry was computed with the Python tool
\texttt{aperture\_photometry} from the \texttt{photutils} module of
\texttt{Astropy}. The errors for the local photometry were estimated in the same way as for global photometry, as described in Section~\ref{subsec:global}. After the variance map was built starting from single-epoch images, we measured the error by integrating inside the local photometric radius.
As an example, a few image tiles are presented in Fig.~\ref{fig:examples_vignettes}, centered on the positions of the SNLS and SDSS host galaxies. The  global photometry ellipse aperture and the local photometry circular aperture are indicated.  In the SDSS cases, the {\em high-deblend} and {\em low-deblend} configuration ellipses are shown in white and red, respectively. The examples illustrate the redshift range  that each survey achieves. A more quantitative comparison between local and global quantities is presented in Section~\ref{subsec:loc_vs_glob}.

\subsection{Host identification}
\label{subsec:host_id}
For the SNLS survey, the supernova host was identified using two criteria: the minimum
distance to the supernova location, and the match between the
photometric redshift of the galaxy (see section
\ref{subsec:sed_fitting}) and the spectroscopic redshift of the
supernova \citep{Howell05,Bronder08,Ellis08,Balland09,Walker11,balland17}.

A normalized elliptical distance $d$ was computed using the \texttt{SExtractor} shape
parameters and Kron elliptical radius, so that $d < 1$ defines the
photometry elliptic aperture of the galaxy.  The supernova host
galaxy was identified as the galaxy that lay closest to the
distance $d$. When the closest galaxy lay at $d>1.8$, the supernova
was declared to have no host.  To unambiguously identify the host, we also required that the photometric
redshift of the closest galaxy was consistent with the spectroscopic redshift of the supernova:
$\Delta z /(1+z)<0.15$.  When more than one galaxy was detected close
to the supernova location within $d<1.8$, the galaxy with the closest
photometric redshift was selected as the host. Ambiguous cases were
flagged as problematic, for instance, when more than one of these close galaxies
met the criterion $\Delta z /(1+z)<0.15$. We also rejected SNIa for which
the identified host might be polluted by a nearby bright star.
These criteria ensure a secure host galaxy
identification. We identified a host galaxy for $\sim$ 87\% of SNLS
SNIa and did not detect a host for 6\%. The remaining 7\% SNIa were either
dubious or problematic cases.

For the SDSS and low-redshift surveys, we identified the host galaxies with a different algorithm, which yielded similar results: the galaxy with the largest overlap
between a circle surrounding the SNIa location and the \texttt{SExtractor} segmentation
map was defined as the parent galaxy of the supernova. When there was no overlap between the SNIa region and a galaxy, we selected the closest object from the segmentation map, as for SNLS. 
%Therefore, the
%\textit{global} photometry corresponds to the flux integrated in the
%ellipse associated to the host galaxy using the Python tool
%\texttt{aperture\_photometry} from the \texttt{photutils} module of
%\texttt{Astropy}. 
Figure \ref{fig:examples_vignettes} illustrates the difficulties encountered when identifying the host galaxies.
 {\em{sn2006hr}} (top right) seems to be occurring in a merger, and {\em{05D3ht}} (bottom right) is just below the normalized elliptic distance limit from its assumed host.
In  these cases, the local color is more reliably related to the SNIa environment than to the stellar mass, which requires a rigorous definition of the host galaxy area.

Host galaxy photometry from SDSS has also been published in S14, with a more sophisticated host identification algorithm. When comparing our own SDSS host ID number, obtained from the SDSS database by matching objects found in single-epoch images, with the number from S14 for the host galaxies that we share, we find that 11\% of them do not match. In appearance, more than 20\% of the hosts do not match S14. The ID numbers are different only because the association with the SDSS database has been
made using single-epoch images. As a result, no change is needed for the corresponding hosts. The large majority of them occur in configurations where the host galaxy is faint ($g\sim21.5$), with a relatively high spectroscopic redshift ($z\sim0.26$). About 70\% of the non-matching cases correspond to configurations where the stacked images are not available, and the host identification was thus performed using single-epoch images. The identification even fails for the remaining 30\% host galaxies that are found in stacked images. This corresponds to only 13 hosts. 
For all non-matching hosts, we selected hosts properties (stellar mass and $U-V$ color) with the identification method S14.
For more details about comparing host properties with S14, see Section~\ref{subsubsec:sako}.
The efficiency of host identification methods has been described
by \cite{gupta06}, where normalized elliptical distances were found to be more accurate. Machine-learning algorithms were also studied in the context of future large SN surveys.

%Indeed, since the association with the SDSS database is performed using single-epoch images, the magnitude depth is not sufficient to extract from the background some of the faintest hosts. Therefore, computing the ($\alpha$,$\delta$) distance between the host from S14 and the host identified by our algorithm conveys additional information.
%For more than half of the non-matching cases, the distance test ensures us that identified hosts are correct. The SDSS IDs are different compared to S14 only because the association with the database is done using single-visit images. As a result, no change is needed here. Concerning the other half, most of SNIa have no available stacked images (66\%). A fraction of them are indeed found in stacked images, but our identification fails. This corresponds to only 16 high redshift SNIa. 

\begin{figure}[!htbp]
  \begin{center}
    \includegraphics[width=0.9\columnwidth]{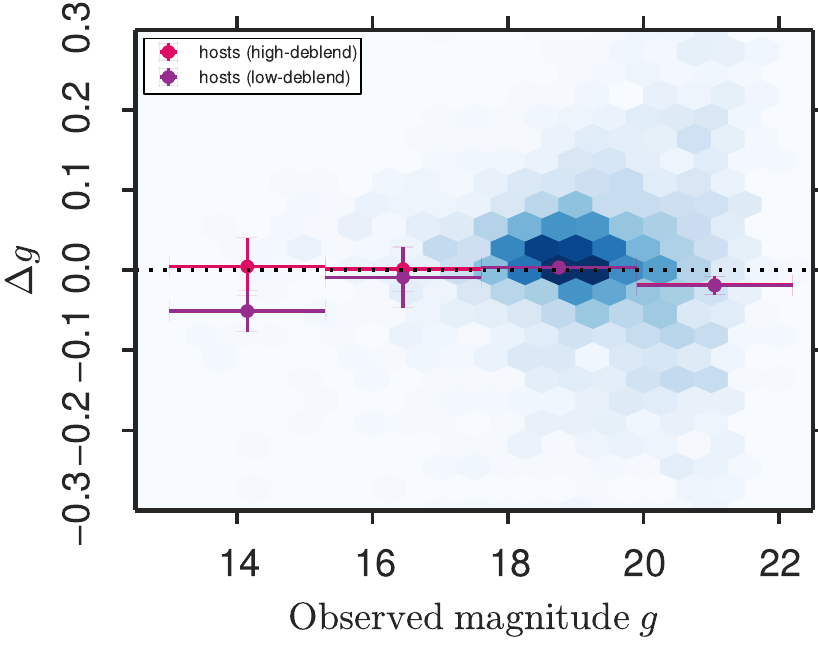}
  \end{center}
  \caption{Density of points (blue hexagonal bins) representing the absolute difference between observed aperture
    magnitudes and SDSS DR12 magnitudes in the $g$ optical band for
    all field galaxies detected in the low-deblending configuration.
    %(cloud of light pink points in the background) and for host
    %galaxies of SNLS SNIa (pink points).
%Field galaxies detected through the low-deblending configuration are shown with a light purple cloud of points, host galaxies being represented by the purple points.    
    Although the dispersion is relatively 
    high, the overall agreement between the two photometries is good
    (see text). When split into
    magnitude bins for host galaxies (pink and purple bins for high- and low-deblending modes, respectively), the two configurations are similar, except for the
    first bin, where the low-deblending case shows less consistency
    with SDSS.}
  \label{fig:comparison_dr12_hosts}
\end{figure}

\subsection{Comparison with SDSS DR12 photometry}
\label{subsubsec:dr12}
Since we estimated the photometry of
hundreds of host galaxies form SDSS images in order to infer their color and stellar
masses, we checked that our measurement was consistent with published
data from SDSS DR12 for all field galaxies detected by
\texttt{SExtractor} in those images.
As described in Section~\ref{subsec:global}, we used
\texttt{SExtractor} in two different configuration modes.
%: a high-deblending mode above $z=0.05$ and a low-deblending mode, so that nearby galaxies below $z=0.05$ are not seen as multiple independent objects instead of galaxies as a whole.
After building the low-deblending and the high-deblending catalogs based on SDSS single images (see Section~\ref{subsec:global}), we matched every source with the SDSS database, restricted to {\em{primary}} objects identified as galaxies with a valid spectroscopic redshift. To ensure that the
comparison was made for the same objects on both sides, we required an association distance shorter than one pixel together with the {\em{clean}} photometry flag from the SDSS database. The magnitude comparison presented here was then performed for 2676 field galaxies using the high-deblending configuration, and 2624 field galaxies using the low-deblending case. These galaxy populations that contain our host galaxies are thus representative of them.

Using the same technique as for the host galaxies, we computed the aperture photometry of all
field galaxies in $ugriz$ bands using a circular aperture for point-like objects
(i.e., detected with a zero Kron radius by
\texttt{SExtractor}) and an elliptical aperture for extended sources
(i.e., detected with a non-zero Kron radius) with the same photometric tool.
Here, we compare our photometry with
different SDSS magnitude definitions. We chose the {\em{model}}
magnitudes to compute the magnitude difference for objects identified
as point-like objects by \texttt{SExtractor} that corresponded to a fixed circular aperture for our photometry.
On the other hand, we took the SDSS {\em{petrosian}} magnitudes when comparing objects seen as extended and corresponding to our elliptical aperture photometry.
Petrosian magnitudes are indeed better adapted for a comparison
of extended source photometry since galaxy fluxes are measured within a circular aperture whose radius is defined by the shape of the azimuthally averaged light profile. The overall result is presented in Fig.~\ref{fig:comparison_dr12_hosts}.

For the $g$-band and taking into account the SDSS magnitude definitions described above, we find a good agreement between the two magnitude estimates. For the high-deblending configuration mode, we obtain $\Delta g=0.008\pm0.080$. For the low-deblending configuration, we have $\Delta g=0.007\pm0.079$, in which error bars represent one standard deviation of the distribution.
In this sample of field galaxies, we can find the subsample corresponding to host galaxies of our SDSS and low-redshift surveys SNIa. Because
the magnitude comparison was computed in the same way as for the whole sample, we obtain a good overall agreement with $\Delta g=-0.001\pm0.076$ using the high-deblending mode and $\Delta g=0.014\pm0.074$ using the low-deblending mode. 

\begin{table}[!htbp]
\centering
\begin{tabular}{ c || c | c }
%\hline
&\multicolumn{2}{ c }{\em{low-deblending}}\\
\hline
Band & Field galaxies & Host galaxies \\
\hline
$u$ & $-0.029\pm 0.309$ & $-0.048\pm 0.307$\\
$g$ & $0.007\pm 0.079$ & $-0.003\pm 0.077$\\
$r$ & $0.019\pm 0.077$ & $0.014\pm 0.060$\\
$i$ & $0.013\pm 0.086$ & $0.009\pm 0.080$\\
$z$ & $-0.004\pm 0.125$ & $-0.027\pm 0.123$\\
%\hline
\end{tabular}
\caption{Mean value and standard deviation for the difference between our observed magnitudes and magnitudes from the single-epoch SDSS catalogs for field galaxies and the subsample of the SNIa hosts. The low-deblending \texttt{SExtractor} configuration file is chosen.}
\label{tab:comparison_dr12}
\end{table}

When we split these results into different magnitude regions, we find
that in the lowest magnitude bin, the two different deblending
settings differ by more than $1\sigma$. This means that the low-blending case is less consistent with SDSS magnitudes than the
first configuration (see the two leftmost bins in Fig.~\ref{fig:comparison_dr12_hosts}). This can be explained by the
fact that for bright (hence large and nearby) galaxies, a higher
deblending threshold should indeed be chosen so that galaxies are not
separated into multiple portions. This corresponds to our low-deblending configuration. In Section~\ref{subsubsec:global_sdss} we illustrated the phenomenon in Fig.~\ref{fig:sex_config} with the white and
red ellipses around nearby host galaxies.
Since we focus on primary galaxies for the purpose of a direct photometry comparison, on occasion we do not consider SDSS {\em{parents}}
for some objects, but objects flagged as {\em{child}} (in principle, subparts of the {\em{parent}} image). For example, a parent object can be a bright star, while an object flagged as its child is a galaxy a few arcseconds away. In the framework of this comparison, we selected only primary galaxies and thus find a slight magnitude difference for the brighter galaxies.

%We believe that a single configuration has been chosen for SDSS DR12 magnitudes with a relatively high deblending, thus our inconsistency for bright galaxies.

Table~\ref{tab:comparison_dr12} lists the offsets we
find between our observed magnitudes and the DR12 magnitudes for the five optical bands for all field galaxies and our subsample of SNIa hosts
using the low-deblending configuration. In the comparison, we  distinguish between extended
and point-like sources from the SDSS point of view, as explained above.

%\begin{equation}
%\begin{array}{l c r}
%\Delta u &=& -0.204\pm 0.333 \\ 
%\Delta g &=& -0.009\pm 0.126 \\
%\Delta r &=& 0.023\pm 0.099 \\
%\Delta i &=& 0.025\pm 0.128 \\
%\Delta z &=& 0.025\pm 0.190 
%\end{array}
%\label{eq:offsets}
%\end{equation}

Except for the $u-$ and $z$-band showing higher dispersion than the
other three, the consistency between SDSS magnitudes and our photometry for field galaxies and parent
galaxies of our SNIa sample is good overall. The remaining differences
lie in the method details of the object extraction.
%Moreover, dispersions are largely reduced once considering only relatively bright galaxies.
In the following, we use our own measurement for the galaxy photometry in order to remain consistent for all surveys.

\subsection{Spectral energy distribution fitting}
\label{subsec:sed_fitting}

All our surveys benefit from precise multiband photometry, from
which we may estimate the galaxy global properties, such as
metallicity, age, color, stellar mass, or star formation rate.

These quantities were derived by fitting a series of template galaxy
SEDs to the $ugriz$ fluxes available for each galaxy. The SED series
was computed using the synthetic evolutive optical galaxies spectra
code \texttt{PEGASE.2} \citep{pegase1,pegase2}. Eight scenarii were provided to
\texttt{PEGASE}, as published in \cite{LeBorgne2002}, corresponding
to a star formation history and an evolution of the gas content
corresponding, at $z=0$ for a given age, to the morphological types E,
S0, and spirals from Sa to Sd. We assumed a \cite{Rana1992} initial
mass function. The synthetic SEDS were computed at 68 different ages
from 0 to 15 Gyr, and the galaxy properties such as the star formation rate (SFR) or the
stellar mass were provided by the code.

The relevant SED was selected via a $\chi^2$ minimization
performed with our own package, which takes into account the extinction
from the Milky Way as estimated using the $E(B-V)$ map from
\cite{planck_dust_13} and allowing for an additional extinction
described by the \cite{Cardelli89} extinction law.
Galaxy properties were then obtained from the best-fitting SED. The
stellar mass was estimated in units of $\mathrm{log}_{10}\mathcal{M}_\odot$ to a
precision better than 11\% for SNLS, better than 23\% for the SDSS survey, and better than 7\% for
the fraction of hosts of the low-redshift sample for which we have
{\em{ugriz}} photometry.

To provide rest-frame colors as closely
interpolated from the observer frame data as possible, we
used 
%Rest-frame color and absolute magnitudes are estimated by fitting the photometric measurement with
a spectral template library
developed for this purpose, \texttt{EGALITE} (Empirical GALactic InTErpolator). As described in
\cite{Kronborg2010}, the building of this library is a two-step
procedure.  We first computed a series of eight spectra using
\texttt{PEGASE}, specifying eight SFR laws proportional to $(t/\tau)\times
\exp(-t/\tau)$ and a given galaxy age $a(\tau)$. The eight synthetic
spectra were indexed using a single parameter that is equivalent to
the mean age of their stellar population $a_\star$ and then interpolated to obtain a continuous spectral sequence
indexed by $a_\star$.
%The initially zero gas metallicity evolves through the successive generations of stars. Extinction is computed using a transfer model for an inclination-averaged disk distribution,the optical depth being estimated from the mass of gas andthe metallicity. 
The continuous spectral sequence was then optimized to refine the data description. For this, the library was trained using the magnitudes of
$\sim 6000$ galaxies in the D3 field catalog, with known
spectroscopic redshift from the DEEP-2 spectroscopic survey
\citep{Davis2003, Davis2007}.  The training procedure corrects the
initial templates by as much as 30\%. Therefore, the galaxy
parameters provided by \texttt{PEGASE} for each of the eight initial
templates, such as the star formation rate or the stellar mass, are no
longer relevant.
The interpolated colors, however, are expected to be closer to the truth because the spectral sequence is specifically trained to describe the data better.

The accuracy of the obtained photometric redshifts was estimated using
spectroscopic data from the VIMOS VLT Deep Survey (VVDS) \citep{Lefevre2004} and the
Cosmic Evolution Survey \citep[COSMOS,][]{Scoville2007}: at limiting
magnitude $i<24$, the rate of catastrophic error $\Delta z/(1+z)>0.15$
is 3.15\%, and the precision is $\sigma[\Delta z/(1+z)] = 0.030$. Photometric redshifts are not used in this analysis.

We fit the host galaxies photometry using \texttt{EGALITE} via a $\chi^2$ minimization performed with our own package involving  the $ugriz$ photometric fluxes that we measured for each galaxy. As described above, we took the Milky Way extinction into account,
and we allowed for an additional
extinction. Observed fluxes $\mathcal{F}$ should be modified compared to the initial fluxes $f$ that we measured with a given measured zero-point $zp$ as
\begin{equation}
\mathcal{F} = f\times e^{-0.4\mathrm{ln}(10)zp}.
\end{equation}
Rest-frame $U-V$ galaxy colors were measured with a mean precision of 13\%, 14\%, and 9.7\%  for SNLS, the SDSS survey, and the
nearby SNIa sample, respectively.
In the following, the spectral sequence
\texttt{EGALITE} is used as a photometry interpolator in order to
compute photometric redshifts and rest-frame local and global colors. 
Stellar masses are estimated using the PEGASE spectral sequence.

\section{Consistent catalog of stellar mass estimates}
\label{sec:mass}
We present here the derivation of the supernova host galaxy stellar masses. Depending on the SNIa sample, stellar masses were derived using different methods applied on different sets of global photometric measurements: SED fitting on $ugriz$ magnitudes measured on SDSS and SNLS imaging data, or the mass  from the Two Micron All Sky Survey near-infrared (NIR) $K$ band. To ensure a consistent host mass catalog,  we intercalibrated these different methods using overlapping data of the surveys. We finally compare the mass catalog to published values for the corresponding host subset.

\subsection{Intercalibration of mass estimates}
\label{subsec:intercalib_snls_sdss}
\begin{verbbox}

SELECT
str(p.ra,13,8) as ra,str(p.[dec],13,8) as dec,
p.modelMag_u,p.modelMag_g,p.modelMag_r,
p.modelMag_i,p.modelMag_z,
p.modelMagErr_u,p.modelMagErr_g,p.modelMagErr_r,
p.modelMagErr_i,p.modelMagErr_z
FROM dr8.PhotoObj AS p JOIN
dbo.fGetObjFromRect(214.042448,215.519408,
52.160128,53.20855) AS b ON p.objID = b.objID
WHERE  (p.type = 3 OR p.type = 6)
\end{verbbox}
We computed the possible bias existing between galaxy stellar
mass estimates from $ugriz$ photometry given by the MegaCam instrument
on the CFHT, and the same mass estimates from similar filters on the
SDSS instrument. For this purpose,
we used the spectral sequence \texttt{PEGASE} and our own fitting code to derive the galaxy stellar mass as described in Section~\ref{subsec:sed_fitting}.
We considered the D3 deep
field of the SNLS survey, located between $214.04^\circ$ and
$215.52^\circ$ in right ascension and between $52.16^\circ$ and
$53.21^\circ$ in declination, which has a large overlap with the
SDSS footprint. We find about 2400 galaxies in common among the
$\sim6000$ galaxies with a spectroscopic redshift in the SNLS D3
catalog and the $\sim20000$ objects found in the same right-ascension and declination area in the SDSS
database%
\footnote{
We used the SDSS CasJobs server at
\url{http://skyserver.sdss.org/CasJobs} with the following \texttt{SQL} command: \theverbbox}
by imposing that galaxy coordinates should match below $10^{-4}$ degrees, or 0.36''.
We applied selection cuts to the outcome of our
query to improve its quality in the following way.
\begin{itemize}
\item We kept galaxies that satisfied $0.1<z<0.8$;
\item We removed galaxies for which the $u$-band magnitude was not properly measured by the SNLS.
\item We discarded all galaxies that lay more than $5\sigma$ away from the mean of
  the distribution in $g$ and $r$ bands (i.e., $\Delta g/\sigma_g<5$
  and $\Delta r/\sigma_r<5$).
\end{itemize}

\begin{figure}[!htbp]
  \begin{center}
     \begin{tabular}{c}
    \includegraphics[width=0.9\columnwidth]{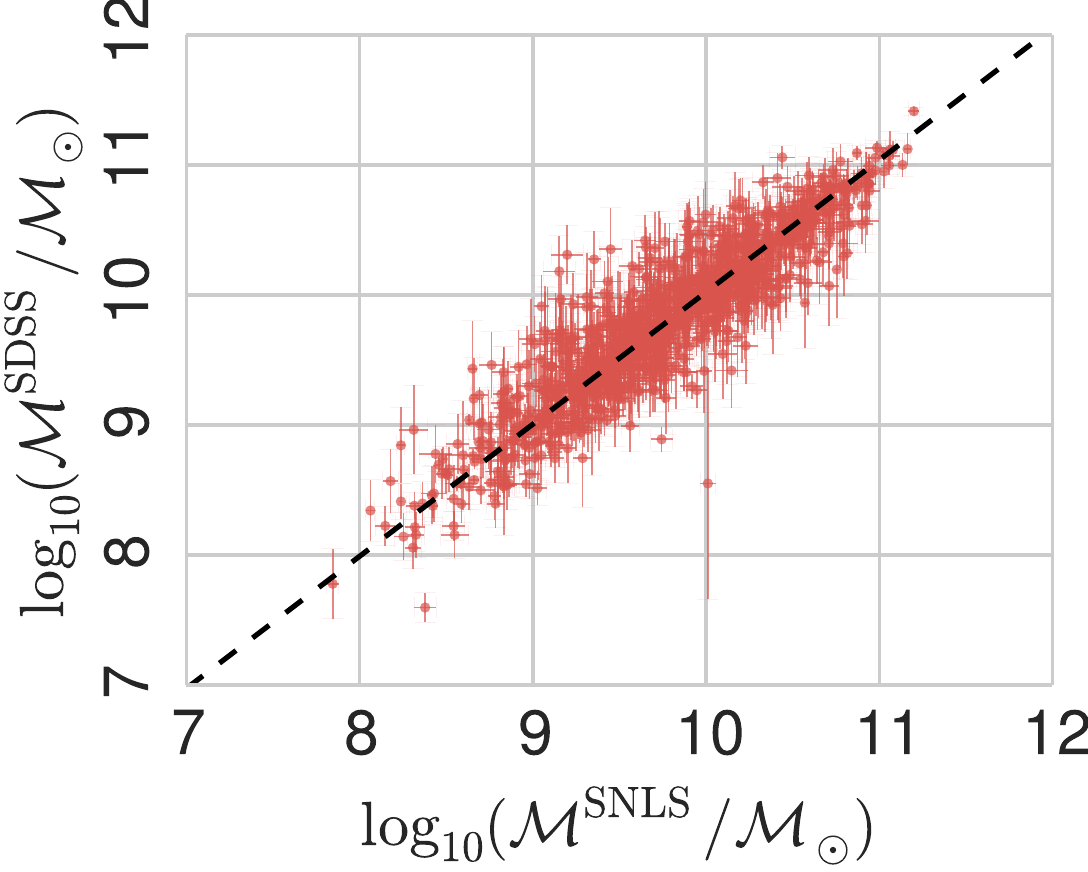} \\
    \includegraphics[width=0.9\columnwidth]{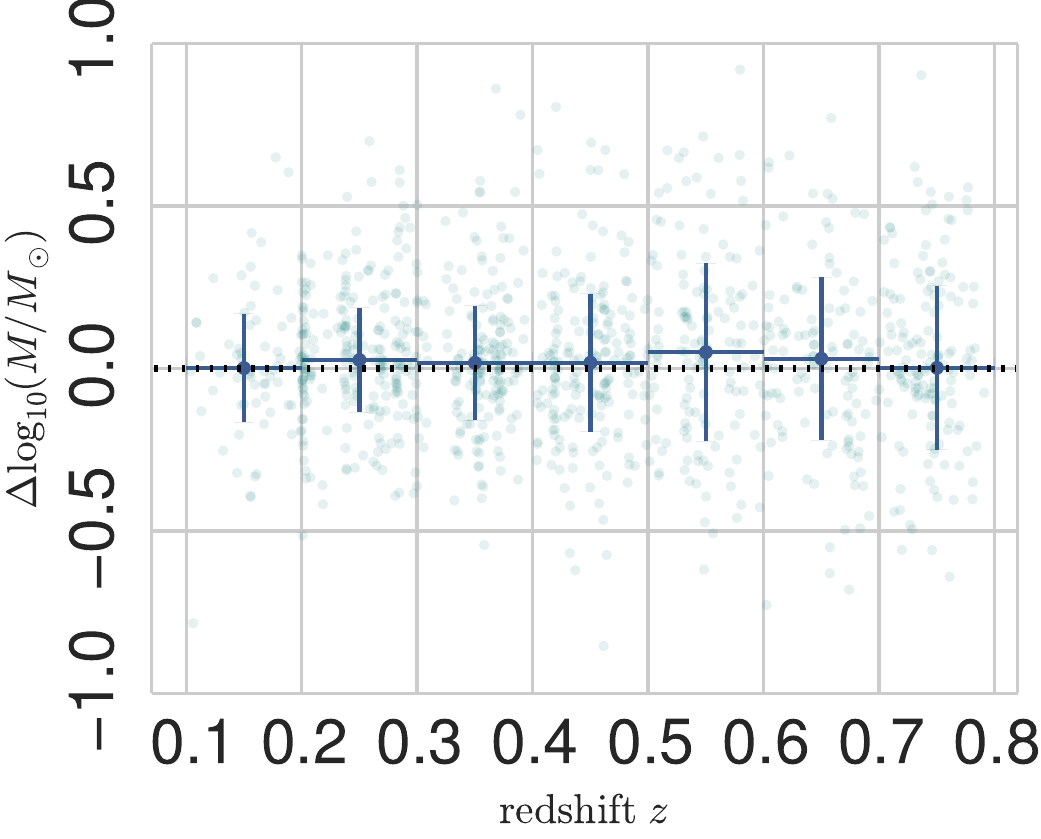}
     \end{tabular}
  \end{center}
    \caption{\textbf{Top:} Distribution of galaxy stellar masses obtained from two different photometries from the SNLS and the SDSS surveys. All galaxies are located in the D3 deep SNLS field. Stellar masses are drawn as red points in the top panel; the resulting weighted linear fit is shown as the dashed black line. \textbf{Bottom:} Redshift dependence of the stellar mass difference. The median rms is 0.21 dex.}
  \label{fig:msdss_vs_msnls}
\end{figure}

%\begin{figure}[!htbp]
%  \begin{center}
%    \includegraphics[width=\columnwidth]{figs/msdss_vs_msnls.pdf}
%  \end{center}
%  \caption{Distribution of galaxy properties for 1700 elements
%    obtained from two different photometries coming from the SNLS and
%    the SDSS surveys. All galaxies are located in the D3 deep SNLS
%    field. Stellar masses are drawn as red points in the top left
%    panel, with the resulting weighted linear fit as the thick dashed black
%    line. The specific star formation rate (sSFR) is represented by
%    the blue points in the top right panel. The bottom left panel
%    shows the galaxy age estimated from both photometries, and the
%    bottom right panel presents the $U-V$ rest-frame colors. Linear
%    relations to the two different plotted variables with a slope of 1
%    are displayed as grey dashed lines.}
%\end{figure}

After these cuts, the sample consisted of about 1700 elements for which
we obtained two different mass estimates based on SNLS and SDSS
photometries. The estimates are displayed in Fig.~\ref{fig:msdss_vs_msnls} as
red points in the top panel. By fitting a linear function to the
logarithmic stellar mass distribution, we obtained a slope of $1.017\pm0.072$ and a
$y$-intercept at $-0.146\pm0.227$. The corresponding fit is drawn as a black
dashed line in Fig.~\ref{fig:msdss_vs_msnls}. As a consequence of the
outcome of the fit, we can equally use SNLS or SDSS photometries
to compute galaxy stellar masses, or any other galactic information. We checked that the redshift
dependence of the difference between stellar masses from SNLS and SDSS
was not significant (see the bottom panel of Fig.~\ref{fig:msdss_vs_msnls}). Moreover, we verified the good correlation between the rest-frame color $U-V$ estimated from SNLS and SDSS observed magnitudes with the \texttt{EGALITE} spectral sequence.
%other global variables (star formation rate, age, rest-frame colors) estimated from SNLS and SDSS observed magnitudes.
%In Fig.~\ref{fig:msdss_vs_msnls} are displayed the correlations between stellar masses obtained from the two different photometries.
%those three other galactic properties.

\subsection{Mass estimates from the 2MASS survey}
\label{subsec:intercalib_2mass}
%\todo{[Sur quel \'echantillon calibrer la loi ? Tester avec la nouvelle photom\'etrie.]}

As described in more detail in section
\ref{subsec:available_host_photo}, some nearby host galaxies
 are not located in the SDSS footprint and have no corresponding
photometry in $ugriz$ bands. Thirty-nine of these have photometry
in the 2MASS All-Sky Data Release from the {\em{Two Micron All Sky Survey,}}  in particular, in the NIR $K$ band. The NIR luminosity traces the emission from low-mass stars that best represents the build-up of stellar material
and thus provides a reliable estimate of the stellar mass of a galaxy because the NIR mass-to-light ratio is nearly independent of star formation history \citep{bell03}. Following the method described in Appendix C of \cite{Betoule14}, we can take advantage of the existing NIR data to infer the host galaxy stellar mass when the galaxy lacks $ugriz$ photometry.

In order to calibrate the relation between galaxy stellar mass and $K$ absolute magnitude, we completed part of the $ugriz$ background galaxy catalog we built on SDSS images as described (see Section~\ref{subsubsec:dr12}), with $K$-band photometry extracted from the SDSS database. We selected objects with a valid spectroscopic redshift and restricted the sample at $z<0.15$. This calibration sample comprises 349 galaxies and has a median redshift of 0.07. We then estimated the stellar mass $\mathcal{M}$ for each galaxy from the $ugriz$ magnitudes we measured on SDSS images with the PEGASE template fit described in Section~\ref{subsec:sed_fitting}.

%In order to calibrate the relation between galaxy stellar mass and $K$
%absolute magnitude, we extract photometric data in $K$-band of 437
%primary objects from the SDSS database\footnote{We consider the 2MASS
%  Kron elliptical aperture magnitude, which is a constant quantity
%  with redshift.}. Those objects actually are the background galaxies
%we detect with \texttt{SExtractor} on the SDSS images containing the
%host galaxies of our SNIa sample.
%Since this section concerns nearby galaxies, we only consider the 327 field galaxies associated to a redshift below 0.15.
%We thus estimate from the SDSS images the $ugriz$
%magnitudes from the same pipeline than for host galaxies, and will be
%able to compare the model stellar mass given by the $K$ absolute
%magnitude and the one we find from SDSS optical bands. 

The $K$
absolute magnitude is defined as
\begin{equation}
M_{\mathrm{K}}=m_{\mathrm{K}}-\mu(z)-k(z),
\label{eq:magk}
\end{equation} 
where $m_{\mathrm{K}}$ is the 2MASS Kron elliptical-aperture magnitude in the $K$ band,
$\mu(z)$ is the distance modulus computed using the galaxy redshift and $H_0=70$ km/s/Mpc. The $K$-correction is defined by $k(z)\sim-2.1z$ as in \cite{bell03}. Given the stellar mass
estimated from the same \texttt{PEGASE} fit based on $ugriz$
magnitudes, we adjusted a weighted linear fit to the $M_{\mathrm{K}}-$stellar
mass distribution $\mathcal{M}$ and obtained
\begin{equation}
\mathrm{log}_{10}(\mathcal{M}/\mathcal{M}_\odot)=-0.405\ (\pm0.003)\ M_{\mathrm{K}}+1.015\ (\pm0.080).
\label{eq:fit_magk}
\end{equation}
We recovered the slope of
$\sim-0.4$ expected from the model of the stellar mass-luminosity ratio from
\cite{bell03}. 
%As we expect a slight change in the relation above depending on redshift, we also fit two other relations above and below $z=0.05$:
%\begin{equation}
%\begin{array}{l c r}
%\mathrm{log}_{10}(\mathcal{M}/\mathcal{M}_\odot)=-0.421\ M_{\mathrm{K}}+0.549 \ (z>0.05)\\
%\mathrm{log}_{10}(\mathcal{M}/\mathcal{M}_\odot)=-0.411\ M_{\mathrm{K}}+0.908 \ (z<0.05)
%\end{array}
%\label{eq:fit_magk_z}
%\end{equation}
%In the following, we will use the adapted $M_{\mathrm{K}}-\mathcal{M}$
%relation depending on the object's redshift. 
Because we know the masses
of these calibration objects from two different estimators, we compared the two
estimated values and obtained a dispersion around the fit of 0.121 dex. 
The relationship between the galaxy mass essentially depends on its absolute $K$ magnitude, but also on its rest-frame $U-V$ color, therefore we observed a color dependence of the fit residuals proportional to $\sim0.3(U-V)$.
We also observed a residual redshift dependence proportional to $z$, which might be due to the limiting $K$ magnitude of the
surveye.
As a consequence, we added in quadrature to the residuals of 0.121 dex a term that takes the redshift bias (0.1) into account, and another that included the color bias (0.2). The total uncertainty is then about 0.25 dex,
and this calibration is precise enough for the purpose of this analysis.
With this calibrated relation between stellar mass and $K$ absolute magnitude in hand, we are able to infer the stellar mass of host galaxies that lack $ugriz$ photometry and complete our catalog of stellar mass estimates in a consistent
way.

%\begin{figure}[!htbp]
%  \begin{center}
%    \includegraphics[width=\columnwidth]{comparison_magk_vs_z_sns.pdf}
%  \end{center}
%  \caption{Redshift evolution of host stellar masses differences between our two estimators (the $K$-band magnitude law and \texttt{PEGASE} fits from SDSS magnitudes). The dashed black line corresponds to the mean difference, with an error bar represented by the grey area.}
%  \label{fig:comparison_magk}
%\end{figure}

In our low-redshift sample, we found 69 parent galaxies from low-redshift SNIa surveys
with photometry from both SDSS and 2MASS, which
we completed with 15 additional intermediate-redshift hosts from the SDSS that are also listed in
the 2MASS extended source catalog. Thus, we can compare stellar mass
estimates from \texttt{PEGASE} fits with $ugriz$ observed
magnitudes using the {\em{low-deblend}} (see Section~\ref{subsec:global}), and stellar mass
estimated with $K$-band absolute magnitudes for this subsample of 84
hosts. The subsample mainly consists of nearby galaxies with a median
redshift of 0.031, of which 80\% were observed in single-epoch images.  We derived the host stellar
masses from Eq.\ref{eq:fit_magk}.  Computing the
absolute stellar mass difference, we found a good overall consistency
of $\Delta\mathrm{log}_{10}(\mathcal{M}/\mathcal{M}_\odot)=0.004\pm0.007$
dex.
%Those results are illustrated in Fig.~\ref{fig:comparison_magk}.
Thus, we can estimate the galaxy stellar mass
from our host catalog either using SDSS observed magnitudes, or with the $K$-band absolute
magnitude, when available.

\subsection{Comparison with public mass catalogs}

\subsubsection{Nearby host galaxies from N09}
\label{subsubsec:neill}

\begin{figure}[!htbp]
  \begin{center}
    \includegraphics[width=\columnwidth]{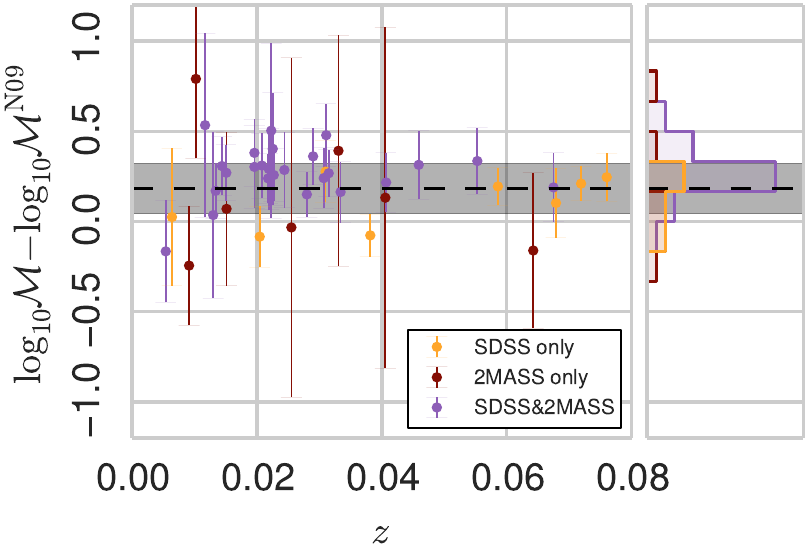}
  \end{center}
  \caption{Redshift evolution of the differences in common host stellar mass between N09 and our
    sample. Galaxies with only SDSS-like photometry are
    represented in orange. Their masses are estimated by fitting the spectral series \texttt{PEGASE} to our observed $ugriz$ magnitudes. Galaxies presenting photometry in the 2MASS
    survey alone are drawn in dark red, their masses are derived from  $K$-band absolute magnitude through Eq.\ref{eq:fit_magk}.
 Stellar mass differences of objects with both types of
    photometry are displayed in purple. For these galaxies, we present masses derived from $ugriz$ observed magnitudes.
    The dashed black line corresponds to the
    mean difference, with the rms represented by the gray area.}
  \label{fig:comparison_neill}
\end{figure}

A catalog of parent galaxy properties of 168 low-redshift SNIa has been published by N09, and it has been
used in many other analyses \citep{Kelly10,Sullivan10,Betoule14}. The
authors used UV imaging with measurements from the homogeneous
magnitude-limited sample {\em{Galaxy Evolution Explorer}} survey \citep[GALEX,][]{Martin05}. Observed magnitudes in the UV range were
obtained by performing surface photometry in elliptical apertures on
sky-subtracted images. For optical photometry, they considered RC3
integrated Johnson UBV magnitudes \citep{dv91} obtained from the
NASA/IPAC Extragalactic Database (NED) and/or SDSS images in the
$ugriz$ bands. For larger hosts, the SDSS catalog was found to be
inaccurate, thus elliptical-aperture photometry was performed on coadded
SDSS images.
Forty-one host galaxies belong to our catalog and the N09 catalog.
%We find in total 41 common host galaxies matching both our catalog and the N09 catalog.
As the photometry is not public, we
directly compared the host masses and distinguished between our two
methods to infer stellar masses (from SED fitting based on $ugriz$
SDSS elliptical aperture magnitudes or from the $K$ optical band
magnitude). The result is shown in Fig.~\ref{fig:comparison_neill}.
Of the 41 common hosts
\begin{itemize}
\item 8 only have SDSS $ugriz$ photometry;
\item 8 only have $K$-band photometry in the 2MASS survey. One
of these corresponds to interacting galaxies and was not identified as the same host in N09. Consequently, the corresponding point does not appear in the figure;
\item 25 have both SDSS $ugriz$ magnitudes and $K$-band magnitude. It is thus equivalent to
  fit the spectral series \texttt{PEGASE} to our $ugriz$ magnitudes or the relation written in
  Eq.\ref{eq:fit_magk}, as demonstrated in section
  \ref{subsec:intercalib_2mass} We used stellar masses derived from $ugriz$ photometry.
\end{itemize}
For the 40 common host galaxies, we find that N09 tends to
underestimate stellar masses by about $0.18\pm0.14$ dex on average compared
to our mass values.
%In the following, we will use host stellar masses derived by our own method.

\subsubsection{Intermediate-redshift host galaxies from S14}
\label{subsubsec:sako}

The data release of the SDSS-II supernova survey has been
compiled by S14 and corresponds to the period between 2005 and 2007. Thus, the overlap with our SNLS 5 year SNIa catalog
is relatively extensive. With their own host identification algorithm
and using a similar SED-fitting code based on a \texttt{PEGASE} spectral template library to obtain galaxy properties, the
authors published magnitudes, redshifts, stellar masses, specific
star-formation rates, and ages for more than 400 host galaxies.
The goal now is to compare the common host galaxy properties derived by S14 with our own estimates that were directly measured in SDSS images.
When
we compared our stellar mass estimates, we focused on objects for which
the host ID was identical to the ID in our sample (see Section~\ref{subsec:host_id}), and for which the global photometry was correctly measured (i.e., we removed galaxies for which the photometry in the $u$-band was not measured). This permits
a comparison of 252 objects.

\begin{figure}[!htbp]
  \begin{center}
    \includegraphics[width=\columnwidth]{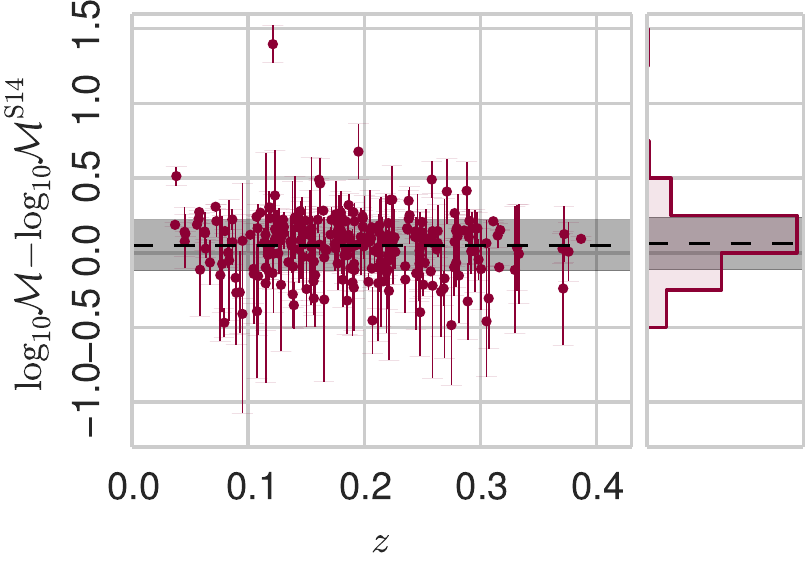}
  \end{center}
  \caption{Redshift evolution of stellar mass differences in common hosts between S14 and our sample for objects with the same identification number (dark red points). This corresponds to intermediate-redshift galaxies detected in the SDSS survey.
%For 74 host galaxies, the difference between the two host stellar mass estimates is important. Those are configurations where the supernova explodes far away from faint high redshift host galaxies which are wrongly identified in our method (light blue points, see text). None of the corresponding SNIa remain after selection cuts.
%Excluding the bad identification cases, 
We find a good agreement between the two mass estimates, with a slight mean overestimation of stellar masses of about 0.07 dex compared to S14 (dashed black line; the gray area represents the rms).}
  \label{fig:comparison_sako}
\end{figure}

The differences between the two stellar mass estimates are represented as red points in Fig.~\ref{fig:comparison_sako}, and they are distributed as a function of redshift. We obtain a mean absolute difference of
$\mathrm{log}_{10}\mathcal{M}-
\mathrm{log}_{10}\mathcal{M}^{S14}=0.072\pm0.011$ dex, which
illustrates the  good agreement with a reasonable dispersion (0.17 dex) between the masses obtained from different host identification algorithms and distinct photometric tools.
%This is done excluding the 74 host galaxies shown as light blue points showing a large mass difference between S14 and our estimation. They correspond to SNIa exploding far away from a faint and relatively high redshift host galaxy ($z\sim0.2$). In that configuration, our host identification method fails.
%As none of the corresponding SNIa enter our final sample after selection requirements, this identification problem does not impact our conclusions.
%When we restrict ourselves to object with the same identification number in SDSS (208 objects), we get a reasonable agreement between S14 and our host stellar mass estimates at the level of 0.05 dex.
We nonetheless find that the S14 mass values are slightly lower than those we obtained with our own photometric method in SDSS images. The difference is not very significant and well below what is required for the analysis presented here.

\section{Environmental dependence of supernova luminosity}
\label{sec:analysis}
We have measured the photometry and local photometry of the host
galaxy inside a 3 kpc radius around the SNIa explosion region. The photometry was transposed into $U-V$ rest-frame color. This section presents the construction of a clean sample using selection requirements based on the local and global measurement quality
and then compares local and global information. Finally, we try to use the local environment as a third standardization variable.

% We present the results of the analysis comparing correlations of
% Hubble diagram residuals with galaxy host properties and local
% environment characteristics. First, cuts are described in
% details. Then, we present how global and local quantity compares and
% what are the significance of the correlations with Hubble diagram
% residuals. Finally, we demonstrate the robustness of our local
% signal.

\subsection{Selection requirements}
\label{subsec:cuts}
To be included in the sample, supernov\ae\ and results of the local photometry of our host galaxies must satisfy several conditions.
%Prior to the photometry measurement, SNIa are fit with the same version of the SALT2 model than the one used in \citep{Betoule14}. We therefore have an estimate of the supernova light-curve parameters.

\begin{table*}[!htbp]
\centering
\begin{tabular}{ c | c c c c c | c }
Requirement
 & CSP
 & CfAIII
 & CfAIV
 & SDSS
 & SNLS
 & All\\
\hline\hline
Available stellar mass of the host
 & 7/7
 & 55/55
 & 34/34
 & 389/389
 & 345/397
 & 830/882 \\
$+\ \sigma_{\mathrm{log}_{10}\mathcal{M}}<0.12$
 & 6/7
 & 51/55
 & 31/34
 & 338/389
 & 309/345
 & 735/830 \\
$+\ \sigma_{C_{\mathrm{L}}}<0.12$
 & 6/6
 & 49/51
 & 30/31
 & 288/338
 & 293/309
 & 666/735 \\
\end{tabular}
\caption{Selection requirements for the analysis focusing on local and global properties of the host. For each requirement and each survey, we show the number of remaining supernov\ae\ together with the total number of available supernov\ae. At the end of the selection process, we had measurements of local environment and host photometry for 666 SNIa.}
\label{tab:cuts}
\end{table*}

%\begin{equation}
%\left\{
%\begin{array}{l c r}
%\sigma_{x_1} &<& 0.5 \\ \vspace{0.15cm}
%z &<& 0.8 \\ \vspace{0.15cm}
%\sigma^{\mathrm{local}}_u &<& 1 \\ 
%\sigma^{\mathrm{local}}_{g,r,i,z} &<& 0.5
%\end{array}
%\right. 
%\label{eq:cuts}
%\end{equation}

First, we required that host galaxy stellar masses were available, which means that the quantity had to be properly derived
through the SED-fitting process. In practice, this requirement discards some high-redshift SNIa in the SNLS survey with bad host identification (discrepant photometric redshift between the SNIa and the host, or contamination by a bright star).
Second, we only selected well-measured host stellar masses and removed the 10\% fraction of SNIa with the largest mass error. This corresponds to a limit of about $\sigma_{\mathrm{log}_{10}\mathcal{M}}\sim0.12$, where $\sigma_{\mathrm{log}_{10}\mathcal{M}}$ is the host mass dispersion.
Finally, we ensured that the local $U-V$ color was well measured
by removing the 10\% fraction of SNIa with the largest error on this parameter. This corresponds to a limit of about $\sigma_{C_{\mathrm{L}}}\sim0.12$, where $\sigma_{C_{\mathrm{L}}}$ is the dispersion of the local color measurement. 

Applying these three requirements together, we retained about 74\% of the 397 parent galaxy
photometries of the SNLS survey, and 293 of them remained in
the analysis sample. Of the 389
hosts from S14 with photometry measured in SDSS images, we retained
a large fraction ($\sim74$\%) of the sample, and 288 elements
remained. Of the hosts
belonging to low-redshift surveys, we discarded 11, thus keeping most of the sample for the analysis. In total, the correlations with Hubble
diagram residuals presented in the following sections include 666
host galaxies of Type Ia supernov\ae\ out of 882, which represents about 76\% of the original sample. They span a
wide redshift range of between $z=0.01$ and $z=1.03$, with a median
redshift of 0.25. More details on the selection requirements for each survey are listed in Table~\ref{tab:cuts}.
The initial selection requirement for the SALT2 fit quality described in Section~\ref{subsec:sn_sample} does not affect our results
because the additional SNIa we included are equally distributed in terms of high and low $U-V$ local color.

\begin{figure}[!htbp]
  \begin{center}
   \begin{tabular}{c}
    \includegraphics[width=0.9\columnwidth]{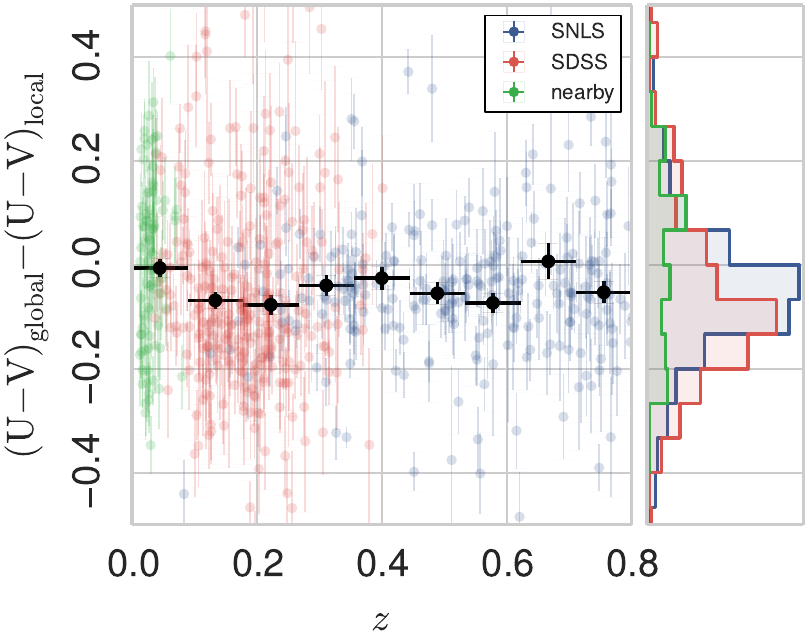} \\
    \includegraphics[width=0.9\columnwidth]{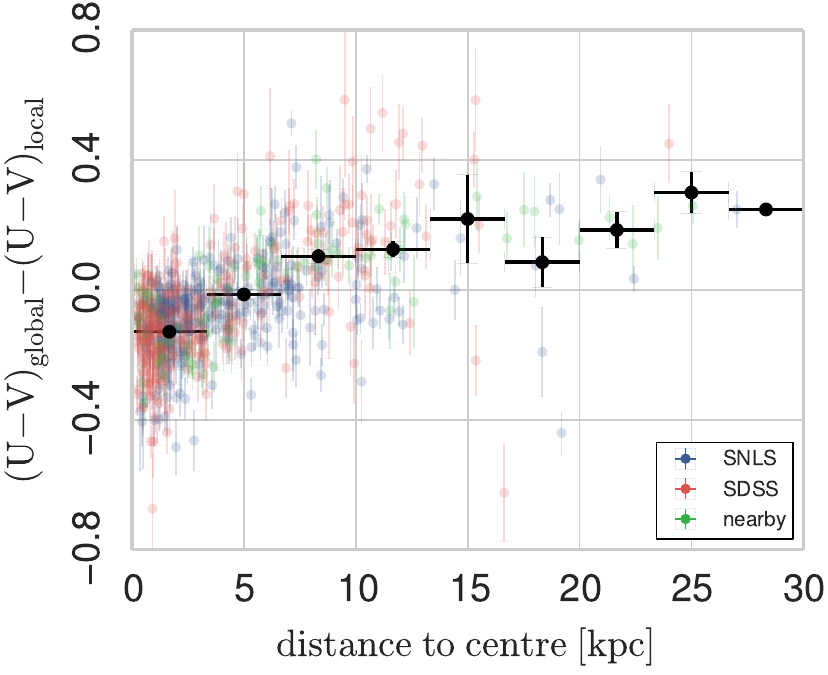}
   \end{tabular}
  \end{center}
  \caption{\textbf{Top:} Difference between U-V rest-frame color of the whole host galaxy and local U-V rest-frame color in a region of 3 kpc around the supernova location for SNLS (blue
    dots), SDSS (red dots), and low-redshift survey (green dots) hosts.
     The difference is displayed as a function
    of redshift. The observed redshift dependence (black bins) indicates that the cosmology is affected when local effects are
included. \textbf{Bottom:} Difference between U-V rest-frame color of the whole host galaxy and local U-V rest-frame color as a function of projected distance to the galactic center in kiloparsec. The color code is the same as for the top plot. The general trend shows that local regions that appear bluer than the host galaxy are located far from the center, whereas locally redder regions are close to the galactic center.
%    $g$-band images of host galaxies of SNIa seen in SDSS images. The top row displays SNIa for which the local surrounding is bluer than the parent galaxy. We notice that they are located far away from the galactic centre. The bottow row shows SNIa for which the local area is redder than their host galaxy. They are situated close to the galactic centre.
}
  \label{fig:global_vs_local}
\end{figure}

\subsection{Comparison of global and local variables}
\label{subsec:loc_vs_glob}

\begin{figure}[!htbp]
  \begin{center}
    \includegraphics[width=0.9\columnwidth]{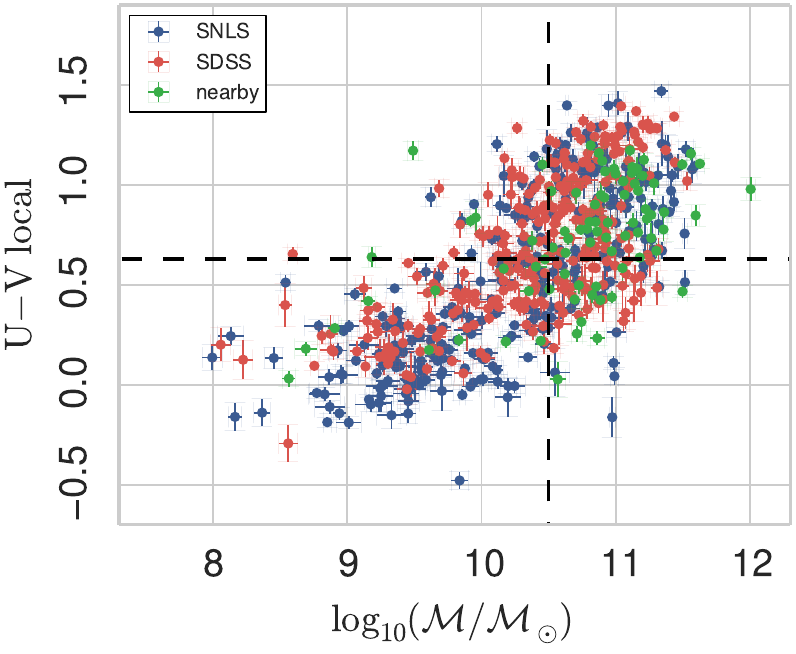}
  \end{center}
  \caption{Local U-V rest-frame color of SNLS (blue dots), SDSS (red
    dots), and low-redshift survey (green dots) hosts in a region of 3
    kpc around the supernova location as a function of the
    stellar mass of the host galaxy (in
    $\mathrm{log}_{10}\mathcal{M}_\odot$). The gray horizontal and vertical
    dashed lines represent the bin separations we used to model the color and mass steps, which correspond to the distribution
median.}
  \label{fig:mass_vs_local}
\end{figure}

%We perform both local and global photometry for the full host galaxies sample and 
Next, we
compare local and global quantities after the selection cuts
described in Section~\ref{subsec:cuts}. In the top panel
of Fig.~\ref{fig:global_vs_local} we displayed the difference between the $U-V$ rest-frame
color of the host
galaxy as a whole (named $U-V$ {\em{global}}) and the $U-V$ rest-frame color estimated within a region of 3 kpc around the supernova
explosion as a function of redshift ($U-V$ {\em{local}}).
The low-redshift
(SNLS and SDSS) surveys are respectively represented by the green (blue and
red) points.
As expected, we find a good correlation between the two quantities on average, %since for most cases the global information is equivalent to the local one.
%(small or distant galaxies)
and we checked that the
difference between global and local rest-frame colors converges to
zero when the ratio between the parent galaxy size and the local
circle area is close to one.  However, we also note that the
difference between the global and local rest-frame colors is
different from zero for a significant number of our host galaxies (40\% of them are more than $2\sigma$ different from the zero difference between global and local colors). These differences indicate that probing the local environment and the host galaxy as a whole do not convey the same physical information.
This can be explained by situations where, for example, a supernova explodes in the old passive bulge of an active spiral galaxy, or in an active region of an old spiral galaxy.
%depending on whether the supernova explodes in the old passive bulge of an active spiral galaxy or in an active region of an old spiral galaxy.
%, and that this effect depends on the distance to the galactic centre.
We find that the color difference is on average different from zero, that it evolves with redshift, and finally, that there is a systematic trend to measure SNIa in locally red regions compared to the host galaxy.

This result can partly be explained by a distance effect to the galactic center, represented in the bottom panel of
Fig.~\ref{fig:global_vs_local}.
For a supernova
exploding in the outskirts of the host galaxy, the explosion area is seen
{\em{bluer}} than its host (positive difference between global and local $U-V$ colors). Conversely, if the explosion region is
close to the galactic center, then the local rest-frame color is
{\em{redder}} than the galaxy as a whole (negative difference between global and local $U-V$ colors). As for the top panel, we observe that the majority of our sample contains SNIa that explode close to the center of their host galaxy.

%A few visual examples on
%SDSS images are presented on the bottom panel of
%Fig.~\ref{fig:global_vs_local}. The top row displays cases where the
%SNIa explosion occurs in the outskirts of the host galaxy, thus
%corresponding to 
%%the area below the dashed grey diagonal line of the plot above.
%positive differences between $U-V$ global and local colors.
%The bottom row shows the opposite situation, where SNIa
%happen to explode very close to the galactic centre. The region is
%seen redder than the host itself, then the corresponding color points
%on the plot above are located below the 
%%dashed grey diagonal line.
%zero level (negative color differences).

In numerous cases, the
rest-frame local and global colors measurements are significantly different
(i.e., points departing from the zero line in the two panels
of Fig.~\ref{fig:global_vs_local}). In some redshift bins, the difference between global and local colors departs from zero with a significance of about $4\sigma$ . This shows that different information might be accessible through the study of local environment, and this has to be
accounted for, as suggested in R13.

The relation between local $U-V$ rest-frame color and the host galaxy
stellar mass is drawn in Fig.~\ref{fig:mass_vs_local} for the three
different survey categories. As expected, we find the most massive
galaxies to be those for which the close environment of the SNIa
explosion is seen as red. This confirms the general trend that
massive galaxies are more passive in terms of star formation than low-mass galaxies \citep{kennicutt98,kauffmann03}. The dashed gray lines in Fig.~\ref{fig:mass_vs_local}
illustrate the bin separation, assuming that both local color and
stellar mass distributions are bimodal (see section
\ref{subsec:hubble_diagram} for more details). A significant
proportion of our sample ($\sim20$\%) belongs to the lower right and upper left parts
of the figure (second and fourth$^{\mathrm{\text{}}}$ quadrants). These specific host galaxies belong to the same mass
bin as hosts in the other quadrants, but they will belong to
different local color bins. Hence, the observed difference between local
environment and host properties as a whole
%$U-V$ rest-frame color and galaxy stellar mass
%regarding Hubble diagram residuals
derives from this fraction of our sample. Investigating these differences
is the purpose of the next section.

\begin{figure}%[!htbp]
  \begin{center}
    \includegraphics[width=0.9\columnwidth]{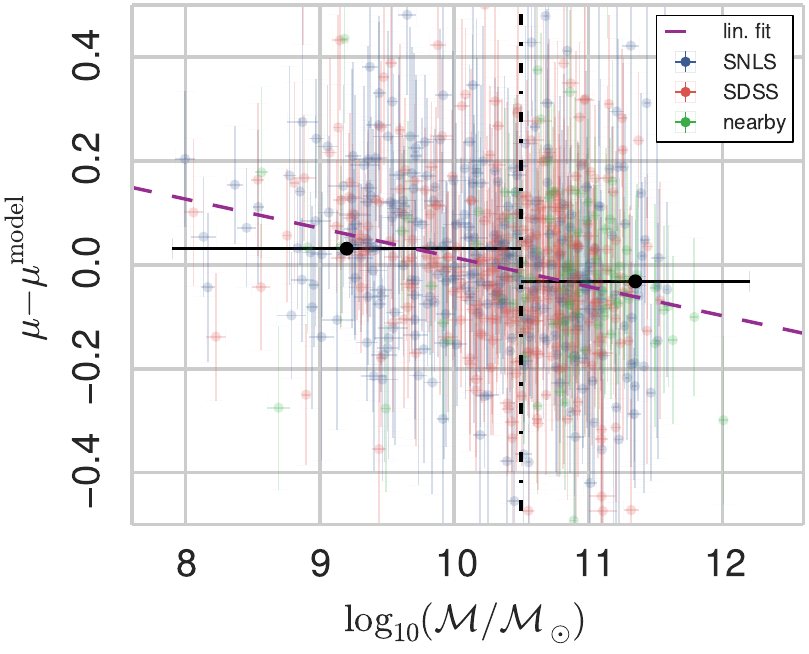}
  \end{center}
  \caption{Correlation of host galaxy stellar mass (in units of $\mathrm{log}_{10}\mathcal{M}_\odot$) with Hubble diagram residuals for the
    SNLS (blue dots), SDSS (red dots), and low-redshift (green dots)
    surveys. The separation between the two mass bins at
    $10^{10.5}\mathcal{M}_\odot$ corresponds to the median mass value of our sample and is shown as the black vertical dash-dotted line. The
    bins are drawn in black, with the weighted mean central values as black dots and
    their dispersion, divided by the square root of the number of elements in the
    bin. The purple dashed line shows the weighted linear
    fit for the mass distribution.}
  \label{fig:correlation_hr_mass}
\end{figure}

\begin{figure}%[!htbp]
  \begin{center}
    \includegraphics[width=0.9\columnwidth]{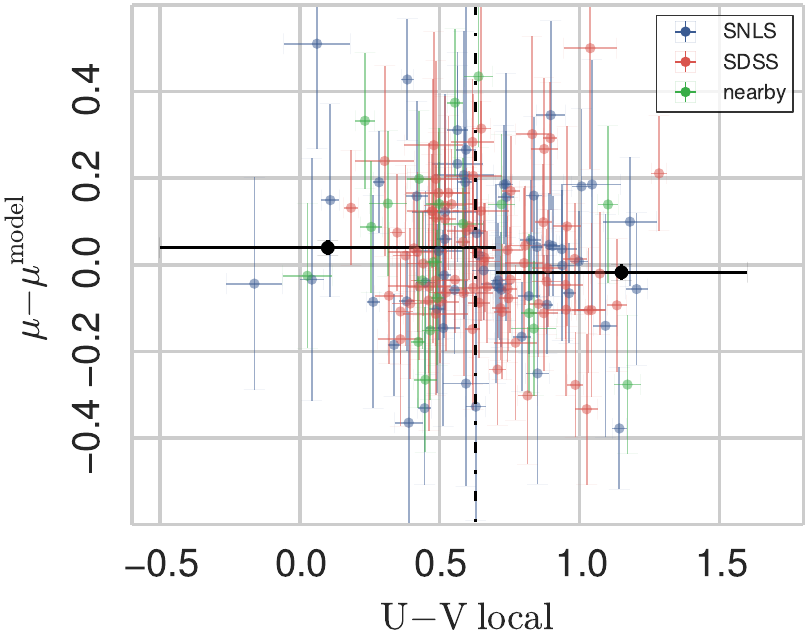}
  \end{center}
  \caption{Correlation of local U-V rest-frame colors (in a region of
    3  kpc around the supernova location) with Hubble diagram
    residuals corrected for stretch and color for the SNLS (blue dots), SDSS (red dots), and
    low-redshift (green dots) surveys for SNIa lying in the$^{\mathrm{\text{}}}$ second and$^{\mathrm{\text{}}}$ fourth quadrants of Fig.~\ref{fig:mass_vs_local}.}
  \label{fig:correlation_quadrants}
\end{figure}

\begin{figure*}%[!htbp]
  \begin{center}
    \includegraphics[width=1.5\columnwidth]{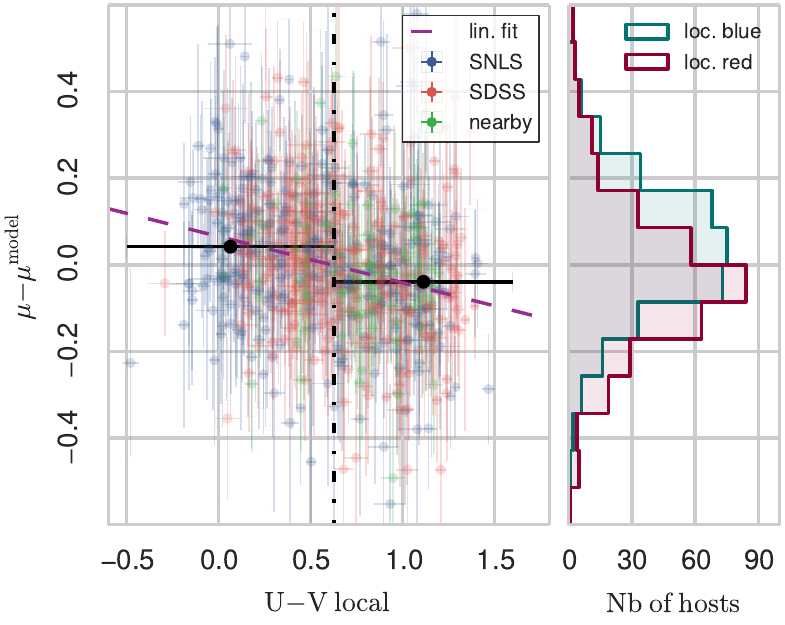}
  \end{center}
  \caption{Correlation of local U-V rest-frame colors (in a region of
    3 kpc around the supernova location) with Hubble diagram
    residuals corrected for stretch and color for the SNLS (blue dots), SDSS (red dots), and
    low-redshift (green dots) surveys. The separation between the two
    local color bins at $\mathrm{U-V}\sim0.63$, corresponding to the median value of the sample, is shown as the black vertical dash-dotted
    line. The bins are drawn in black, with the weighted mean values as black
    dots and their dispersion, divided by the square root of the number of
    elements in the bin. The purple dashed line shows the weighted linear
    fit for the entire color distribution.
%The top histogram displays the repartition of the 441 elements in terms of local color, together with the fit to a bimodal distribution represented by the solid red line.
    The histogram on the right
    draws the Hubble diagram residual distribution of {\em{blue}}
    SNIa regions (for local color values below the separation in $\mathrm{U-V}$), associated with star-forming
    regions, together with the equivalent distribution of {\em{red}}
    regions (for local color values above the separation in $\mathrm{U-V}$), associated with more passive areas.}
  \label{fig:correlation_hr_local}
\end{figure*}

\subsection{Third parameter and correlations with Hubble diagram residuals}
\label{subsec:hubble_diagram}

At this stage, we can estimate distance moduli and Hubble diagram residuals, defined
as the difference between the observed distances and the distances predicted by the best-fit cosmology (while simultaneously fitting and applying the brighter-slower and brighter-bluer relations).
%, corresponding to SNIa for which we estimated the local color around the explosion, and the host galaxy color and stellar mass.
%Hubble diagram residuals thus reflect the intrinsic luminosity scatter among the supernova sample.
The distance estimator assumes that supernov\ae\ with identical color, shape, and galactic environment on average have the same intrinsic luminosity at all redshifts. This yields a standardized distance modulus $\mu=5\mathrm{log}_{10}(d_L/10\mathrm{pc})$\begin{equation}
\mu = m_B^*-(M_B-\alpha\times x_1+\beta\times c)
,\end{equation}
where $m_B^*$ is the observed peak magnitude in rest-frame $B$ band, and $\alpha$, $\beta,$ and $M_B$ are nuisance parameters. Hubble diagram residuals thus represent the difference between this quantity and the cosmological model prediction $\mu_{\mathrm{model}}(d_L(z;\Omega_m))=5\mathrm{log}_{10}(d_L(z;\Omega_m)/10\mathrm{pc})$ computed for a fixed value of $H_0=70\  \mathrm{km.}\mathrm{s.}^{-1}\mathrm{Mpc}^{-1}$, assuming a Friedmann-Lema\^itre-Robertson-Walker geometry. $\Omega_m$ is defined as the total matter density in the Universe. As a result, the residuals reflect the intrinsic luminosity scatter of the supernova sample. However, the following results do not depend on the cosmological model, as explained in Section~\ref{subsec:cosmo_dependence}.

In practice, we performed a cosmological fit 
%using an updated version of the SALT fitter used in \cite{Betoule14} \todo{[Marc: d\'etails dans une section s\'epar\'ee ?]}, 
that computes distances based on blinded fluxes allowing for a third variable that standardizes SNIa and minimizes the function\begin{equation}
\chi^2=\left[\hat{\mu}-\mu_{\mathrm{model}}(d_L(z;\Omega_m))\right]^\dagger\mathcal{C}^{-1}\left[
\hat{\mu}-\mu_{\mathrm{model}}(d_L(z;\Omega_m))\right]
,\end{equation}
where $\mathcal{C}=\mathcal{C}_{\mathrm{stat}}$ is the covariance matrix of $\hat{\mu}$ rewritten in matrix notation. It is obtained from the error propagation of light-curve fit uncertainties.

The third variable was taken into account in the cosmological fit through a step function for the absolute magnitude $M_B$, assuming a bimodal distribution for all types of host properties (local color, host global color, and stellar mass of the host). Hence SNIa were divided into two groups of locally red regions or locally blue regions, but also red host galaxies or blue host galaxies and low-mass hosts or high-mass hosts.
We followed the procedure described in \cite{Sullivan10}, which defines a magnitude step written as 
%It writes as follows:
\begin{equation}
M_B=\left\{
\begin{array}{l c r}
M_B &\mathrm{if}&\ X<X_{\mathrm{lim}}\\
M_B+\Delta M_B &\mathrm{if}&\ X>X_{\mathrm{lim}}
\end{array}
\right.
\label{eq:step_mag}
,\end{equation}
where $X$ is any type of variable related to host galaxy properties and $X_{\mathrm{lim}}$ is the bimodal distribution limit.

For each variable, the separation between the groups, $X_{\mathrm{lim}}$, corresponds to the median value of the sample.
%We concretely fit a bimodal gaussian distribution function and estimate in each case the best fit for the separation between its two peaks.
SNIa were then separated in half between locally red
and blue regions for $U-V\sim0.63$. Equivalently, SNIa were found in
two groups of red and blue host galaxies with a separation estimated
at $U-V\sim0.62$.
%However, the bimodal gaussian fit distribution does not converge regarding the host stellar mass distribution.
We
find the bin limit for the stellar mass of the host at $10^{10.5}\mathcal{M}_\odot$, which is different from previous studies, for which the separation was arbitrarily chosen at $10^{10}\mathcal{M}_\odot$.  In
Figs.~\ref{fig:correlation_hr_local} and \ref{fig:correlation_hr_mass}
we display the Hubble diagram residuals {\em{corrected for stretch and color}} as a function of stellar mass of the
host and local $U-V$ color for SNIa detected by the SNLS (blue), SDSS (red),
and low-redshift surveys (green).
Fig.~\ref{fig:correlation_quadrants} is the same as Fig.~\ref{fig:correlation_hr_local}, but only considering the second$^{\mathrm{\text{}}}$ and fourth$^{\mathrm{\text{}}}$ quadrants of Fig.~\ref{fig:mass_vs_local}.
We separated the final sample into two bins at the median. The separation is represented by
the dotted black lines, and the bin positions by black points with
error bars corresponding to the dispersion in the given bin, divided
by the square root of the number of elements in it.  A linear
evolution of Hubble diagram residuals with our local and global
variables were also computed and are shown as the purple dashed line (see Section~\ref{subsec:alternative_correlations}). Next,
we interpret the observed Hubble diagram dependence on color or mass
as environmental effects that can be absorbed in the cosmological fit as a third standardization variable.
%the significance of the $\Delta M_B$ variable from the cosmological fit, computed along with distance moduli and Hubble diagram residuals.

\begin{table*}[!htbp]
\centering
\begin{tabular}{ c | c | c | c | c }
Survey/$z$ range& Nb of SNIa & $\Delta M_B$ Local color (mag) & $\Delta M_B$ Host color (mag) & $\Delta M_B$ Host stellar mass (mag)\\
\hline\hline
Nearby
 & 85
 &$-0.0491\pm0.0462$ ($1.1\sigma$)
 &$-0.0401\pm0.0454$ ($0.9\sigma$)
 &$-0.0235\pm0.0430$ ($0.5\sigma$)\\
SDSS
 & 288
 &$-0.0877\pm0.0189$ ($4.6\sigma$)
 &$-0.0526\pm0.0190$ ($2.8\sigma$)
 &$-0.0604\pm0.0188$ ($3.2\sigma$)\\
SNLS
 & 293
 & $-0.0993\pm0.0205$ ($4.8\sigma$)
 & $-0.0917\pm0.0202$ ($4.5\sigma$)
 & $-0.0882\pm0.0205$ ($4.3\sigma$)\\
\hline
$z<0.1$
 & 123
 &$-0.0534\pm0.0323$ ($1.7\sigma$)
 &$-0.0119\pm0.0313$ ($0.4\sigma$)
 &$-0.0260\pm0.0310$ ($0.8\sigma$)\\
$0.1<z<0.5$
 & 350
 &$-0.1172\pm0.0171$ ($6.9\sigma$)
 &$-0.0975\pm0.0171$ ($5.7\sigma$)
 &$-0.0834\pm0.0168$ ($5.0\sigma$)\\
$z>0.5$
 & 193
 &$-0.0586\pm0.0259$ ($2.3\sigma$)
 &$-0.0556\pm0.0258$ ($2.2\sigma$)
 &$-0.0702\pm0.0262$ ($2.7\sigma$)\\
\hline
All
 & 666
 & $-0.0909\pm0.0130$ ($7.0\sigma$)
 & $-0.0689\pm0.0130$ ($5.3\sigma$)
 & $-0.0704\pm0.0128$ ($5.5\sigma$)\\
\end{tabular}
\caption{Significance of the third parameter in the cosmological fit standardizing light-curves of SNIa as a function of survey type and redshift range for the different galactic properties, assuming a bimodal distribution.
%$\mathcal{M}$ stands for host stellar mass, $C_{\mathrm{G}}$ for $U-V$ galaxy total color and $C_{\mathrm{L}}$ for $U-V$ local color.
The significance considering the combined sample is listed in the last row.}
\label{tab:nb_sigma}
\end{table*}

When we performed a cosmological fit using the stellar mass of
the host as a third variable, we find
\begin{equation}
\vert\Delta M_B\vert=0.070\pm0.013,
\end{equation}
which corresponds to a significance of $5.5\sigma$. This value is
consistent with the most recent results obtained by \cite{Betoule14}
with different samples and different selection criteria.
%though the sample selection here is different.
Applying
the same method to the host galaxy $U-V$ color, we obtain
\begin{equation}
\vert\Delta M_B\vert=0.069\pm0.013,
\end{equation}
thus a change in $\Delta M_B$ at the $5.3\sigma$ confidence
level. Considering the {\em{local}} $U-V$ color as the third standardization parameter, we obtain
\begin{equation}
\vert\Delta M_B\vert=0.091\pm0.013,
\end{equation}
which means that the change in $M_B$ magnitude with a color step
is at a
significance of about $7.0\sigma$. Its amplitude is compatible with similar local age-bias measurements from R13 and R15.

The details of this study for the
different surveys and for various redshift ranges are shown in Table
\ref{tab:nb_sigma}. We observe a general increase in significance when the local $U-V$ rest-frame color is considered as a third standardization parameter compared to the host global color or stellar mass.
The physical meaning can be
understood when we assume that galaxy color can be related to other properties
(star formation rate, metallicity, or age), and that the galaxy stellar mass is
very likely a proxy for these properties.
%Therefore, more information is linked to environmental effects.
The local $U-V$ color is the
most highly correlated variable for high- and intermediate-redshift surveys, whereas global color is slightly more significant for low-redshift surveys.
When
we split the sample into redshift bins below $z=0.1$, between $z=0.1$ and $z=0.5$, and above $z=0.5,$ we
find 
that the local color is the most significant third variable for all redshift intervals,
and that it is comparable to the host stellar mass variable at high redhsifts.
The local signal is particularly stronger at intermediate redshifts, where we see a departure from the zero line in the top panel of Fig.\ref{fig:global_vs_local}.
We note that for each of the significance calculations, we considered the median value of the total sample
as a separation between the bins.
Last but not least, when all redshifts are combined, a clear preference for local color is visible, as described above.

In Section~\ref{sec:photo} we described the high quality of the photometry we derived in detail, which leads to precise measurements of the  stellar masses and local $U-V$ colors of the hosts. As a result, a small fraction of SNIa are likely to change bin within errors. Of the SNIa that constitute the basic sample,   7.8\% are within $1\sigma$ of the local color bin separation, and 5.6\% are within $1\sigma$ of the mass bin separation.
To precisely characterize the effect of SNIa changing bin, we randomly drew Gaussian-distributed local colors (or host stellar masses). With 1000 realizations, the local (mass) step amplitude is shifted at $1\sigma$ significance by $0.00391$ mag ($0.00383$ mag). This variation is approximately three times less important than the errors that are due to uncertainties on the distance measurement.

When we restrict the SNIa from our basic sample to the SNIa that are in common with the SNIa of the JLA analysis \citep{Betoule14}, we find 589 matches (see Section~\ref{subsec:sn_sample} for the difference between the two samples). Without applying any selection requirement on the common sample and cutting the sample in half with respect to the stellar masses of the hosts and local $U-V$ colors, we measure a local color step of $-0.075\pm0.012$ when this variable is considered as a third standardization parameter. If the stellar mass of the host is used instead, the mass step is found to be $-0.077\pm0.012$. These values are
%obtained placing the bin separation at the sample median, and are 
consistent with the step amplitudes reported in Table~\ref{tab:nb_sigma}.

\begin{figure}%[!htbp]
  \begin{center}
    \includegraphics[width=0.9\columnwidth]{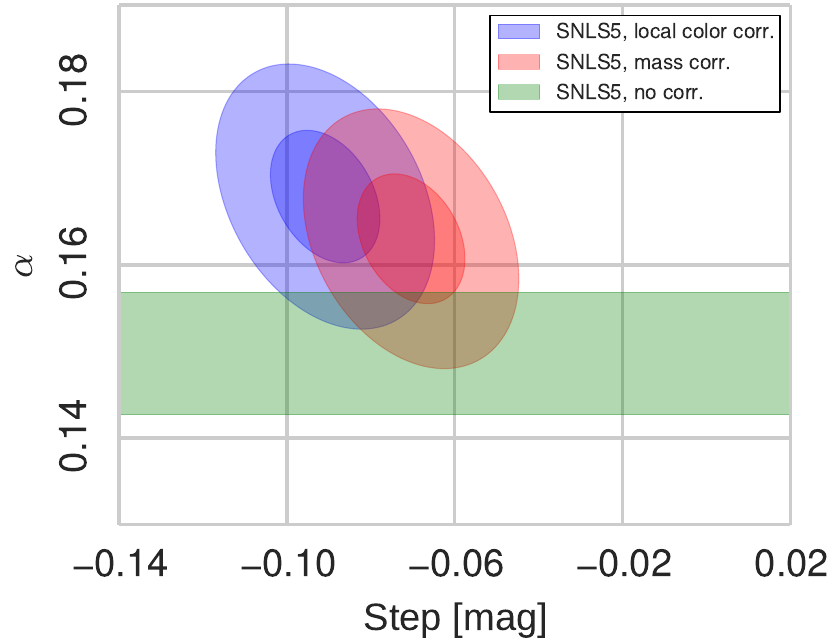}
  \end{center}
  \caption{Covariance of $\alpha$ and the absolute magnitude step when three standardization parameters are used (blue contours for local color correction, and red contours for host stellar mass correction). The green rectangle corresponds to the case where two standardization parameters are used. Contours correspond to 68\% and 95\% confidence levels.}
  \label{fig:alpha_covariance}
\end{figure}

As expected from the relation between $x_1$ and the absolute magnitude, values of the cosmological fit nuisance parameters depend on the method that is used to estimate the magnitude steps. The covariance of the nuisance parameter associated with the stretch factor ($\alpha$) with the magnitude step $\Delta M_B$ is shown in Fig.~\ref{fig:alpha_covariance}. The central value when the stellar mass of the host is used as a third-parameter standardization is $\alpha=0.16301\pm0.0075$, and $\alpha=0.16788\pm0.0076$ when the local $U-V$ color is chosen. These values are higher when
they are compared to JLA estimates because different initial selection cuts, validated on simulations, were used in our analysis (see Section~\ref{subsec:sn_sample}). However, this difference in $\alpha$ between our sample and JLA has little effect ($<<1\sigma$) on the amplitude of both local color and mass steps when $\alpha$ is a free fit parameter.

\begin{figure*}%[!htbp]
  \begin{center}
        \begin{tabular}{cc}
    \includegraphics[width=0.85\columnwidth]{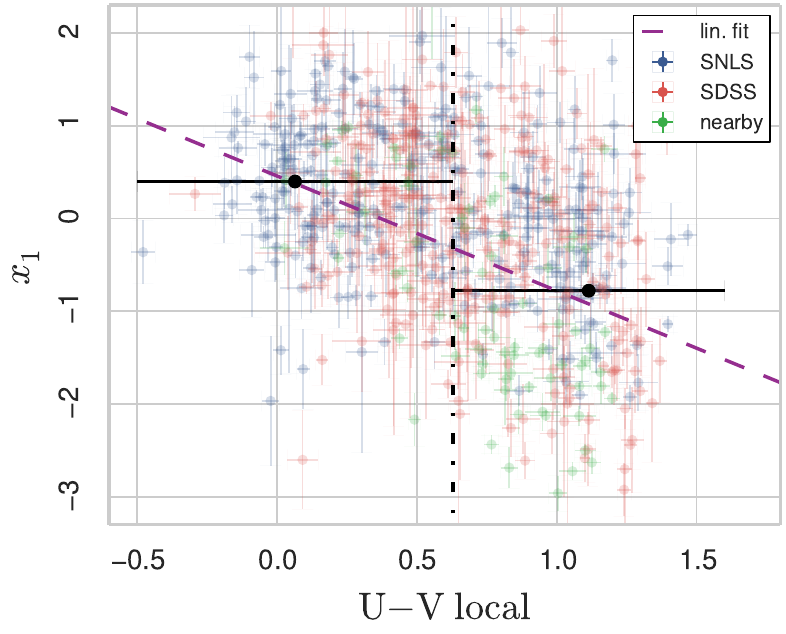}&
    \includegraphics[width=0.85\columnwidth]{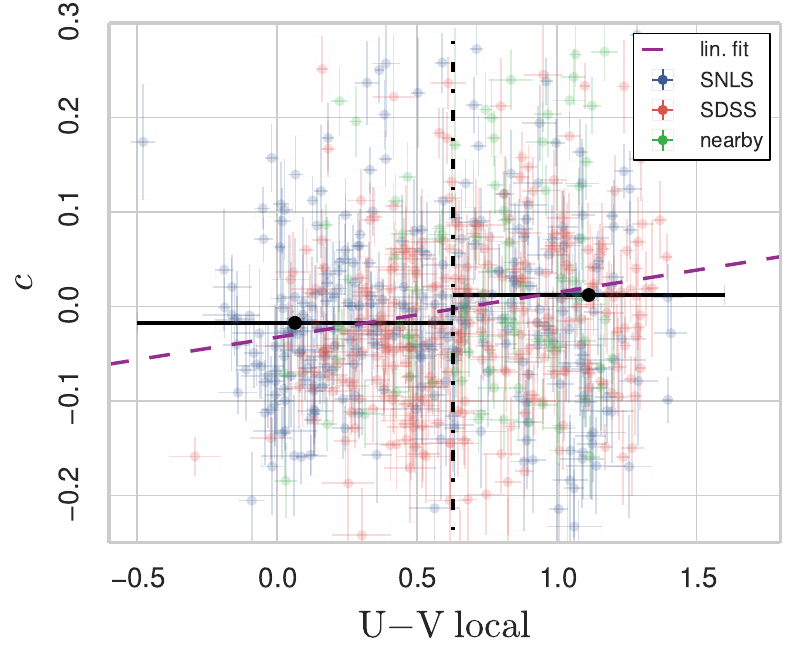}\\    
    \includegraphics[width=0.85\columnwidth]{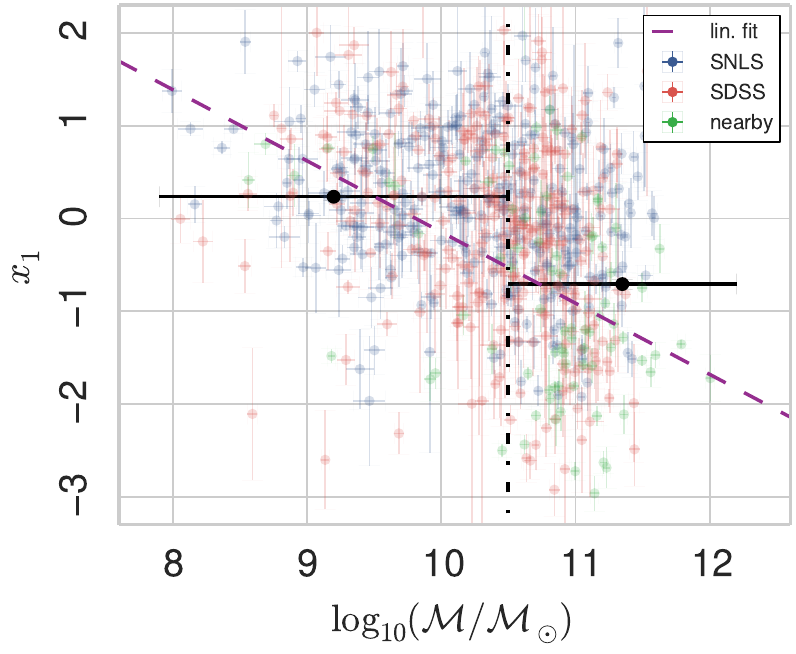}&
    \includegraphics[width=0.85\columnwidth]{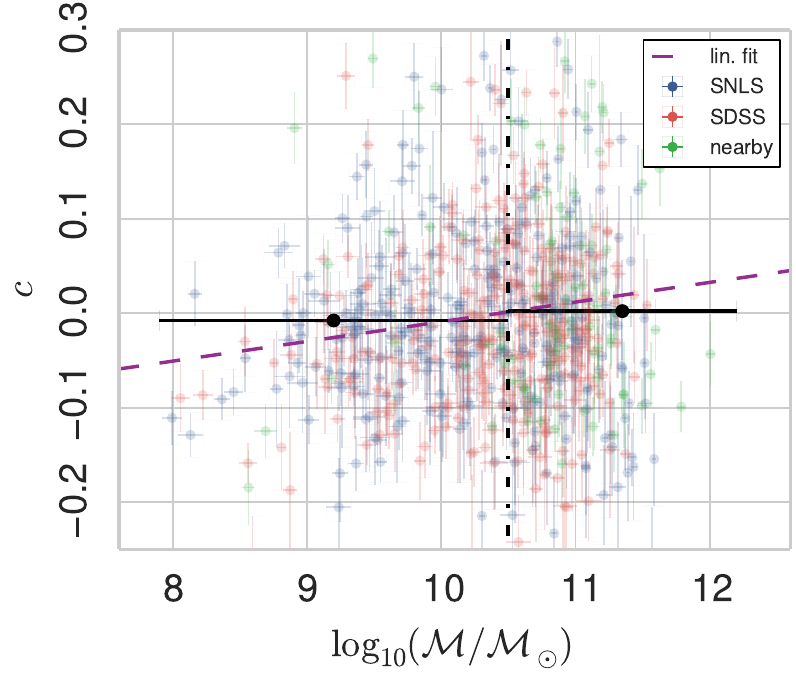}\\     
        \end{tabular}
  \end{center}
  \caption{Correlations between supernova properties (stretch and color) as a function of local $U-V$ color (first row) and host galaxy stellar mass (second row). The bins are drawn in black, with the weighted mean central values as black dots and the bin dispersion. Linear fits of the data are shown as dashed purple lines.}
  \label{fig:correlations_x1_c}
\end{figure*}

In Fig.~\ref{fig:correlations_x1_c} we display the correlations between SNIa properties (stretch and color) as a function of local $U-V$ color and stellar mass of the host galaxy. As has
been noted in R13, R15, and in previous studies, we find a strong correlation of the two variables with the stretch factor $x_1$
\citep{branch96,Howell09,Neill09,sullivan06}, and a small color dependence on local environment and host mass \citep{Sullivan10, Childress13}. Although we note a strong dependence of light-curve decline rate with environment variables, this cannot explain the significance of a third SALT2 parameter that takes local color into account because the cosmological fit was performed
after correcting for stretch.

\subsection{Varying the local radius}
\label{subsec:more_local}
\begin{figure}%[!htbp]
  \begin{center}
    \includegraphics[width=0.9\columnwidth]{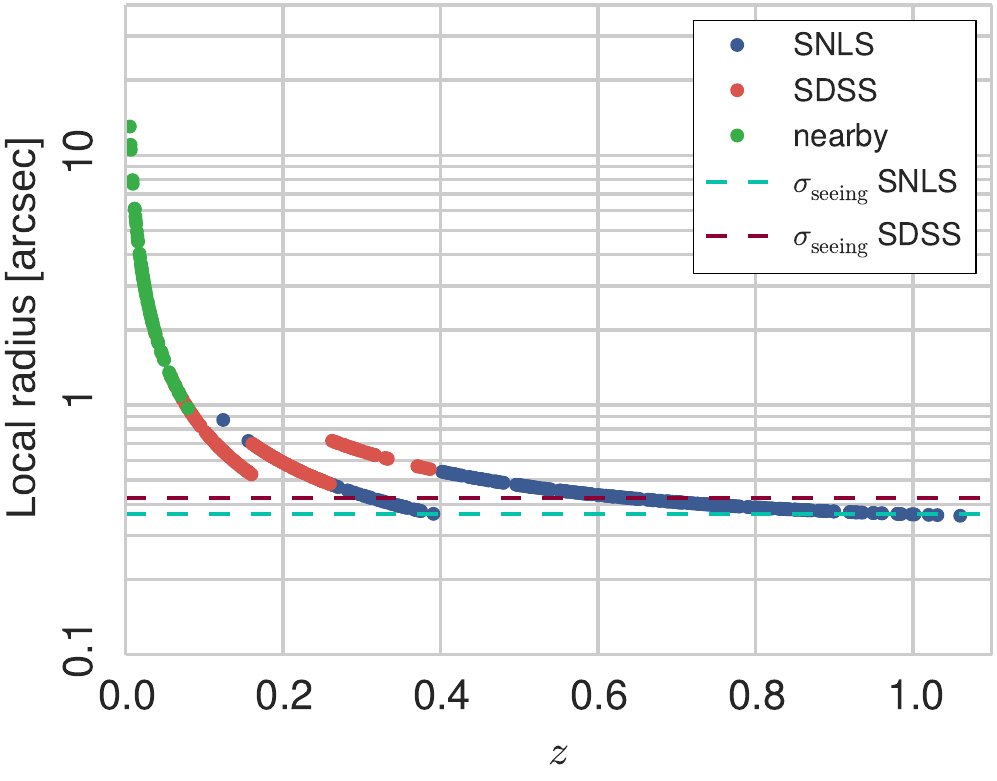}
  \end{center}
  \caption{Evolution of the local photometric radius of 1.5, 2, and 3  kpc with redshift in arcseconds for
    host galaxies of SNIa observed by the SNLS (blue), SDSS (red), and low-redshift
    surveys (green points). The $1\sigma$ seeing size of SNLS and SDSS images are illustrated with dashed lines using the same color code. A $1\sigma$ seeing corresponds to a diameter of about 85\% of the FWHM.}
  \label{fig:radius_vs_z_local}
\end{figure}

The basis of this work has been chosen so that the local probed physical size remains 3 kpc in radius at all redshifts. Therefore, the obtained results are consistent.
%By probing the same physical size of 3 kpc at all redshifts, we show consistent results and that is the reason why this analysis is presented as the baseline of this work.
For redshifts that are low enough for the local photometric radius to be still larger than the resolution of a given survey, however,
we can probe smaller physical sizes and check whether it increases the significance of the local color as a third standardization parameter. In practice, we measured photometry in a circle with
a radius of 2 kpc for $z<0.4$ in SNLS. In the SDSS and low-redshift surveys, we probed a physical size of 2 kpc for $0.16<z<0.26$ and of 1.5 kpc when the redshift verified the condition $z<0.16$ (see Fig.~\ref{fig:radius_vs_z_local}, the equivalent of Fig.~\ref{fig:radius_vs_z}). Performing the same cosmological fit using the new local colors as a third variable on 620 events (the selection requirements cause a slight change), we find $\vert\Delta M\vert=0.087\pm0.014$.
We conclude that the standardized supernova brightness still displays a dependence on local environment at the $6.2\sigma$ confidence level. This is very similar to the significance we find using a 3 kpc local radius at all redshifts.
A comparable size test has been performed in R15 and yielded similar results.

\begin{figure}%[!htbp]
  \begin{center}
    \includegraphics[width=0.9\columnwidth]{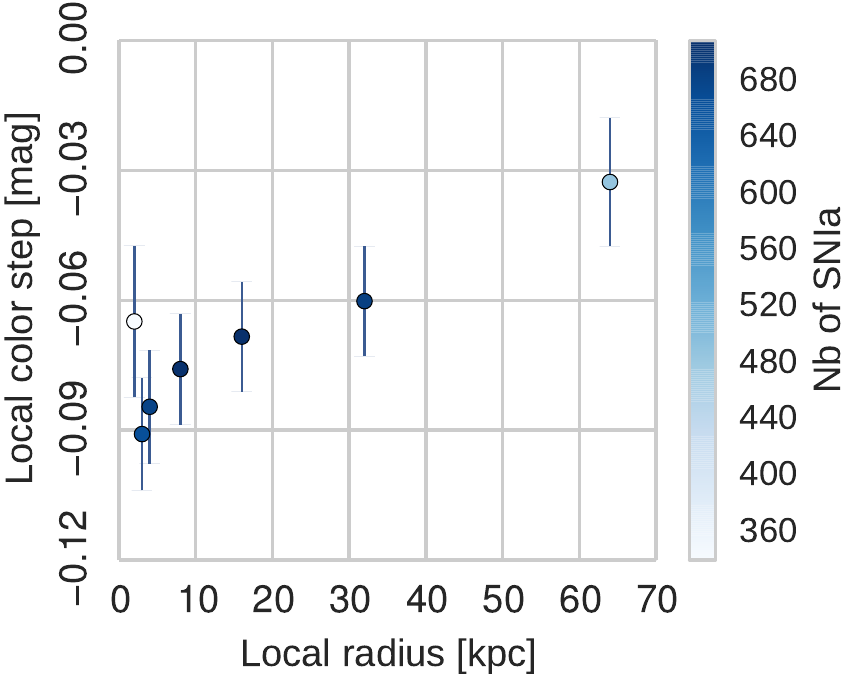}
  \end{center}
  \caption{Evolution of the local step amplitude (in magnitude) as a function of the local radius size (in kpc). Points with error bars are colored as a function of the number of SNIa that belong to the corresponding samples.}
  \label{fig:size_dependence}
\end{figure}

We modified values of the local radius from 3 kpc to 2, 4, 8, 16, 32, and 64 kpc and analyzed the local color step effect as a function of radius. The local step radius dependence is represented in Figure~\ref{fig:size_dependence}, where points are colored as a function of the number of SNIa in the corresponding samples. For each radius, we applied the selection requirements described in Section~\ref{subsec:cuts}, and to compute the local step, we used the method explained in Section~\ref{subsec:hubble_diagram}. Within the cosmological fit, we split the obtained local $U-V$ colors into two bins using the basic sample median as a separation. Between radii of 3 and 32 kpc, we find a smooth transition from local to global, with values compatible with mass or global $U-V$ steps. When the radius is smaller than 3 kpc, the regions have a physical size smaller than $1\sigma$ seeing. This means that local $U-V$ are measured with larger error bars than in the basic case, and the selection requirement on local color ($\sigma_{C_{\mathrm{L}}}<0.12$) removes more SNIa. The local step is less important in this configuration. As a result, we cannot probe regions smaller than 3 kpc at all redshifts without significantly decreasing the sample.
For large radii ($r=64$ kpc), the flux errors increase and prevent \texttt{EGALITE} from finding good $U-V$ colors. Since fewer SNIa are present in this situation, the local step is also less important. It is thus important to note that the error bars for dark blue points are correlated (the sample contains the same objects), whereas the error bars for $r=2$ kpc and $r=64$ kpc are estimated with different SNIa in the sample. 

\section{Reliability tests}
\label{sec:robust}

\subsection{Alternative methods for estimating correlations}
\label{subsec:alternative_correlations}
%When we perform the cosmological fit estimating distance moduli with stretch and color standardization, we also compute residuals using the local $U-V$ color or the host stellar mass (or the host color) as a third variable.
In the basic analysis, we studied the significance of a magnitude step for a cosmological fit including a three-variable standardization with host properties. Alternatively, we might also estimate Hubble diagram residuals corrected for stretch and color, and verify the correlation with any other variable related to host properties.
Therefore, we have access to correlations between Hubble diagram residuals and host properties {\em{a posteriori}} and can verify whether the conclusions we drew beforehand are still valid. Assuming a bimodal behavior of the local $U-V$ color and of the stellar mass distribution of the host, we can split each distribution into two bins and thus estimate the {\em{a posteriori}} statistical significance of Hubble diagram residuals with respect to these two variables. In practice, this corresponds to the significance of the difference between the two black bins in Figs.~\ref{fig:correlation_hr_mass} and \ref{fig:correlation_hr_local}. With this other method, we find a $7.1\sigma$ significance for the correlation between Hubble diagram residuals computed with the regular standardization and the local color, which confirms the result obtained with a three-parameter standardization in the cosmological fit. When the stellar mass of the host is chosen instead, the correlation is present at a significance level of $5.4\sigma$. The stellar mass of the host is still less strongly correlated to the
Hubble diagram residuals than the color inside the area around the supernova explosion. 

With this technique, it is also possible to obtain the remaining signal after the Hubble diagram residuals are estimated using a standardization with three variables. In particular,  the correlation between the residuals {\em{corrected using the stellar mass of the host as a third variable}} and the local color shows a $\delta\mu$ difference that is significant at $4.6\sigma$. This means that correcting for the stellar mass of the host does not capture all the physical information, as expected. On the other hand, when we correct distance moduli with local color in addition to stretch and color effects, we find a correlation between the corresponding Hubble diagram residuals and the stellar mass of the host that is half as significant ($2.3\sigma$). This shows that more information is taken into account when distance moduli are corrected using the local environment of Type Ia supernov\ae.

Until now, correlations were computed assuming that the stellar mass of the host or the local $U-V$ color distributions are separated into two distinct bins. This was taken into account in the cosmological fit computing Hubble diagram residuals with a step function for the absolute magnitude (see Eq.\ref{eq:step_mag}). Alternatively, the importance of an additional variable that standardizes SNIa light curves by computing weighted linear fits for the Hubble diagram residuals can be described as corresponding to a two-variable standardization, rather than a step function. The weighted linear regressions are represented as the dashed purple line in Figs.~\ref{fig:correlation_hr_mass} and \ref{fig:correlation_hr_local}. We find a slope of $-0.107\pm0.046$ when the Hubble diagram residuals are distributed against local color, and a slope of $-0.056\pm0.033$ for the host mass. Therefore,
%as the fit is weighted and the error bars rather large,
the deviation from a zero slope is not highly significant ($2.3\sigma$ for local color and $1.7\sigma$ for the stellar mass of the host). Since slopes can be biased by outliers, we also used the Theil-Sen estimator, which is minimally affected by them \citep{theil92,sen68}. We found better constraints on slopes with compatible central values than for the least-squares estimator: $-0.093\pm0.015$ for the local color, and $-0.055\pm0.009$ for the stellar mass
of the host.
%In addition to the fact that the local color variable is still the one with a more significant deviation from the zero slope, we note that the slope is significantly steeper for this same variable.
%Indeed, the slope value regarding the variable probing the close environment of supernov\ae\ is more than $3\sigma$ steeper than for the host mass.
We can still conclude that studying weighted linear fits to the data also tends to prove that a stronger dependence of the Hubble diagram residuals on local color is observed. Moreover, the step function appears to describe the effect better.

\subsection{Removing supernov\ae\ from the sample}
\begin{figure}%[!htbp]
  \begin{center}
   \begin{tabular}{c}
    \includegraphics[width=0.9\columnwidth]{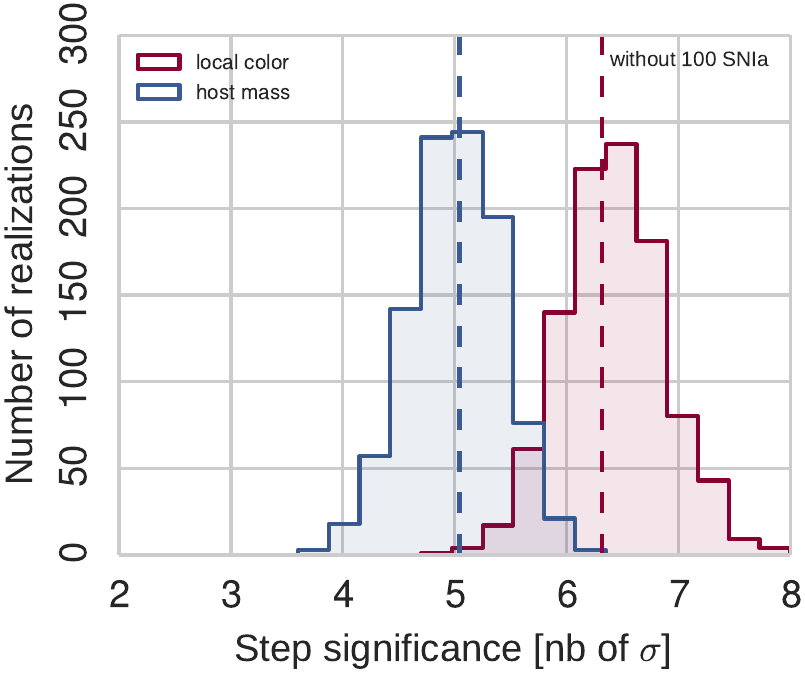}\\
    \includegraphics[width=0.9\columnwidth]{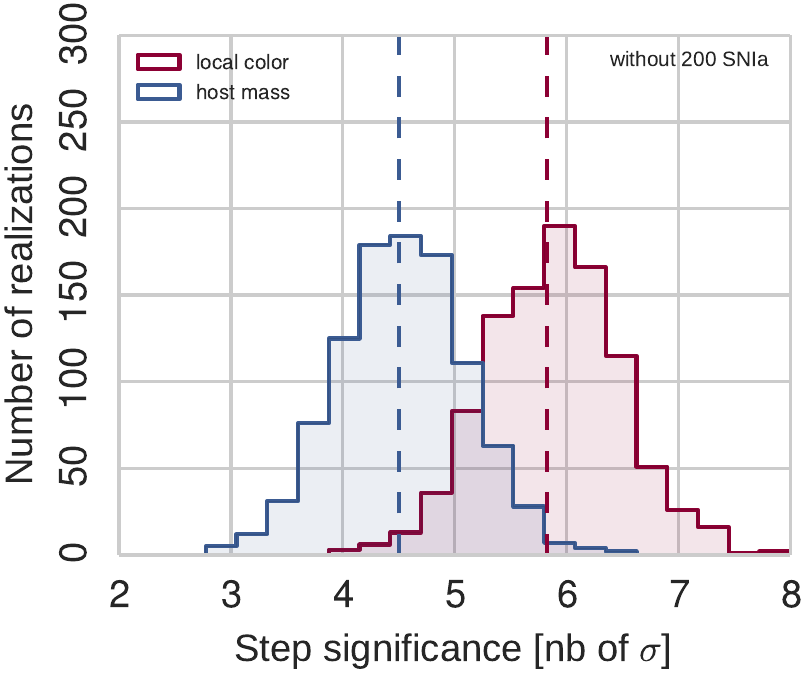}\\
    \includegraphics[width=0.9\columnwidth]{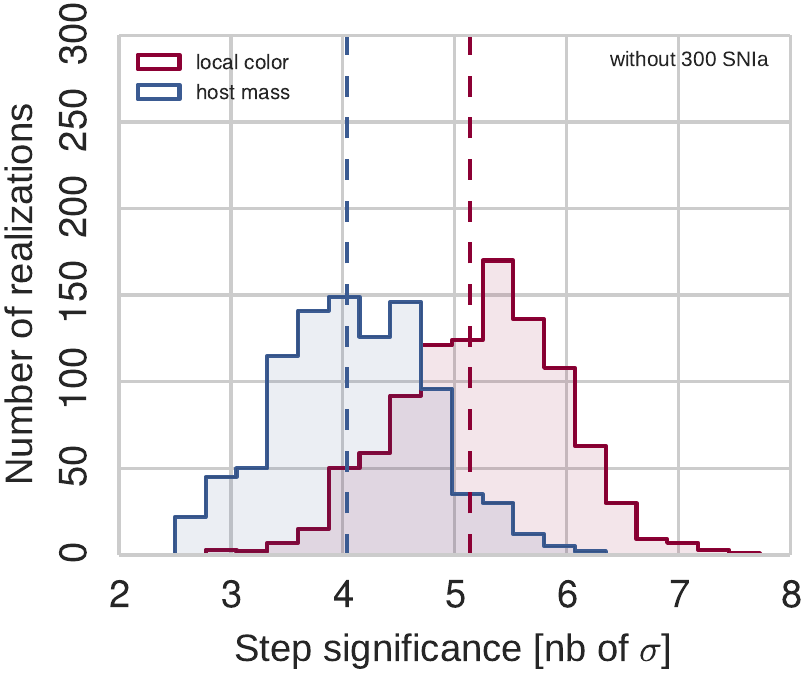}\\    
   \end{tabular}
  \end{center}
  \caption{Histograms of the step significance of the local $U-V$ color (dark red) and stellar mass  of the host (dark blue) used as a third standardization parameter in the cosmological fit. Correlations are estimated by removing 100 (top panel), 200 (middle panel), and 300 (bottom panel) random supernov\ae\ from the initial sample of events and repeating the process a thousand times. The expected significances, assuming that it scales with $\sqrt{\mathcal{N}}$ ($\mathcal{N}$ is the total number of supernov\ae\ in the sample) are shown as the dashed lines with the same color code.
  }
  \label{fig:test_removing_sn}
\end{figure}

As a reliability test, we aimed to verify that the signal we observed with local color and stellar mass of the host was not due to the presence of a subsample of Type Ia supernov\ae\ that drives this effect and pushes toward high significance. In order to test this, we randomly selected 100, 200, and 300 SNIa and removed them from the sample. Then we computed the significance of a third variable in the cosmological fit with a magnitude step $\Delta M_B$. We repeated the random process a thousand times in total for the three configurations and the two variables. The corresponding significance histograms are shown in Fig.~\ref{fig:test_removing_sn}.

First, the observed signal that shows a correlation between the
Hubble diagram residuals and local color (or stellar mass of
the host) is still present when part of the initial sample is missing, thus proving that the results are not dependent on the number of events and are reliable regarding the random disappearance of some supernov\ae. Moreover, the histograms are compatible with the significance we expect with a sample containing 100, 200, and 300 fewer events (dashed lines in Fig.~\ref{fig:test_removing_sn}). These expected values were estimated assuming that the significance scales as the square root of the number of events in the sample. Considering that we obtained a $7\sigma$ confidence level for the basic case with local color as a third standardization parameter using 666 events, we then expect $6.4\sigma$, $5.9\sigma,$ and $5.2\sigma$ with only 566, 466, and 366 events, respectively, provided that the change in significance is only due to the decrease in statistics and not to another effect that decreases the local color signal itself.
Since the histograms are consistent with the expected significances, we are confident that the signal we obtain is reliable and does not disappear when part of the sample is removed. We also note that within each subsample, the local color step is on average more significant than the mass step. Removing very many SNIa (half of the sample) enlarges the overlap between step significances for local color and stellar mass of the host.

\subsection{Local environment from SED fitting in magnitudes}
As explained in Section~\ref{subsec:sed_fitting}, we fit a series of template galaxy SEDs using interpolated fluxes from \texttt{EGALITE} codes in order to obtain the local $U-V$ rest-frame color.
This ensures higher precision on colors, since objects with low surface brightness are easily handled.
On the other hand, global properties such as color and stellar masses of the host are derived with \texttt{EGALITE} and \texttt{PEGASE} starting from observed magnitudes. For the purpose of a consistency check, we used \texttt{EGALITE} as an SED fitting code, taking as inputs observed magnitudes and estimating local galactic properties by finding the best-fit spectrum (through a $\chi^2$ minimization).

%\begin{equation}
%\mathcal{F} = f\times e^{-0.4\mathrm{ln}(10)zp}.
%\end{equation}

After verifying that the local $U-V$ colors were consistent with those obtained when the SED fit was performed using observed fluxes, we built a new analysis sample. We discarded from our basic sample the few events for which the local measurement error was below the limit defined in Section~\ref{subsec:cuts} for local colors that were directly estimated by interpolating fluxes, but above the limit when magnitudes were used. Thus we obtain in Table~\ref{tab:cuts_mag} the equivalent of Table~\ref{tab:cuts}, using the same definition for the selection requirements.
%Numbers are represented in Table~\ref{tab:cuts_mag}.

We then performed the same cosmological fit using the local $U-V$ color as a third variable. We find that its significance is $6.9\sigma$ with $\Delta M_B=-0.090\pm0.013$ ($5.6\sigma$ for the stellar mass), which is compatible with the results we obtained by interpolating fluxes to fit local galactic properties.
Furthermore, using the SED fitting 
code considering magnitudes, we also computed the local colors estimated
inside 2 kpc for the SNLS and SDSS surveys and 1.5 kpc for the SDSS (see details in Section~\ref{subsec:more_local}). We obtained a similar significance of $6.5\sigma$ (with $\Delta M_B=-0.081\pm0.013$) for the local color step.

This shows that using a flux interpolator to derive local environment effects around the SNIa location is a comparable approach to the approach that uses observed magnitudes as inputs of the estimation of local galactic properties. Both methods yield similar results, although fitting in flux allows including the uncertainties better.

\begin{table}
\centering
\begin{tabular}{ c | c c c | c }
Requirement
 & Nearby
 & SDSS
 & SNLS
 & All\\
\hline\hline
Available $\mathcal{M}$
 & 96/96
 & 389/389
 & 345/397
 & 830/882 \\
$+\ \sigma_{\mathrm{log}_{10}\mathcal{M}}<0.12$
 & 88/96
 & 338/389
 & 309/345
 & 735/830 \\
$+\ \sigma_{C_{\mathrm{L}}}<0.12$
 & 86/88
 & 285/338
 & 288/309
 & 659/735 \\
\end{tabular}
\caption{Selection requirements when we use \texttt{EGALITE} as an SED fitting code taking observed magnitudes into account. For each survey category, we show the number of remaining supernov\ae\ together with the total number of available supernov\ae.}
\label{tab:cuts_mag}
\end{table}

\subsection{Effect of calibration uncertainties, blinding, and cosmological assumptions}
\label{subsec:cosmo_dependence}
%Since we measure photometry with the same optical bands of the same instruments, we do not expect a major influence of relative calibration of flux measurements on the significance of the mass step or the color step. We verified that adding a covariance matrix taking into account calibration uncertainties in the error budget only corresponds to 4.8\% of the mass step statistical significance, and 8.5\% of the color step significance. Thus, it is negligible for the purpose of this analysis.

Calibration uncertainties are not expected to have a strong influence on the results presented above because the two considered populations (locally red/blue, massive/light host) are present in all redshift bins. All of the information thus comes from the comparison of SNIa in the same redshift bin, that is, brightness measurements conducted with the same instrument and in the same effective filters.

For the same reason, we do not expect an effect of the blinding and of the assumed cosmological model.
In our analysis, and in particular, in the cosmological fit producing Hubble diagram residuals corrected for a third variable in addition to stretch and color, we assumed a model with a flat Universe including evolving dark energy $w(z)$. Even when we are blind to the cosmological parameters, we can still verify whether the color step significance is independent of the cosmological model, as it should be. For a model with time-independent dark energy ($w(z)$ is constant), we obtain an increase in $\Delta M_B$ of 1\%, thus raising the color step significance by 1.1\%. When using the $\Lambda$CDM concordance model with dark energy as a cosmological constant ($w=-1$), the increase in $\Delta M_B$ is also at the 1\% level. The significance of the color step will then be on the same order of magnitude. Assuming the basic cosmological model, variations in $\Delta M_B$ can be found when additional cosmological priors are used. Considering the most recent constraints on dark energy from {\em{Planck}} and BOSS \citep{planck_de,boss_de}, we note an increase in absolute magnitude of 0.6\%, which corresponds to a significance modification of 0.8\%. Therefore, we can conclude that cosmological assumptions do not affect our analysis.

\subsection{Interplay with the selection bias}
Concerns have been raised in \cite{scolnic16} that selection bias  affects supernov\ae\ with different brightnesses differently. Specifically, caution must be exerted when
dividing supernov\ae\ into two groups with {\emph{a priori}} different mean
brightness because selection effects will affect the mean
brightness of each group differently.
In the present case, such an effect is expected because $x_1$ is
strongly correlated to both local color and stellar mass of the
host. As a
consequence, segregation into two bins according to one of the latter
variables forms two groups with different $x_1$ distributions
and hence different effective brightnesses, which are differently affected by selection bias. 
We computed a worst-case estimate of this effect using a simulation of the SNLS survey, based on the rate measurement of \cite{perrett12},
color and stretch intrinsic distribution measured in \cite{scolnic16}, and assuming a selection function giving the probability of a
supernova of true apparent magnitude $m$ to be selected as $p(m) =
max(0.4 * (0.5 + arctan((24.3 - m) / 0.2)), 0)$. We found that the
difference in selection bias that would affect the perfectly
segregated $x_1 < 0$ and $x_1 > 0$ bins is negligibly small
($2\pm1\cdot10^{-4}$ mag). Therefore, we did not consider correcting
for it.

%In the present case, such an effect is expected, due to $x_1$ being strongly correlated to both local color and galaxy stellar mass (among other things). The magnitude of the effect can be easily determined from a simulation creating a random SNLS-like survey and introducing selection biases for 10000 realizations of the sample. When splitting the SNIa sample into two groups of high and low brightness, we measure in total an effect of $2\times10^{-4}$ mag at $1.5\sigma$ confidence level. It is thus perfectly negligible, so we did not consider correcting for it.

\subsection{Inclusion of SNIa with faint hosts}

\begin{table}
\centering
\begin{tabular}{ c | c c c | c }
Requirement
 & Nearby
 & SDSS
 & SNLS
 & All\\
\hline\hline
$\sigma_{C_{\mathrm{L}}}<0.12$
 & 93/96
 & 307/389
 & 337/397
 & 737/882 \\
\end{tabular}
\caption{Unique selection requirement imposed when we include events with faint host galaxies in the sample. For each survey
category, we show the number of remaining supernov\ae\ together with the total number of available supernov\ae.}
\label{tab:check_cuts}
\end{table}

The selection requirements are described in details in Section~\ref{subsec:cuts}. We imposed that a stellar mass needs to be properly measured and that our sample keeps the 90\% best measured  host stellar masses and local $U-V$ rest-frame colors. As another consistency check, we aimed at including more diverse situations in the analysis sample, for which the host is faint and the host properties not easily measurable. However, the local measurement was still possible.
In Table~\ref{tab:check_cuts} we present the numbers in each survey, using a single requirement on the quality of the local color measurement derived through flux interpolation with \texttt{EGALITE}. The new sample includes more than 60 SNIa, which correspond to negative local fluxes or local fluxes close to zero.
With 737 SNIa in total, we performed a cosmological fit using three standardization parameters and found a
significance of $6.6\sigma$ for the local $U-V$ color variable (with $\Delta M_B=-0.083\pm0.013$) and $5.8\sigma$ for the host stellar mass variable ($\Delta M_B=-0.071\pm0.012$). In this configuration where faint hosts are included, we also found that local color is the most significant third parameter.
Without any selection cut (the whole SNIa sample was used), we found $\Delta M_B=-0.075\pm0.012$ with the local $U-V$ color and $\Delta M_B=-0.077\pm0.012$ with the host mass.

\begin{figure}%[!htbp]
  \begin{center}
   \begin{tabular}{c}
    \includegraphics[width=0.9\columnwidth]{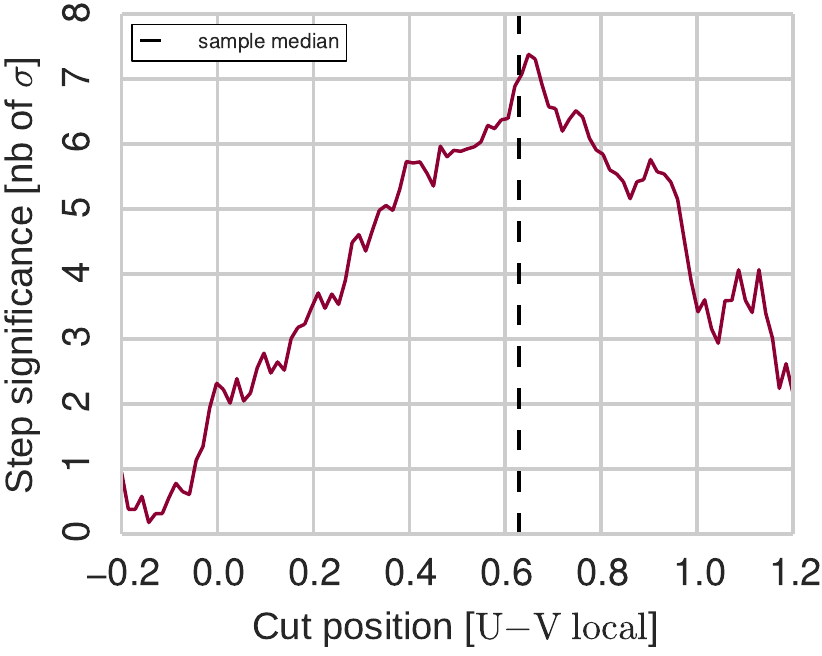} \\
    \includegraphics[width=0.9\columnwidth]{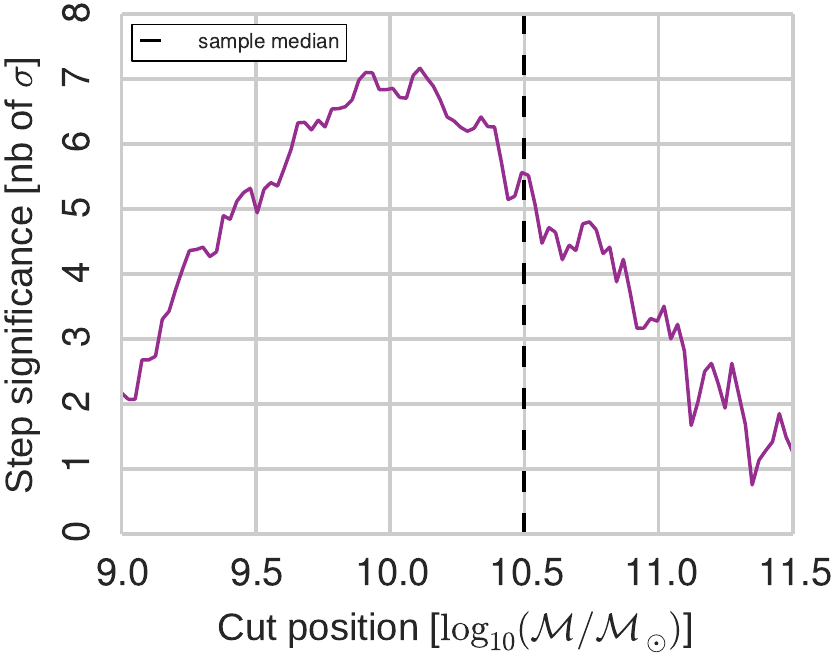}
   \end{tabular}
  \end{center}
  \caption{\textbf{Top:} Evolution of local color step significance as a function of the position of the local $U-V$ color cut. The position of the sample median, chosen for the basic analysis, is shown as the dashed black line. \textbf{Bottom:} Same as the top panel, but for the stellar mass of the host.
}
  \label{fig:cut_dependence}
\end{figure}

\subsection{Look-elsewhere effect}
In the entire analysis, we presented magnitude step significances assuming a bimodal distribution of the third standardization parameter (stellar mass or color of the host, or local color) with a separation corresponding to the median of this distribution. Therefore, we did not aim at maximizing the step significance. When we changed the local color bin separation in order to maximize the significance of the color step, the limit in local color was slightly larger than the median value at $U-V\sim0.65$ with a magnitude step significant at $7.4\sigma$. On the other hand, maximizing the mass step significance led to a mass limit of about $10^{10.1}\mathcal{M}_\odot$, corresponding to a magnitude step of $7.2\sigma$ significance. In this case, the separation does not lie close to the median value, since 31\% of our sample is found below the new mass limit, and 69\% above. This new limit is closer to the mass bin separation of $10^{10}\mathcal{M}_\odot$, chosen in all previous studies on host galaxies. In particular, \cite{Childress13} found an apparent Hubble residual transition between $10^{9.8}$ and $10^{10.4}\mathcal{M}_\odot$. Choosing this mass separation yields comparably significant mass and color steps.
The top and bottom panels of Fig.~\ref{fig:cut_dependence} show the evolution of step significance as a function of the cut position when local $U-V$ colors and host stellar masses are used as a third standardization parameter. The sample median lies close to the maximum local color step significance, whereas the most significant mass step implies a bin separation at $10^{10.1}\mathcal{M}_\odot$.

However, the correlations between the Hubble diagram residuals derived from a three-parameter standardization and our host variables yield additional information. When residuals are corrected for local $U-V$ color, the remaining {\em{maximum}} mass step significance is of $3.8\sigma$ (i.e., the maximum correlation between stellar mass and corrected Hubble diagram residuals using the most significant local step). Conversely, the remaining {\em{maximum}} local color step after mass correction as a third variable is of $5\sigma$ (i.e., the maximum correlation between local $U-V$ color and corrected Hubble diagram residuals using the most significant mass step). Hence, maximizing the significance still shows that scatter in the Hubble diagram is better accounted for
%more information is taken into account
when the supernova brightness is corrected for local environment.
It is also worth noting that the local $U-V$ color does not fully account for the remaining scatter either. A better proxy, or a combination of proxies, linked to properties of the local environment of SNIa, is yet to be found.

\begin{table*}
\centering
\begin{tabular}{c|c|c|c|c|c}
%Test
 & Nb of
 & Local color step
 & Host mass step
 & corr. for local color
 & corr. for mass \\
 Test&SNIa&(mag)&(mag)&(mag)&(mag)\\
\hline\hline
Basis
 & 666
 & $-0.0909\pm0.0130$ ($7.0\sigma$)
 & $-0.0704\pm0.0128$ ($5.5\sigma$)
 & \ldots
 & \ldots \\
\hline
JLA common
 & 589
 & $-0.0759\pm0.0139$ ($5.5\sigma$)
 & $-0.0719\pm0.0134$ ($5.3\sigma$)
 & \ldots
 & \ldots \\
Smaller radius
 & 620
 & $-0.0872\pm0.0143$ ($6.2\sigma$)
 & \ldots
 & \ldots
 & \ldots \\
Correl. a posteriori
 & 666
 & $-0.0811\pm0.0115$ ($7.1\sigma$)
 & $-0.0628\pm0.0112$ ($5.4\sigma$)
 & $-0.0264\pm0.0116$ %($2.3\sigma$)
 & $-0.0526\pm0.0115$ \\%($4.6\sigma$) \\
Linear fit
 & 666
 & $-0.1067\pm0.0458$ ($2.3\sigma$)
 & $-0.0561\pm0.0334$ ($1.7\sigma$)
 & \ldots
 & \ldots \\
Fit with mag.
 & 659
 & $-0.0899\pm0.0130$ ($6.9\sigma$)
 & $-0.0724\pm0.0128$ ($5.6\sigma$)
 & \ldots
 & \ldots \\
With faint hosts
 & 737
 & $-0.0833\pm0.0126$ ($6.6\sigma$)
 & $-0.0706\pm0.0122$ ($5.8\sigma$)
 & \ldots
 & \ldots \\
No selection cuts
 & 882
 & $-0.0745\pm0.0119$ ($6.2\sigma$)
 & $-0.0768\pm0.0116$ ($6.6\sigma$)
 & \ldots
 & \ldots \\
Max. significance
 & 666
 & $-0.0972\pm0.0132$ ($7.4\sigma$)
 & $-0.0944\pm0.0132$ ($7.2\sigma$)
 & $-0.0459\pm0.0121$ %($3.8\sigma$)
 & $-0.0571\pm0.0115$ \\%($5.0\sigma$) \\
Radius 2 kpc
 & 341
 & $-0.0649\pm0.0175$ ($3.7\sigma$)
 & \ldots
 & \ldots
 & \ldots \\
Radius 4 kpc
 & 684
 & $-0.0846\pm0.0131$ ($6.5\sigma$)
 & \ldots
 & \ldots
 & \ldots \\ 
Radius 8 kpc
 & 708
 & $-0.0759\pm0.0128$ ($5.9\sigma$)
 & \ldots
 & \ldots
 & \ldots \\ 
Radius 16 kpc
 & 710
 & $-0.0684\pm0.0127$ ($5.4\sigma$)
 & \ldots
 & \ldots
 & \ldots \\ 
Radius 32 kpc
 & 688
 & $-0.0602\pm0.0127$ ($4.7\sigma$)
 & \ldots
 & \ldots
 & \ldots \\
Radius 64 kpc
 & 487
 & $-0.0327\pm0.0148$ ($2.2\sigma$)
 & \ldots
 & \ldots
 & \ldots \\
\end{tabular}
\caption{Summary of all tests performed throughout the paper with the corresponding number of SNIa in the sample, the local $U-V$ color step, the host stellar mass step, and step values when corrected for one or the other effect.}
\label{tab:all_checks}
\end{table*}

\section{Summary and perspectives}
\label{sec:discussion}
As elaborated in Section~\ref{sec:introduction}, many studies in the literature have found compelling proof of a dependence of Type Ia supernov\ae\ properties on their close or distant environment, concerning a few hundred events at most
that exploded in the nearby Universe. In this work, we measured the photometry of the host galaxy and local fluxes in regions with a radius a 3 kpc for 882 supernov\ae\ that are distributed in a wide redshift range, from $z=0.01$ to $z=1.1$. We consistently measured photometry from SNLS and SDSS images, and gathered additional photometry from the 2MASS catalog when needed, in order to complete our sample of host galaxy properties. On SDSS images, we used a single method to find hosts and define the region where galaxy photometry should be estimated, and we applied the same technique for all supernov\ae\ discovered by SDSS or low-redshift surveys and all field galaxies corresponding to images where these supernov\ae\ are found. With this new set of measurements, we find good consistency with public SDSS galaxy photometry.

As a consequence, we publish here a consistent and large catalog of host properties of supernov\ae\ spanning this redshift range. This catalog contains intercalibrated stellar masses of the hosts, $U-V$ colors of the host galaxies, and the local $U-V$ color estimated with radii of 1.5, 2, and 3 kpc around the location of the supernova explosion. When we compared the {\em{global}} and {\em{local}} properties of our sample of host galaxies, we found 
%as expected that for high redshift supernov\ae\ it is on average equivalent to measure the 
a correlation between the
color of the whole galaxy and the color measured in a more local region. At high, intermediate, and low redshift, a significant fraction of the sample deviates from a null difference between local and global color. We note that SNIa locations situated in the outskirts of their host galaxy tend on average to be locally bluer than their host, and that SNIa exploding regions that lie
close to the galactic center show redder local colors than the host itself. This means that the global properties of host galaxies do not capture all the information, and that local effects that
are due to stellar population evolution need to be taken into account to constrain cosmology with Type Ia supernov\ae, as has
first been claimed in R13 and R15.

The remaining dispersion in the sample luminosity of about 0.15 magnitude after a two-step standardization, although rather low considering that these extremely energetic events occur at different moments of the history of the Universe, needs to be reduced in order to set new competitive constraints on the nature of dark energy. For many years, it was established that Hubble diagram residuals also correlate with SNIa environment. In this work we envisage the possibility that in addition to stretch and color, a third variable related to the close environment of supernov\ae\ must be used in light-curve cosmological fitters. 
After selecting SNIa based on their data quality (light curve and host photometry),
we find that the
local $U-V$ color is well suited to be chosen as a third standardization parameter.
When the sample is split at the median local $U-V$ color, the corresponding magnitude step is significant at $7\sigma$.
Using a global variable instead, such as the stellar mass or the $U-V$ color of the host with the same cut location, leads to a statistically less significant magnitude step.
%and that it is more significant
%than other global variables such as the host stellar mass or the host $U-V$ color. 
When we probed larger galactic regions, the
%significance is of $6.8\sigma$, 
color step effect was still present, although less significant.
This adds evidence that local host galaxy information might be the key parameter for reducing dispersions in the Hubble diagram. %can be reduced including local effects.

In order to verify the reliability of the observed signal,
we performed a variety of tests that are summarized in Table~\ref{tab:all_checks}, and we investigated all possible modifications of our initial choices.
First, we randomly reshuffled part of the sample and randomly removed a significant part of it. In all cases, we found that the local environment and the mass steps were still present. We also used magnitudes instead of interpolating fluxes to derive the properties of the host galaxies and noted no meaningful change in the significance of the third parameter needed for the light-curve standardization. Then, we checked alternative methods that did not rely on the assumption of a bimodal distribution for local color and stellar masses of the host, which led to the same conclusions, and we still obtained a high significance of local environment effects. Moreover, when we corrected distance moduli with a mass step, we still found room for an additional local effect at the $4.6\sigma$ confidence level. The opposite exercise proved that this statement is not reciprocal. 
When the mass or color bin separation was modified in order to maximize the step significance, we found similar effects when the mass separation was chosen far from the median value. However, this other approach also confirmed that more information is taken into account when standardizing with the local color.
Moreover, a posteriori linear fits of color and stellar mass distribution point in the same direction and also highlight the dependence of Hubble diagram residuals on SNIa local environment. 
We also demonstrated that calibration uncertainties, cosmology, and selection cuts do not have a major influence on the conclusions we draw in this work.
Moreover, a study focusing on SNIa spectra from SNLS also mentions an effect on host stellar mass that is seen even stronger using local $U-V$ color \citep{balland17}. Therefore, spectroscopic analyses of the SNIa environment are also very promising for on-going and future surveys.

After the correlation with either the mass or the local color of the host galaxy was taken into account, the correlation with the other parameter remained significant.
This might indicate that neither the mass nor the local color of the host galaxy is the key parameter for modeling supernov\ae\ luminosity in addition to stretch and color. 
While it seems logical that supernova physics depends on local galactic characteristics, it should be emphasized that the mass and the rest-frame color are themselves related to several features of the stellar history of the galaxy (metallicity, past star formation rate, etc.). It therefore needs to be considered that this key standardization parameter, if it exists, is in this sense multi-factorial.

Although this analysis is blind to the actual values of cosmological parameters, using {\em{Planck}} survey priors on the dark energy equation of state, we have access to the expected order of magnitude of the change in $w$ and $\Omega_m$ when correcting for local color. The change is $\Delta w\sim0.01$ and $\Delta\Omega_m\sim0.003$ when compared to a two-parameter standardization, considering a varying equation-of-state parameter. A comparison of the local color correction and the host mass correction yields $\Delta w\sim0.006$ and $\Delta\Omega_m\sim0.002$. This is important enough to be taken into account in current and future surveys, since their expected statistic precision is close to the parameter shift we estimate.
When cosmological constraints using the SNLS five-year sample are unblinded and published, it will be possible to characterize the exact change and improvement in the measurement of the dark energy equation-of-state parameter and the total matter density. 
For the SNLS five-year analysis, the local $U-V$ color should be taken into account in the standardization, using only a selection requirement on the local color measurement quality. The number of SNIa in the SNLS five-year cosmological sample is then expected to be larger than the number published in this work, since we constructed our sample by applying quality cuts for both local color and stellar mass in order to show comparative results.

With ongoing SNIa surveys such as the DES and future surveys such as the LSST, larger SNIa samples will be obtained ($\mathcal{O}(10^4)$ for the LSST, depending on the efficiency of spectroscopic follow-up and of supernova classification), with increasing precision. In order to derive competitive constraints and shed new light on the nature of the dark energy that drives the accelerated expansion of the Universe, it will be crucial to measure and include effects of the local environment of Type Ia supernov\ae: the local $U-V$ color, or any better proxy of the stellar population age; this will provide further insights on the progenitor type.

\section*{Acknowledgements}
This paper is based on observations obtained with MegaPrime/MegaCam, a joint project of CFHT and CEA/DAPNIA, at the Canada-France-Hawaii Telescope (CFHT) which is operated by the National Research Council (NRC) of Canada, the Institut National des Sciences de l'Univers of the Centre National de la Recherche Scientifique (CNRS) of France, and the University of Hawaii.

It is based
on observations obtained at the Gemini Observatory, which is operated
by the Association of Universities for Research in Astronomy,
Inc., under a cooperative agreement with the NSF on behalf of
the Gemini partnership: the National Science Foundation (United
States), the Science and Technology Facilities Council (United
Kingdom), the National Research Council (Canada), CONICYT
(Chile), the Australian Research Council (Australia), Minist\'erio da
Ci\^{e}ncia, Tecnologia e Inovac\~{a}o (Brazil) and Ministerio de Ciencia, Tecnolog\'ia e Innovaci\'on Productiva (Argentina). 
Observations were also obtained with FORS1 and FORS2 at the Very Large Telescope on Cerro Paranal, operated by the European Southern Observatory, Chile (ESO Large Programs 171.A-0486 and 176.A-0589).

This publication makes use of data products from the Two Micron All Sky Survey, which is a joint project of the University of Massachusetts and the Infrared Processing and Analysis Center/California Institute of Technology, funded by the National Aeronautics and Space Administration and the National Science Foundation.

Funding for SDSS-III has been provided by the Alfred P. Sloan Foundation, the Participating Institutions, the National Science Foundation, and the U.S. Department of Energy Office of Science. The SDSS-III web site is \url{http://www.sdss3.org/}.
SDSS-III is managed by the Astrophysical Research Consortium for the Participating Institutions of the SDSS-III Collaboration including the University of Arizona, the Brazilian Participation Group, Brookhaven National Laboratory, Carnegie Mellon University, University of Florida, the French Participation Group, the German Participation Group, Harvard University, the Instituto de Astrofisica de Canarias, the Michigan State/Notre Dame/JINA Participation Group, Johns Hopkins University, Lawrence Berkeley National Laboratory, Max Planck Institute for Astrophysics, Max Planck Institute for Extraterrestrial Physics, New Mexico State University, New York University, Ohio State University, Pennsylvania State University, University of Portsmouth, Princeton University, the Spanish Participation Group, University of Tokyo, University of Utah, Vanderbilt University, University of Virginia, University of Washington, and Yale University.

Parts of this research were conducted by the Australian Research Council Centre of Excellence for All-sky Astrophysics (CAASTRO), through project number CE110001020.

We thank Micka\"{e}l Rigault, Adam Riess, Dan Scolnic, and David Jones for useful discussions.
%__________________________________________________________________
\bibliographystyle{aa}
\bibliography{biblio,bib_host_sn}
%__________________________________________________________________

\end{document}